\documentclass[aps,prl,superscriptaddress,twocolumn,showpacs]{revtex4-1} 

\usepackage[colorlinks=true, linkcolor=blue, citecolor=blue, urlcolor=blue]{hyperref}
\usepackage{amsmath,amssymb,mathrsfs}
\usepackage{latexsym}
\usepackage{graphicx} 
\usepackage{epstopdf}
\usepackage{graphicx,epstopdf,color}
\usepackage{amsfonts}
\usepackage{bm}
\usepackage{bbm}
\usepackage{multirow}

\def\ie{\emph{i.e.},\ }
\def\eg{\emph{e.g.}\ }
\def\ea{\emph{et al.}}

\newcommand{\pd}{{\phantom{\dagger}}}
\newcommand{\ps}{{\phantom{*}}}

\begin{document}

\title{Unconventional superconductivity in the extended Hubbard model:\\
Weak-coupling renormalization group}

\author{Sebastian Wolf}
\affiliation{School of Physics, University of Melbourne, Parkville, VIC 3010, Australia}
\author{Thomas L.\ Schmidt}
\affiliation{Physics and Materials Science Research Unit, University of Luxembourg, L-1511 Luxembourg}
\author{Stephan Rachel}
\affiliation{School of Physics, University of Melbourne, Parkville, VIC 3010, Australia}
 \pagestyle{plain}

\begin{abstract}
We employ the weak-coupling renormalization group approach to study unconventional superconducting phases emerging in the extended, repulsive Hubbard model on paradigmatic two-dimensional lattices. Repulsive interactions usually lead to higher-angular momentum Cooper pairing.
By considering not only longer-ranged hoppings, but also non-local electron-electron interactions, we are able to find superconducting solutions for all irreducible representations on the square and hexagonal lattices, including extended regions of chiral topological superconductivity.
For the square, triangular and honeycomb lattices, we provide detailed superconducting phase diagrams as well as the coupling strengths which quantify the corresponding critical temperatures 
depending on the bandstructure parameters, band filling, and interaction parameters.
We discuss the sensitivity of the method with respect to the numerical resolution of the integration grid and the patching scheme. Eventually we show how to efficiently reach a high numerical accuracy. 
\end{abstract}


\maketitle

\section{Introduction} 
Superconductivity is amongst the oldest known quantum liquids on earth. More than 100 years after its discovery it has maintained its fascination to experimental and theoretical researchers. Most elemental superconductors such as Pb and Hg are well explained within the framework of the celebrated Bardeen--Cooper--Schrieffer (BCS) theory\,\cite{bardeen_theory_1957} where the pairing of electrons into Cooper pairs is due to electron-phonon interaction.
There are, however, several materials that behave differently. In particular, for the important material classes of copper-oxide\,\cite{bednorz_possible_1986} and iron-based\,\cite{kamihara_iron-based_2006} high-temperature superconductors, phonon-mediated pairing is very unlikely. Also topological superconductors, superconducting analogs of quantum Hall-type systems, seem to have an {\it unconventional} pairing mechanism\,\cite{kallin_chiral_2016,sato_topological_2017,mackenzie_superconductivity_2003}. Instead, strong electron correlations which arise from repulsive interactions are believed to be crucial for the formation of Cooper pairs in these systems.

An attractive effective interaction which is needed for the formation of Cooper pairs can be caused by a repulsive interaction, as first pointed out by Kohn and Luttinger\,\cite{kohn_new_1965}. 
Usually, an attractive interaction between electrons yields a superconducting state exhibiting $s$-wave symmetry (\ie angular momentum $\ell=0$), which provides a full superconducting gap and nodes are absent.
Contrary to that, superconducting states stemming from repulsive interactions typically prefer higher angular momentum pairing ($\ell > 0$) symmetries.
These are labeled, for instance, as $p$-, $d$-, $f$-wave symmetric corresponding to $\ell=1,2, 3$, respectively. 
Higher angular momentum pairing causes sign changes in the superconducting gap leading to nodes\,\cite{kohn_new_1965}, which are absent for $s$-wave superconductors.
Quite generally, superconducting states can be fully classified by virtue of the irreducible representations (irreps) of the symmetry group of the underlying lattice\,\cite{annett_symmetry_1990}. That is, the superconducting gap function {\it must} transform according to one of the irreps; and the basis functions of the irreps correspond to $s$-, $p$-, $d$-wave or higher $\ell$ symmetry. These symmetries refer to the ``orbital'' wavefunction in real or momentum space. The total Cooper pair wavefunction, which is always antisymmetric under exchange of electrons, is a product of this orbital and the spin wavefunctions. Since  two electrons are involved, only spin singlet (totally antisymmetric) or spin triplet pairing (totally symmetric) is possible. Correspondingly, the orbital wavefunction must be either symmetric ($\ell=0, 2, 4, \ldots$) with even parity or antisymmetric ($\ell = 1, 3, 5, \ldots$) with odd parity.

Nodes in the superconducting gapfunction correspond to zeros in the condensation energy, \ie they reduce it. Intuitively, one can understand this as the mentioned energy gain is proportional to the number of Cooper pairs or the average size of the gap.
An opportunity to get rid of the nodes naturally presents itself when the superconducting groundstate is degenerate. For the scenario of degenerate gapfunctions, say, $\psi_{1}$ and $\psi_{2}$, one can avoid nodes by forming complex superpositions, $\psi_{1}\pm i\psi_{2}$. The resulting superconducting gap and the condensation energy are node-free. However, the system has to choose one of the chiralities, either  $\psi_{1}+ i\psi_{2}$ or $\psi_{1}- i\psi_{2}$, and by doing so break spontaneously time-reversal symmetry. The resulting time-reversal broken superconductor is chiral\,\cite{kallin_chiral_2016}: it features chiral edge modes circulating around the sample edge and it is characterized by a topological invariant, the first Chern number also known as TKNN invariant\,\cite{thouless_quantized_1982} (which reveals the mentioned connection to the quantum Hall effect). This exotic state is referred to as topological  superconductivity.

Traditionally, a superconductor is considered to be ``topological'' when it has the ability to trap Majorana zero modes at its vortex cores. For instance, part of the activities to understand the superfluid phases of He-3 are motivated by this idea. It turns out that the number of Majorana zero modes is given by the Chern number; thus only odd Chern numbers are of interest since pairs of real Majorana zero modes can recombine into complex Dirac fermions. One can show that odd parity, \ie spin triplet, superconductors possess odd Chern numbers. That is the reason why there is so much interest in identifying the so-called $p+ip$ or $f+if$ superconducting states in a real material. More recently, a second notion of ``topology'' has been established: more than ten years after the discovery of topological insulators, there is also considerable interest in superconductors which are gapped in their bulk but carry gapless edge modes. Indeed, chiral superconductors are the superconducting analogues of topological insulators (or, to be more precise, of Chern / quantum Hall insulators). Both notions of topology are frequently used in the literature. It is also worth emphasizing that the $p+ ip$ superconductor (with Chern number $C=1$) fits into both categories in contrast to the chiral $d+ id$ state ($C=2$). In this paper, we will distinguish between even and odd parity (or, alternatively, spin singlet and triplet) superconductors; but we will refer to all chiral states as topological superconductors, regardless what their parity or angular momentum is.

Here we study three paradigmatic two-dimensional (2D) lattices and investigate their unconventional superconducting ground states caused by local and non-local Coulomb interactions. While there are many works in the literature\,\cite{shankar_renormalization-group_1994,husslein_quantum_1996,zanchi_superconducting_1996,hlubina_phase_1999,zanchi_weakly_2000,halboth_renormalization-group_2000,honerkamp_magnetic_2001,binz_weakly_2003,raghu_superconductivity_2010,raghu_effects_2012,nandkishore_chiral_2012,kiesel_sublattice_2012,cho_band_2013,platt_functional_2013,maharaj_particle-hole_2013,nandkishore_superconductivity_2014,platt_spin-orbit_2016,vucicevic_trilex_2017,cao_chiral_2018}, we are not aware of a systematic investigation of the phase spaces involving both longer-ranged hoppings and longer-ranged repulsive interactions. We restrict ourselves to consider only one orbital per site. Most materials are  described by multiple orbitals, even after ``downfolding'' them to an effective tight-binding model. Nonetheless, as a first guess, the one-orbital scenario might give a reasonable answer. Moreover, there is recent interest in certain adatom systems such as $X$/Ge(111) and $X$/Si(111) ($X$=Sn,Pb)\,\cite{carpinelli_direct_1996,custance_low_2001,brihuega_intrinsic_2005,li_magnetic_2013,hansmann_long-range_2013,adler_correlation-driven_2018,tresca_chiral_2018} with the surface atoms forming a triangular net; Sn/Si(111) has recently even been claimed to show superconductivity at low temperatures\,\cite{ming_realization_2017}. These systems are well-described by one-band Hubbard models. 

In this paper, we use a weak coupling renormalization group (WCRG) approach, developed by Raghu and coworkers\,\cite{raghu_superconductivity_2010,raghu_effects_2012,cho_band_2013} built up on earlier work by Kohn and Luttinger\,\cite{kohn_new_1965}, to study the various superconducting phases that may arise in paradigmatic 2D lattices from repulsive interaction between the electrons.
The paper is organized as follows: in the next section, we give a thorough description of the WCRG method. Then we study the  superconducting phase diagrams for the square, triangular, and honeycomb lattices as paradigms for 2D systems. In the subsequent section, we will discuss how the numerical resolution influences the groundstates found within the WCRG method. In particular, we resolve some ambiguities reported in the litreature. This leads us to discuss how to efficiently reach a high numerical accuracy. The paper ends with a Conclusion.

%
%
\section{Method}\label{sec:method}

We use the WCRG approach to calculate the superconducting instabilities that arise in the Hubbard-model with repulsive interactions. The Hamiltonian of an arbitrary 2D lattice without spin-orbit coupling is given by
 \begin{align}
 \label{eq:hubbard_hamiltonian}
  H&=H_{0}+H_{\text{int}}, \\
  H_{0}&=-\sum_{n=1}^{n_r}t_{n}\sum_{\left<i,j\right>_{n}}\sum_{\sigma}c_{i\sigma}^{\dagger}c_{j\sigma}^{\pd}\ ,\\
  H_{\text{int}}&=\sum_{m=0}^{m_r}U_{m}\sum_{\left<i,j\right>_{m}}\sum_{\sigma,\sigma'}c_{i,\sigma}^{\dagger}c_{j,\sigma'}^{\dagger}c_{j,\sigma'}^{\pd}c_{i,\sigma}^{\pd},
 \end{align}
where $t_{n}$ and $U_{n}$ denote the $n$-th nearest-neighbor hopping and Coulomb interaction amplitudes, respectively, and $U_0$ the local Hubbard interaction.
$c_{i,\sigma}^{\dagger}$ is the creation operator of an electron on site $i$ with spin $\sigma$ and $\left<i,j\right>_{n}$ are $n$-th nearest neighbor sites. $n_r$ is the number of unit cells and $m_r$ the range of non-local interactions.

The bandstructure $E(n,\vec{k})$ is obtained by diagonalizing the Bloch matrix $h(\vec{k})$ of the non-interacting system $H_{0}$:
\begin{align}\label{def:u_nj}
 E(n,\vec{k})=&-\sum_{i,j}u^{*}_{i,n}(\vec{k})h_{ji}(\vec{k})u_{n,j}(\vec{k})\\
 \label{eq:general_disp_rel}
 =&-\sum_{\nu=1}^{n_r}t_{\nu}\varepsilon_{n,\nu}(\vec{k}),
\end{align}
where $u_{n,i}(\vec{k})$ is the $i$-th element of the $n$-th eigenvector of $h(\vec{k})$ and $n$ is the band-index. 
In the following, we will drop $n$ for the case of one-band models (square and triangular lattices).
The amplitude of the $m$-th neighbor interaction in momentum space, $V_{m}$, can be conveniently written as
\begin{align}\label{eq:Vn}
 V_{m}&(n_{1},\vec{k}_{1};n_{2},\vec{k}_{2};n_{3},\vec{k}_{2}-\vec{q};n_{4},\vec{k}_{1}+\vec{q})=U_{m}\sum_{i,j}h_{m,ij}(\vec{q})\nonumber\\
 &\times u_{n_{1},i}^{*}(\vec{k}_{1})u_{n_{2},j}^{*}(\vec{k}_{2})u_{n_{3},j}^{\ps}(\vec{k}_{2}-\vec{q})u_{n_{4},i}^{\ps}(\vec{k}_{1}+\vec{q}),
\end{align}
where we used $h(\vec{k})=\sum_{m}t_{m}h_{m}(\vec{k})$. For the case of a single orbital per site $h_{0}(\vec{k})$ is the unit matrix and in the one-band case \eqref{eq:Vn} simplifies to
\begin{equation}\label{eq:Vn_singleband}
 V_{m}(\vec{q})=U_{m}\varepsilon_{m}(\vec{q}).
\end{equation}

In this paper, we only consider onsite and nearest-neighbor interactions with amplitudes $U_{0}$ and $U_{1}$, respectively. For the sake of clarity, we will neglect $U_{1}$ for the time being and discuss its implication later.

We define an orbital factor $M$ as
\begin{align}\label{newM-factor}
 M&(n_{1},\vec{k}_{1};n_{2},\vec{k}_{2};n_{3},\vec{k}_{3};n_{4},\vec{k}_{4})=\\
 &=\sum_{i}u_{n_{1},i}^{*}(\vec{k}_{1})u_{n_{2},i}^{*}(\vec{k}_{2})u_{n_{3},i}^{\ps}(\vec{k}_{3})u_{n_{4},i}^{\ps}(\vec{k}_{4}).\nonumber
\end{align}
For $V_{0}$ and only a single orbital per site we can thus write
\begin{align}
 V_{0}&(n_{1},\vec{k}_{1};n_{2},\vec{k}_{2};n_{3},\vec{k}_{2}-\vec{q};n_{4},\vec{k}_{1}+\vec{q})=\\
 &=U_{0}\,M(n_{1},\vec{k}_{1};n_{2},\vec{k}_{2};n_{3},\vec{k}_{2}-\vec{q};n_{4},\vec{k}_{1}+\vec{q}).
\end{align}

Following the pioneering work of Kohn and Luttinger\,\cite{kohn_new_1965}, the only relevant instability in a weakly interacting fermionic system is superconductivity, \ie the formation of Cooper pairs. Thus we consider only two-particle scattering processes. Since we do not take spin-orbit coupling into account, the spin of the particles is conserved during these scattering processes. There are only two distinct spin configurations: scattering of particles with equal or opposite spin corresponding to $S=1$ and $S=0$ superconductivity, respectively, where $S$ denotes the total spin of the Cooper pair. Thus we will omit the spin index in the following.

In the remainder of this section, we discuss the method in detail including a derivation of the important equations. The reader who is only interested in the physical results may skip this part and continue reading in the next section (``Results'') where at the beginning the main aspects of the method are briefly summarized.

\begin{figure}[t]
 \centering
 \begin{tabular}{cc}
  \includegraphics[width=0.33\columnwidth]{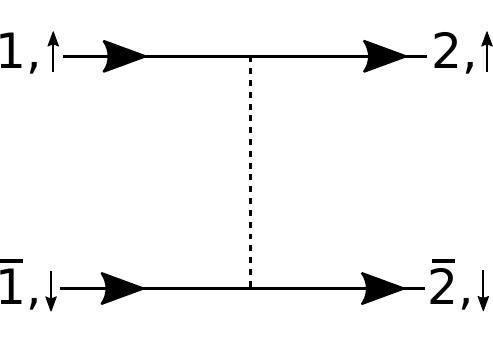} & \hspace{8mm}
  \includegraphics[width=0.33\columnwidth]{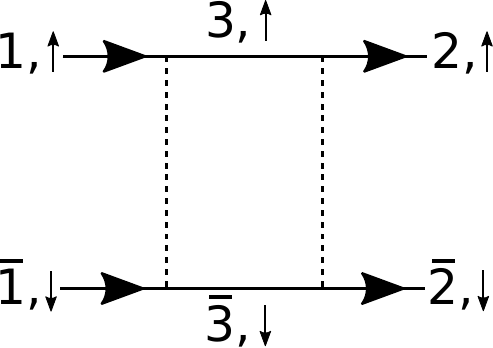} \\[1mm]
  (1) & \hspace{8mm} (2a) \\[4mm]
  \includegraphics[width=0.33\columnwidth]{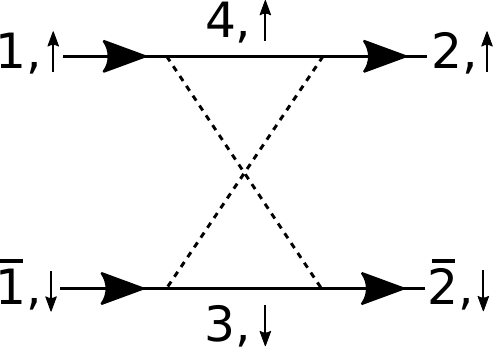} & \hspace{8mm}
  \includegraphics[width=0.33\columnwidth]{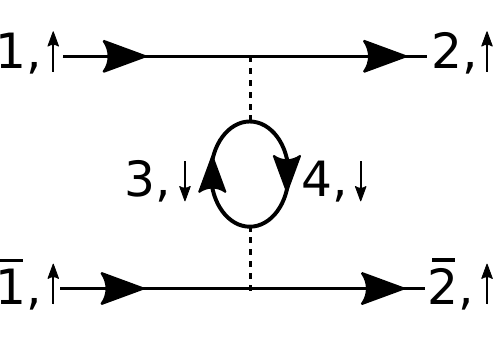} \\[1mm]
  (2b) & \hspace{8mm} (2c)
 \end{tabular}
\caption{\label{fig:feynman_diagrams}Relevant Feynman diagrams for on-site interaction in the Hubbard Hamiltonian \eqref{eq:hubbard_hamiltonian}, for the cases considered in this paper, when treated in the weak coupling limit. 
Diagrams 1, 2a, and 2b refer to the singlet channel and diagram 2c to the triplet channel. Solid lines represent electron propagators and dashed lines interactions. For clarity, the spins are explicitly given as up and down arrows.}
\end{figure}

We follow the discussion of Raghu \ea\,\cite{raghu_superconductivity_2010}. 
The WCRG approach is limited to only the description of Cooper pairing and to scattering processes with vanishing total momentum, \ie the scattering of particles with opposite momenta $\vec{k}_{1}$ and $-\vec{k}_{1}$ in band $n_{1}$ into states with momenta $\vec{k}_{2}$ and $-\vec{k}_{2}$ in band $n_{2}$. These processes are described by the two-particle vertex function $\Gamma(n_{2},\vec{k}_{2};n_{1},\vec{k}_{1})$, which is shown in the form of Feynman diagrams in Fig.\,\ref{fig:feynman_diagrams}. The diagrams can be expressed in terms of the static particle-hole susceptibility (Lindhard function), $\chi_{\rm ph}(\vec{k})$, given by
\begin{align}
 \chi_{\rm ph}(\vec{k}_{1})=-\sum_{n_{1},n_{2}}\int\frac{\text{d}^{2}k_{2}}{(2\pi)^{2}}X_{\rm ph}(n_{1},\vec{k}_{1}+\vec{k}_{2};n_{2},\vec{k}_{2}),
\end{align}
with
\begin{align}
\label{eq:chi_integrand}
 X_{\rm ph}(n_{1},\vec{k}_{1};n_{2},\vec{k}_{2})=\frac{f(E(n_{1},\vec{k}_{1}))-f(E(n_{2},\vec{k}_{2}))}{E(n_{1},\vec{k}_{1})-E(n_{2},\vec{k}_{2})}\ ,
\end{align}
where $f(E)$ is the Fermi distribution, and the static particle-particle susceptibility, $\chi_{\rm pp}(\Omega_{0})$, given by
\begin{align}
 \chi_{\rm pp}(\Omega_{0})=\sum_{n_{1}}\int_{|E|>\Omega_{0}}\frac{\text{d}^{2}k_{1}}{(2\pi)^{2}}X_{\rm pp}(n_{1},\vec{k}_{1})
\end{align}
with
\begin{align}
 X_{\rm pp}(n_{1},\vec{k}_{1})=\frac{1-2f(E(n_{1},\vec{k}_{1}))}{-2E(n_{1},\vec{k}_{1})}\ .
\end{align}
Note that the cut-off $\Omega_{0}$ is needed in the integration, as it is otherwise logarithmically divergent in the limit of zero temperature, $T\rightarrow0$, which indicates the appearance of an instability in the particle-particle channel (Cooper instability). Usually, for conventional superconductivity, the cut-off is chosen according to the phonon bandwidth, \ie the typical energy scale of the interaction that drives the Cooper pairing mechanism, and which is linked to the critical temperature $T_{c}$. This scheme, however, is only valid for attractive interactions. Since we consider a repulsive interaction in this work, we cannot use it. As we will show in the following, using the WCRG approach, we will find the appropriate cut-off $\Omega^{*}$, which is linked to an attractive effective interaction $V_{\text{eff}}$, which arises within the renormalization process.

For the remainder of the paper, we will use the following short notation of the arguments,
\begin{align}
 &1\equiv n_{1},\vec{k}_{1},\hspace{3mm} \bar{1}\equiv n_{1},-\vec{k}_{1}.
\end{align}
Thus we can write for the two-particle vertex function
\begin{equation}
 \Gamma(n_{2},\vec{k}_{2};n_{1},\vec{k}_{1})=:\Gamma(2,1).
\end{equation}
The two-particle vertex function can be split into the singlet and triplet channels by symmetrizing and antisymmetrizing $\Gamma$ in momentum space, respectively:
\begin{align}\label{anti-symmetrize}
 \Gamma^{s,t}(2,1)=\frac{1}{2}\big[\Gamma(2,1)\pm\Gamma(\bar{2},1)\big].
\end{align}
%
%
%
Here, the label $s$ ($t$) refers to the symmetric (antisymmetric) singlet (triplet) channel. Note that it is not the singlet that is symmetric but the $k$-dependent wavefunction such that the total Cooper pair wave function remains antisymmetric.
The weak coupling limit offers the advantage that we are able to expand the vertex function in powers of the on-site interaction, $U_{0}$, which yields
\begin{equation}
 \Gamma(2,1)=\sum_{n=1}^{\infty}U_{0}^{n}\Gamma^{(n)}(2,1).
\end{equation}
Here, $n$ denotes the number of interaction lines (dashed) in the corresponding Feynman diagrams of $\Gamma^{(n)}$, shown in Fig.\,\ref{fig:feynman_diagrams}, which are given by
\begin{align}
\label{eq:Gamma_1}
 \Gamma^{(1)}(2,1)&=M(2,\bar{2},\bar{1},1), \\
\label{eq:Gamma_2a}
 \Gamma^{(2a)}(2,1)&=-\int_{3}X_{\rm pp}(3)M(2,\bar{2},\bar{3},3)M(3,\bar{3},\bar{1},1),\\
 \Gamma^{(2b)}(2,1)&=-\int_{3'}X_{\rm ph}(3,4)M(2,3,\bar{1},4)M(4,\bar{2},3,1)
\label{eq:Gamma_2b}\\
 \Gamma^{(2c)}(2,1)&=\int_{3'}X_{\rm ph}(3,4)M(2,4,3,1)M(3,\bar{2},\bar{1},4)
\label{eq:Gamma_2c}
\end{align}
where we introduced the short notations for the integrals:
\begin{align}
\label{eq:short_int_1}
 \int_{3}\equiv\sum_{n_{3}}\int\frac{\text{d}k_{3}^{2}}{(2\pi)^{2}},\hspace{2mm} \int_{3'}\equiv\sum_{n_{4}}\int_{3}.
\end{align}


Momentum conservation yields $\vec{k}_{4}=\vec{k}_{1}+\vec{k}_{2}+\vec{k}_{3}$ for $\Gamma^{(2b)}$ and $\vec{k}_{4}=\vec{k}_{1}-\vec{k}_{2}+\vec{k}_{3}$ for $\Gamma^{(2c)}$. Note that since we only considered on-site interaction so far, the interaction lines in the diagrams can only connect electron lines of opposite spin.

\begin{figure}[t]
 \centering
 \includegraphics[width=0.35\columnwidth]{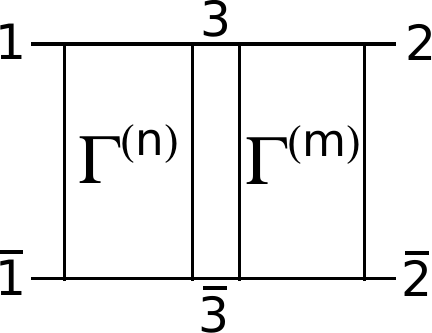}
\caption{\label{fig:feynman_gamma-gamma}General Feynman diagram that produces an additional logarithmic divergence, independent of the behavior of $\Gamma^{(n)}$ and $\Gamma^{(m)}$.}
\end{figure}

It is important to note that all diagrams containing $X_{\rm pp}(n,\vec k)$ in their integrand show a logarithmic divergence. As discussed before, this is directly connected to the respective Feynman diagram being of the form as shown in Fig.\,\ref{fig:feynman_gamma-gamma}, 
which can formally be written as
\begin{align}
\label{eq:concatenation}
 \Gamma(2,1) &= \int_{3}\Gamma^{(m)}(2,3)X_{\rm pp}(3)\Gamma^{(n)}(3,1)\\
 &=: \, \Gamma^{(m)}(2,3)*\Gamma^{(n)}(3,1).
\end{align}
However, if a vertex function cannot be written as the convolution of two vertex functions with $X_{\rm pp}$, then it must be divergence-free. As examples, let us consider the diagrams shown in Fig.\,\ref{fig:feynman_diagrams}: diagram 1 is not of the form shown in Fig.\,\ref{fig:feynman_gamma-gamma} and thus divergence-free. Diagram 2a, on the other side, can be obtained by gluing together two $\Gamma^{(1)}$ diagrams and writing according to \eqref{eq:concatenation}, $\Gamma^{(2a)} = \Gamma^{(1)}*\Gamma^{(1)}$. It produces a logarithmic divergence. Diagrams 2b and 2c are not of the form $\Gamma^{(m)}*\Gamma^{(n)}$ and also divergence-free.
In the following, we use the convention that we add a tilde on top of $\Gamma$ if the vertex function is divergence-free. For instance, $\Gamma^{(1)}\equiv \tilde\Gamma^{(1)}$, $\Gamma^{(2b)}\equiv \tilde\Gamma^{(2b)}$, $\Gamma^{(2c)}\equiv \tilde\Gamma^{(2c)}$ etc.
From this discussion we learn that we can build the full vertex function $\Gamma(2,1)$ just from non-divergent parts, $\tilde{\Gamma}(2,1)$, by a convolution with $X_{\rm pp}$, which yields
\begin{align}
 \Gamma=&\sum_{n=1}^{\infty}U_{0}^{n}\tilde{\Gamma}^{(n)}+\sum_{m=1}^{\infty}U_{0}^{m}\tilde{\Gamma}^{(m)}*\sum_{n=1}^{\infty}U_{0}^{n}\tilde{\Gamma}^{(n)}+\dots\\
 \label{eq:dyson_gamma}
 =&\,\tilde{\Gamma}+\tilde{\Gamma}*\tilde{\Gamma}+\tilde{\Gamma}*\tilde{\Gamma}*\tilde{\Gamma}+\dots=\tilde{\Gamma}+\tilde{\Gamma}*\Gamma.
\end{align}
From here, we can obtain the RG flow equation by evaluating the integral in Eq.\,(\ref{eq:concatenation}):
\begin{align}
\label{eq:conc_integral_1}
 &\int_{3}\tilde{\Gamma}^{(m)}(2,3)X_{\rm pp}(3)\tilde{\Gamma}^{(n)}(3,1)\\
 &=\int_{\hat{3}}\int_{\xi_{n_{3}}^{-}}^{\xi_{n_{3}}^{+}}\text{d}\xi_{n_{3}}\frac{\rho(\xi_{n_{3}})}{\rho_{n_{3}}}\frac{1-2f(\xi_{n_{3}})}{-2\xi_{n_{e}}}\tilde{\Gamma}^{(m)}(2,3)\tilde{\Gamma}^{(n)}(3,1),\nonumber
\end{align}
where $\xi_{n_{i}}^{-(+)}$ denote the bottom (top) of the $n_{i}$-th band with respect to the Fermi level and $\rho_{n_{i}}$ the density of states (DOS) of band $n_{i}$ at the Fermi level. The short notation for the integral is defined as
\begin{equation}
 \int_{\hat{3}}\equiv\sum_{n_{3}}\rho_{n_{3}}\int\frac{\text{d}\hat{k}_{3}}{S_{n_{3},F}}\frac{\bar{v}_{n_{3},F}}{v_{F}(\hat{3})}.
\end{equation}
Here, $v_{F}(\hat{i})$ denotes the Fermi velocity of band $n_{i}$ at momentum $\hat{k}_{i}$, $S_{n_{i},F}$ is the total length of the Fermi surface of band $n_{i}$ and
\begin{equation}
 \frac{1}{\bar{v}_{n_{i},F}}=\int\frac{d\hat{k}_{i}}{S_{n_{i},F}}\frac{1}{v_{F}(\hat{i})}.
\end{equation}
In the limit of zero temperature the Fermi distribution becomes the Heaviside step-function, $\theta(\xi)$, \ie
\begin{equation}
 \lim_{T\rightarrow0}f(\xi)=1-\theta(\xi)=\begin{cases}
           1, & \text{if }\xi<0,\\
           0, & \text{otherwise}.
          \end{cases}
\end{equation}
Substituting this limit in Eq.\,(\ref{eq:conc_integral_1}) lets us split the energy integral into parts over positive and negative energies, and thus introduce a small cut-off $\epsilon_{0}>0$:
\begin{widetext}
\begin{align}
\label{eq:conc_integral_2}
 \int_{3}\tilde{\Gamma}^{(m)}(2,3)X_{\rm pp}(3)\tilde{\Gamma}^{(n)}(3,1)=-\int_{\hat{3}}\Bigg(\int_{\xi_{n_{3}}^{-}}^{-\epsilon_{0}}\frac{\text{d}\xi_{n_{3}}}{2|\xi_{n_{3}}|}\frac{\rho(\xi_{n_{3}})}{\rho_{n_{3}}}+\int_{\epsilon_{0}}^{\xi_{n_{3}}^{+}}\frac{\text{d}\xi_{n_{3}}}{2|\xi_{n_{3}}|}\frac{\rho(\xi_{n_{3}})}{\rho_{n_{3}}}\Bigg)\tilde{\Gamma}^{(m)}(2,3)\tilde{\Gamma}^{(n)}(3,1)\ .
\end{align}
\end{widetext}
Here the cut-off should be chosen sufficiently small such that all quantities that converge in the limit $\xi\rightarrow0$ can be approximated by their values at the Fermi level for $|\xi|\leq\epsilon_{0}$.

The next step is to calculate the change in $\Gamma$ when the cut-off is reduced to $\epsilon$, \ie
\begin{equation}
 \Delta\Gamma=\Gamma_{\epsilon}-\Gamma_{\epsilon_{0}},
\end{equation}
where $(\epsilon_{0}-\epsilon)/\epsilon_{0}\ll1$. Using Eqs.\,(\ref{eq:dyson_gamma}) and (\ref{eq:conc_integral_2}), we have
\begin{align}
 \Delta\Gamma(2,1)=&-\int_{\hat{3}}\int_{\epsilon}^{\epsilon_{0}}\frac{\text{d}\xi_{n_{3}}}{|\xi_{n_{3}}|}\tilde{\Gamma}(2,\hat{3}){\Gamma}(\hat{3},1)\\
 \label{eq.30}=&-\ln\left(\frac{\epsilon_{0}}{\epsilon}\right)\int_{\hat{3}}\tilde{\Gamma}(2,\hat{3}){\Gamma}(\hat{3},1)\\
 \label{eq.31}=&-\ln\left(\frac{\epsilon_{0}}{\epsilon}\right)\int_{\hat{3}}\tilde{\Gamma}(2,\hat{3})\tilde{\Gamma}(\hat{3},1)\\
 &+\ln^{2}\left(\frac{\epsilon_{0}}{\epsilon}\right)\int_{\hat{3}\,\hat{4}}\tilde{\Gamma}(2,\hat{3})\tilde{\Gamma}(\hat{3},\hat{4})\tilde{\Gamma}(\hat{4},1)\nonumber\\
 &-\ln^{3}\left(\frac{\epsilon_{0}}{\epsilon}\right)\int_{\hat{3}\,\hat{4}\,\hat{5}}\tilde{\Gamma}(2,\hat{3})\tilde{\Gamma}(\hat{3},\hat{4})\tilde{\Gamma}(\hat{4},\hat{5})\tilde{\Gamma}(\hat{5},1)\nonumber\\
 &\pm\dots.\nonumber
\end{align}
Note that \eqref{eq.30} and the first line of \eqref{eq.31} differ by the tilde on top of $\Gamma(\hat 3,1)$.
From here we obtain the RG flow equation
\begin{equation}
\label{eq:RG_flow}
 \frac{\partial\Gamma(2,1)}{\partial\ln(\epsilon_{0}/\epsilon)}=\frac{\partial\Delta\Gamma(2,1)}{\partial\ln(\epsilon_{0}/\epsilon)}=-\int_{\hat{3}}\Gamma(2,\hat{3})\Gamma(\hat{3},1).
\end{equation}

Due to the onsite interaction being repulsive, the nodeless $s$-wave solution is suppressed. Thus, we only need to focus on the subspace of nodal superconducting states, \ie those with sign-changing form factors of the gap-function. We achieve that by splitting $\Gamma$ as follows:
\begin{align}
 &\Gamma=\Gamma_{0}+\Gamma_{1}+\Gamma_{01}, \\
 &\Gamma_{0}=U_{0}\tilde{\Gamma}^{(1)}+U_{0}^{2}\tilde{\Gamma}^{(1)}*\tilde{\Gamma}^{(1)}+U_{0}^{3}\tilde{\Gamma}^{(1)}*\tilde{\Gamma}^{(1)}*\tilde{\Gamma}^{(1)}+\dots, \nonumber\\
 &\Gamma_{1}=U_{0}^{2}\tilde{\Gamma}^{(2)}+U_{0}^{3}\tilde{\Gamma}^{(3)}+U_{0}^{4}\left(\tilde{\Gamma}^{(2)}*\tilde{\Gamma}^{(2)}+\tilde{\Gamma}^{(4)}\right)+\dots, \nonumber\\
 &\Gamma_{01}=U_{0}^{3}\left(\tilde{\Gamma}^{(1)}*\tilde{\Gamma}^{(2)}+\tilde{\Gamma}^{(2)}*\tilde{\Gamma}^{(1)}\right)+\dots. \nonumber
\end{align}
Here, $\Gamma_{0}$ is the subspace of the nodeless $s$-wave, $\Gamma_{1}$ the one of nodal superconductivity and orthogonal to $\Gamma_{0}$, and $\Gamma_{01}$ describes the interaction between these two. Since we are in the weak-coupling limit, we take only the lowest relevant order in $U_{0}$. With the nodeless $s$-wave being suppressed, one obtains for the singlet channel
\begin{equation}
 \Gamma^s(2,1)\approx U_{0}^{2}\tilde{\Gamma}^{s,(2)}(2,1)=U_{0}^{2}\Gamma^{(2b)}(2,1)\ 
 \end{equation}
and for the triplet channel
\begin{equation}
\Gamma^t(2,1)\approx U_{0}^{2}\tilde{\Gamma}^{t,(2)}(2,1)=U_{0}^{2}  \Gamma^{(2c)}(2,1)\ .
\end{equation}
The diagram $\Gamma^{(2a)}$ is not present as it is not divergence-free. 

We now transform the integral over the Fermi surface into a sum over discrete points suited to be numerically evaluated. Since we work in the limit $T\to 0$, we can restrict the scattering processes to the ones where the electrons start and end on the Fermi surface. Thus, we can rescale $\Gamma$ with a momentum dependent scaling factor given by
\begin{equation}
\label{eq:gmat}
 g(\hat{2},\hat{1}):=\sqrt{\rho_{n_{2}}\rho_{n_{1}}\frac{\bar{v}_{n_{2},F}\bar{v}_{n_{1},F}}{v_{F}(\hat{2})v_{F}(\hat{1})}\frac{\ell(\hat{2})\ell(\hat{1})}{S_{n_{2},F}S_{n_{1},F}}}\Gamma(\hat{2},\hat{1})\ ,
\end{equation}
where $\ell(\hat{i})$ is the length associated with the discrete Fermi surface point with index $i$. 
We substitute an orthonormal eigensystem of $g$, \ie
\begin{align}
 &\sum_{n_{1}}\sum_{\hat{k}_{1}}g(\hat{2},\hat{1})\psi_{\nu}(\hat{1})=\lambda_{\nu}\psi_{\nu}(\hat{2}),\\
 &\sum_{n_{1}}\sum_{\hat{k}_{1}}\psi_{\nu}(\hat{1})\psi_{\eta}(\hat{1})=\delta_{\nu\eta},
\end{align}
in Eq.\,\eqref{eq:RG_flow}. The RG flow equation then becomes
\begin{equation}
\label{eq:flow_lambda}
 \frac{\partial\lambda_{\nu}}{\partial\ln(\epsilon_{0}/\epsilon)}=-\lambda_{\nu}^{2},
\end{equation}
where, due to the rescaling of $\Gamma$, the eigenvalues $\lambda_{\nu}$ coincide with the eigenvalues of $g$. Thus, the RG flow equation is solved by a simple diagonalization of $g$. Also, we see from Eq.\,(\ref{eq:flow_lambda}) that all $\lambda_{\nu}$ renormalize independently and that Eq.\,\eqref{eq:flow_lambda} is solved by
\begin{equation}
 \lambda_{\nu}=\frac{\lambda_{\nu}^{0}}{1+\lambda_{\nu}^{0}\ln(\epsilon_{0}/\epsilon)},
\end{equation}
where $\lambda_{\nu}^{0}$ is the eigenvalue obtained by using the cut-off $\epsilon_{0}$. Once the cut-off $\epsilon$ is reduced to
\begin{equation}
 \epsilon^{*}=\epsilon_{0}\,e^{-1/\lambda_{\nu}^{0}},
\end{equation}
$\lambda_{\nu}$ diverges. The leading instability is given by the first appearing divergence, \ie by the most negative eigenvalue $\lambda_{\text{min}}$ and $\epsilon^{*}$ is identified with the critical temperature. Note that $\lambda_{\nu}^{0}$, when taken to the appropriate order in $U_{0}$, depends on the initial cut-off, $\epsilon_{0}$, such that it cancels out\,\cite{raghu_superconductivity_2010}, \ie
\begin{equation}
 T_{c}\sim e^{-1/\lambda_{\text{min}}},
\end{equation}
which yields for the effective interaction
\begin{equation}
 V_{\text{eff}}=\frac{\lambda_{\text{min}}}{\rho}.
\end{equation}
That only negative eigenvalues are relevant in the RG flow reflects the general idea that only an attractive effective interaction between electrons can lead to superconductivity. However, we see that this may arise under renormalization, starting from a repulsive onsite interaction in a many-body system in the same way a matrix with only positive entries can have negative eigenvalues.

The eigenvector $\psi_{\text{min}}$ which corresponds to $\lambda_{\text{min}}$ describes the discrete form-factor of the gap function of the resulting superconducting instability along the Fermi surface. 
Superconducting states can be fully characterized by the irreps of the point group associated with the lattice structure\,\cite{annett_symmetry_1990}. That is, $\psi_{\text{min}}$ must transform according to one of the irreps of the relevant point group (see also discussion in the next section).

If there is no dependence on spin other than the selection of the diagrams due to the Pauli exclusion principle (\ie there is no spin-orbit coupling) and only Hubbard interactions within the same orbital are considered, we identify via Eqs.\,(\ref{eq:Gamma_2b}) and~(\ref{eq:Gamma_2c}) the following relation:
\begin{equation}
 \Gamma^{(2b)}(2,1)=-\Gamma^{(2c)}(\bar{2},1)\ .
\end{equation}
That is, all triplet states are already contained in the singlet channel. Symmetrization (anti-symmetrization) according to \eqref{anti-symmetrize} removes the triplet (singlet) states.

The reader should note that the above discussion about what diagrams contribute to what channel is valid for the single-band case discussed in this paper. Multi-orbital scenarios naturally involve additional inter-orbital interactions as well as Hund's coupling. Then the above discussion needs to be extended, \eg the diagram 2c from Fig.\,\ref{fig:feynman_diagrams} contributes to both singlet and triplet channels. Moreover, new additional diagrams must be included\,\cite{cho_band_2013}.

\subsection{Limit $k_{4}\rightarrow k_{3}$}
Special attention has to be taken when handling the integral of $\Gamma$, Eqs.\,(\ref{eq:Gamma_2b}) and (\ref{eq:Gamma_2c}), in the limit of small momentum transfer, $k_{4}\rightarrow k_{3}$, and for $T\rightarrow0$, in which the integrand function of the particle-hole susceptibility, $X_{\rm ph}$, yields for equal band indices, $n_{3}=n_{4}$:
\begin{align}
 \lim_{k_{4}\rightarrow k_{3}}\lim_{T\rightarrow0}X_{\rm ph}(3,4)&= \lim_{k_{4}\rightarrow k_{3}}\frac{\theta(E(4))-\theta(E(3))}{E(4)-E(3)}
  =\delta(E)
\end{align}
where $\delta(E)$ denotes the Dirac delta-distribution. Consequently, Eq.\,(\ref{eq:Gamma_2c}) becomes
\begin{align}
 \lim_{k_{4}\rightarrow k_{3}}\lim_{T\rightarrow0}\Gamma_{n_{3}=n_{4}}^{(2b)}(2,1)=
 \int_{\hat{3}}M(2,\hat{3},\bar{1},\hat{3})M(\hat{3},\bar{2},\hat{3},1).
\end{align}
The importance of this limit lies in the correct numerical handling of the integrand approaching the delta-distribution. In contrast, an incorrect handling might possibly result in
\begin{align}
 &\lim_{k_{4}\rightarrow k_{3}}\lim_{T\rightarrow0}\Gamma_{n_{3}=n_{4}}^{(2b)}(2,1)=0,
\end{align}
producing different (incorrect) effective couplings $V_{\text{eff}}$.

\subsection{Longer range interaction}

We now consider also nonlocal interactions; in particular, we include a nearest neighbor Coulomb repulsion to the Hamiltonian,
\begin{align}
 H_{\text{int}}&=U_{0}\sum_{i}\sum_{\sigma\neq\sigma'}c_{i\sigma}^{\dagger}c_{i\sigma'}^{\dagger}c_{i\sigma'}^{\pd}c_{i\sigma}^{\pd} \\
 &+U_{1}\sum_{\left<i,j\right>}\sum_{\sigma,\sigma'}c_{i\sigma}^{\dagger}c_{j\sigma'}^{\dagger}c_{j\sigma'}^{\pd}c_{i\sigma}^{\pd}.\label{eq:NN-rep}
\end{align}
The changes that arise in the vertex function are given by replacing the on-site interaction $U_{0}$ with the full interaction $U$, and by possibly adding new diagrams. The latter comes from the fact that non-local interactions may also connect particle lines with equal spin in the Feynman diagrams.

Since we only consider terms up to order $U_{0}^{2}$, we take $U_{1}\propto U_{0}^{2}$ for simplicity. As a consequence, the longer range interactions only appear in $\Gamma^{(1)}$ which is linear in $U$. To be precise, this choice implies that in the limit $U_{0}\rightarrow0$, we change $U_{1}$ such that
\begin{equation}
\label{eq:def-alpha}
\alpha =  \frac{U_1 W}{U_0^2}
\end{equation}
is kept constant. Here, $W$ is the bandwidth of the noninteracting system $H_{0}$.
Thus no new diagrams have to be considered. For single-band systems, we have $M=1$. Thus, the relevant terms of the vertex function are then given by
\begin{equation}\label{eq:Gamma-for_U1}
 \Gamma^{s}(2,1)\approx U_{0}^{2}\left(\Gamma^{(2b)}(2,1)+\alpha\frac{\varepsilon_{1}(\vec{k}_{2}-\vec{k}_{1})}{W}\right)
\end{equation}
for the singlet channel. 
Due to the momentum dependence [see Eqs.\,\eqref{eq:Vn} and \eqref{eq:Vn_singleband}], longer ranged interactions are handled slightly differently in these vertex functions. 

Multiple bands arise only due to spin-orbit coupling (not considered in this paper) or multiple orbitals.
The honeycomb lattice with its two-atomic unit cell is the only multiband system we will discuss in this paper.
Using Eq.\,\eqref{eq:Vn}, the orbital factor as defined in Eq.\,\eqref{newM-factor} needs to be modified for the case of longer-ranged interactions,
\begin{align}\label{eq:Mm}
 M_{1}&(1,2,3,4)=\\
 &\sum_{ij}h_{1,ij}(\vec{k}_{2}-\vec{k}_{3})u_{n_{1},i}^{*}(\vec{k}_{1})u_{n_{2},j}^{*}(\vec{k}_{2})u_{n_{3},j}^{\ps}(\vec{k}_{3})u_{n_{4},i}^{\ps}(\vec{k}_{4}).\nonumber
\end{align}
 Thus, we obtain the vertex function as
\begin{equation}
 \Gamma^{s}(2,1)\approx U_{0}^{2}\left( \Gamma^{(2b)}(2,1)+\alpha\frac{M_{1}(2,\bar{2},\bar{1},1)}{W} \right)
\end{equation}
for the singlet channel.
The corresponding expression for the triplet channel, $\Gamma^t(2,1)$, is the same after replacing $\Gamma^{(2b)}$ by $\Gamma^{(2c)}$.
The factor $\alpha$ will be used in the following to control the nearest-neighbor interaction strength.

%

%
%
\section{Results}\label{sec:results}

In the following, we will discuss the extended Hubbard model for three paradigmatic 2D lattices and the corresponding superconducting ground states. All results obtained within the WCRG method are asymptotically correct, \ie for interactions $U_0 \to 0$ which we refer to as ``weak-coupling limit.'' We further emphasize that our analysis is limited to the Cooper channel with vanishing total momentum of the Cooper pairs\,\cite{raghu_superconductivity_2010}; the limitation to superconductivity is reasonable as this is the only weak-coupling instability of the Fermi sea. Other competing instabilities such as magnetism can also occur at weak interactions due to nesting (\eg on the square lattice with $t_2=0$) but this is not a generic feature. Nesting causes these strong-coupling phenomena to emerge at weak couplings, thus we neglect them here. Although the results of the WCRG method are only exact in the limit $U\rightarrow0$, we will argue below (see the section ``Summary of results and discussion'') that they provide a guiding principle for stronger-correlated phases and materials.

What can the method do -- and what cannot? The WCRG only finds superconducting solutions, as mentioned before. Moreover, it {\it always} finds a superconducting instability. In contrast to mean-field theories, where ``you only can get what you put in'', the WCRG method finds the leading instability without any bias towards one superconducting state or another. The underlying lattice determines the point group (in this paper, $D_4$ for the square and $D_6$ for the hexagonal lattices) and superconducting instabilities {\it must} transform according to an irrep of this point group. Details about the irreps will be discussed in the subsections about the specific lattices. How meaningful it is to find a superconducting ground state is dictated by the corresponding coupling strength $V_{\rm eff}$ which is a measure of the critical temperature $T_c$. If $V_{\rm eff}$ is practically zero, so is $T_c$ -- apparently it is not of much interest to discuss superconductors with a $T_c$ of a few $\mu$K. Thus we will in addition to the superconducting ground states also show the effective coupling strengths $V_{\rm eff}$. Moreover, a superconducting instability can be fragile if another superconducting solution is close-by or even almost degenerate. In this case, tiny changes of parameters can cause a phase transition from one superconducting phase to another; more experiment-oriented, small variations in strain, pressure, or temperature might have such effects.
Thus we will also investigate the relative difference between the second-lowest eigenvalue and the lowest eigenvalues $\lambda_{\rm min}$ of the normalized vertex function $g$.

As mentioned before, the WCRG was introduced by Raghu \ea\,\cite{raghu_superconductivity_2010} based on early work of Kohn and Luttinger\,\cite{kohn_new_1965}. Given the interest to study unconventional superconductivity on 2D lattices, some of our results have been derived previously. For the sake of completeness, we will show them nevertheless. Moreover, these results will allow direct comparison with Ref.\,\onlinecite{raghu_superconductivity_2010} (and other papers in the literature) and also serve for benchmarking purposes. In particular, the results for the square lattice for $\alpha=0$ will be vital for the discussion in the section about ``Numerical development and performance analysis''.

Given the longstanding puzzle in understanding the pairing mechanism of the copper-oxide superconductors\,\cite{keimer_quantum_2015,annett_symmetry_1990}, there is enormous interest in $d$-wave superconductors. In the past years there has been, however, a growing interest in topological superconductivity. Besides the fundamental interest in this exotic state of matter, the discovery of topological and Chern insulators\,\cite{kane_$z_2$_2005,bernevig_quantum_2006,konig_quantum_2007}, the electronic cousins of topological superconductors, has renewed the research interest in the past decade. The by far most important aspect of topological superconductors is, however, the ability to trap Majorana zero modes at their vortex cores or at their boundaries -- then referred to as chiral Majorana modes -- to other superconducting or normal-state systems including vacuum. Topological superconductivity naturally occurs due to the following mechanism. Superconducting instabilities could be degenerate -- this might either happen due to fine-tuning of parameters or generically for two-dimensional irreps such as $E$ on the square lattice. Whenever there is a degeneracy, a quantum mechanical system can choose an arbitrary superposition within this subspace, including complex superpositions of the type $\Delta(\vec k) \sim \psi_1 \pm i \psi_2$. With the exception of an ordinary $s$-wave state, all higher-angular momentum superconducting states possess nodes in the superconducting gap functions. These zeros lead to a decrease of the condensation energy which is proportional to $|\Delta(\vec k)|^2$. The aforementioned complex superposition causes, however, a full gap and thus maximizes the condensation energy. That is, for purely energetic reasons, nature might prefer gap functions of the form $\Delta(\vec k) \sim \psi_1 \pm i \psi_2$ if $\psi_1$ and $\psi_2$ are degenerate.
The price nature has to pay for this energy maximization is to choose a chirality (either $\psi_1 + i \psi_2$ or $\psi_1 - i \psi_2$) and to break time-reversal symmetry spontaneously. The result is a chiral, topological superconductor which can be thought of as a quantum Hall or Chern insulator of Bogoliubov quasiparticles. Like quantum Hall insulators, this chiral superconductor is surrounded by one or several chiral edge modes and is characterized by  a topological invariant, the first Chern number or TKNN invariant\,\cite{thouless_quantized_1982}. In addition to this phenomenology which is more or less similar to quantum Hall systems, topological superconductors can host Majorana zero modes which have been claimed to be useful for future topological quantum computer technologies. The WCRG method can only detect the aforementioned degeneracies which are generically present for all two-dimensional irreps; for the square lattice (triangular/hexagonal lattices) these are irrep $E$ (irreps $E_1$ and $E_2$). Using the energetic argument from above, we simply assume that the resulting state will be a chiral topological superconductor. In the following we will discuss the superconducting instabilities for paradigmatic single-band Hubbard models on the square, triangular, and honeycomb lattices. For the remainder of this paper, we set the lattice spacing $a\equiv1$.

%
%
\subsection{Square lattice}

\begin{figure}[b]
\centering
\begin{tabular}{c|c|ccc}
\hline\\[-12pt]
A$_{1}$    &   {\small $x^{2}+y^{2}$}   &   \raisebox{-0.49\height}{\includegraphics[height=40pt]{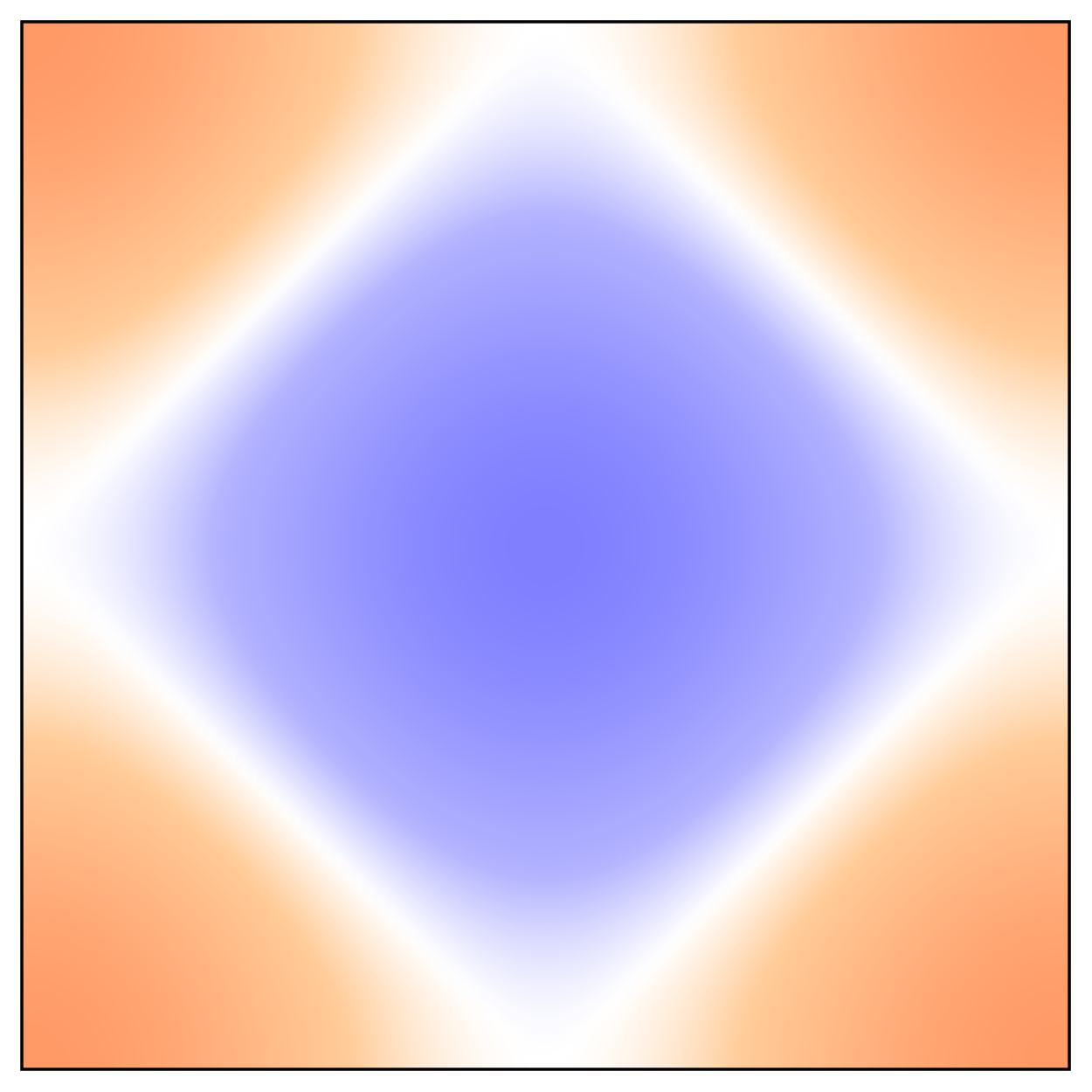}} &   \raisebox{-0.49\height}{\includegraphics[height=40pt]{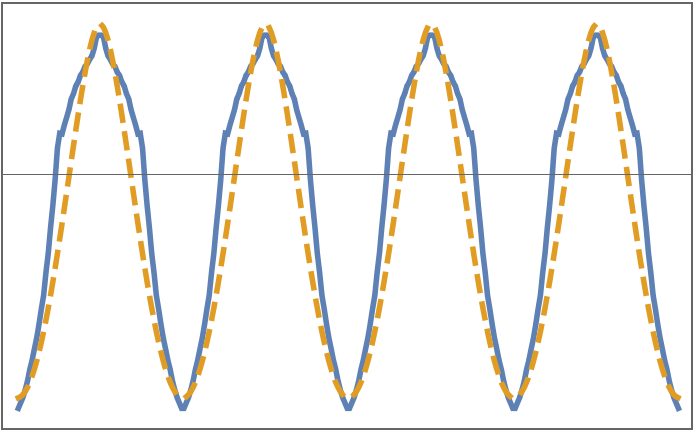}}    &   \raisebox{-0.49\height}{\includegraphics[height=40pt]{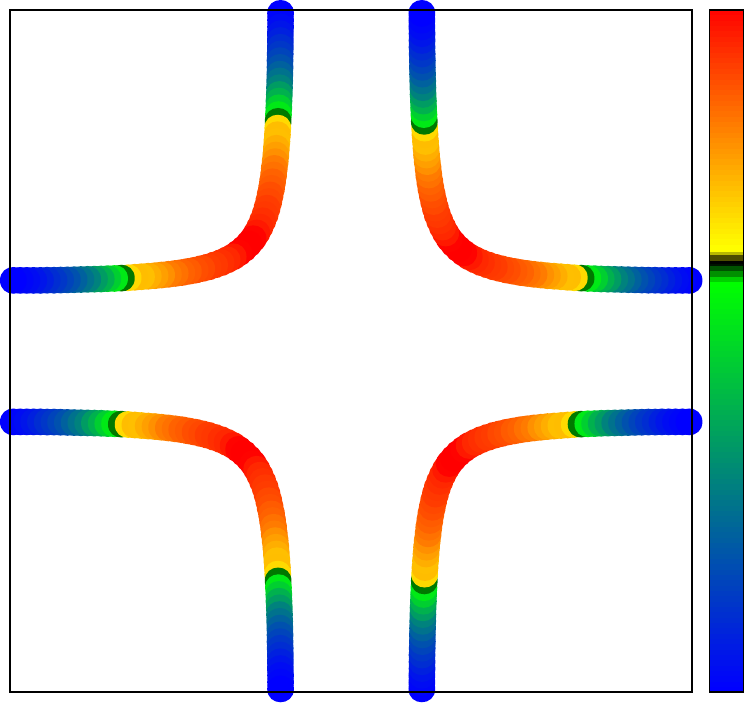}} \\[4pt]
 \hline\\[-12pt]
 A$_{2}$    &   {\small $xy(x^{2}-y^{2})$}   &   \raisebox{-0.49\height}{\includegraphics[height=40pt]{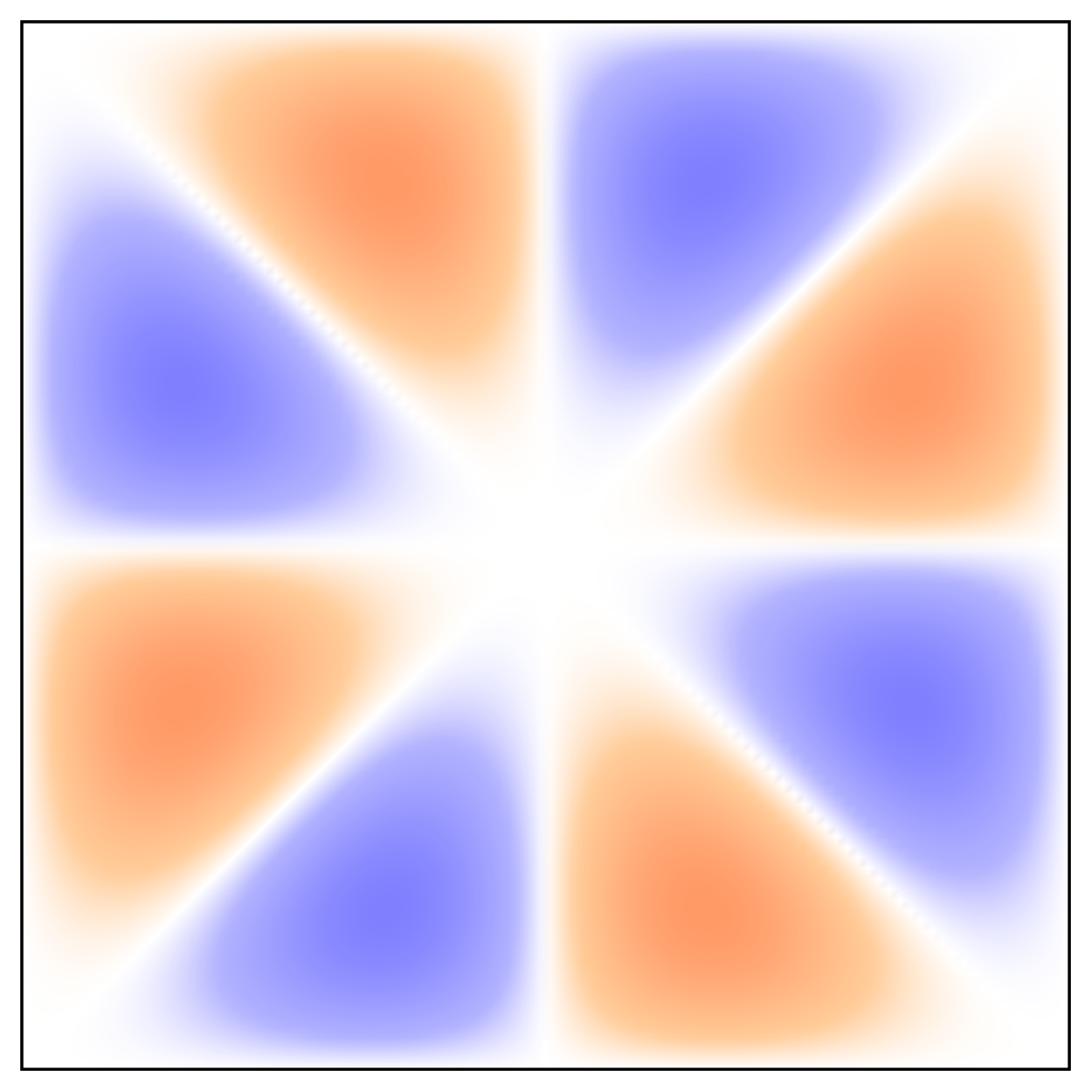}} &   \raisebox{-0.49\height}{\includegraphics[height=40pt]{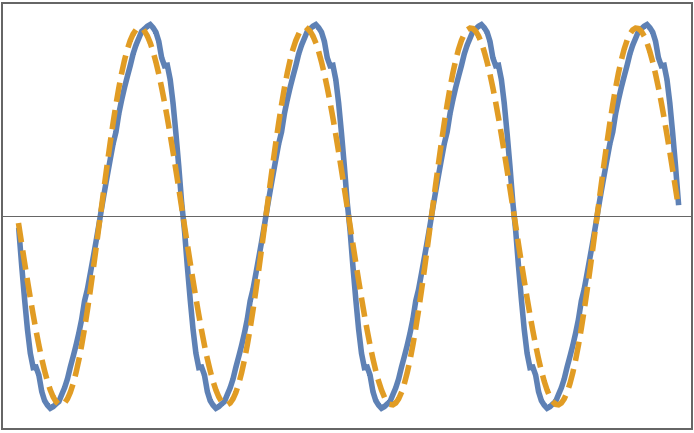}}    &   \raisebox{-0.49\height}{\includegraphics[height=40pt]{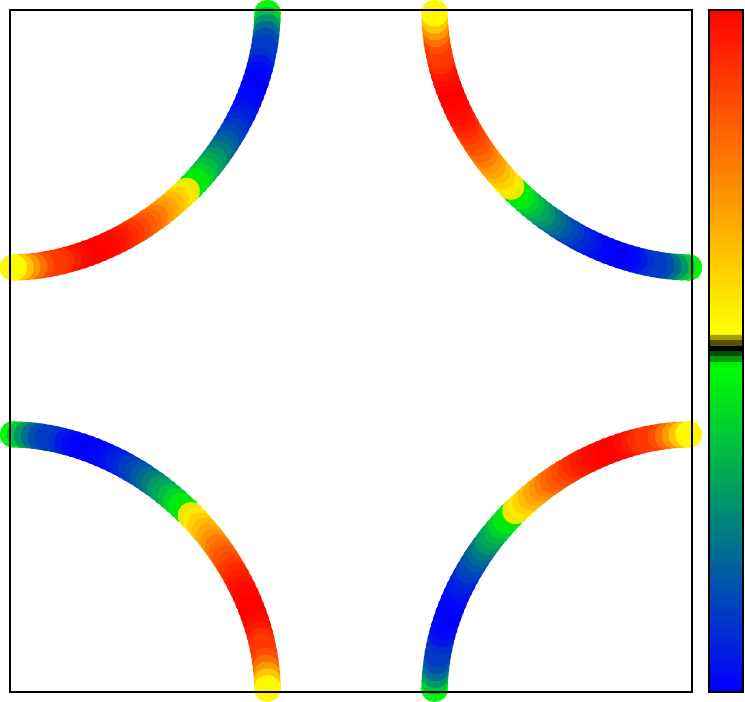}} \\[4pt]
 \hline\\[-12pt]
 B$_{1}$    &   {\small $x^{2}-y^{2}$}   &   \raisebox{-0.49\height}{\includegraphics[height=40pt]{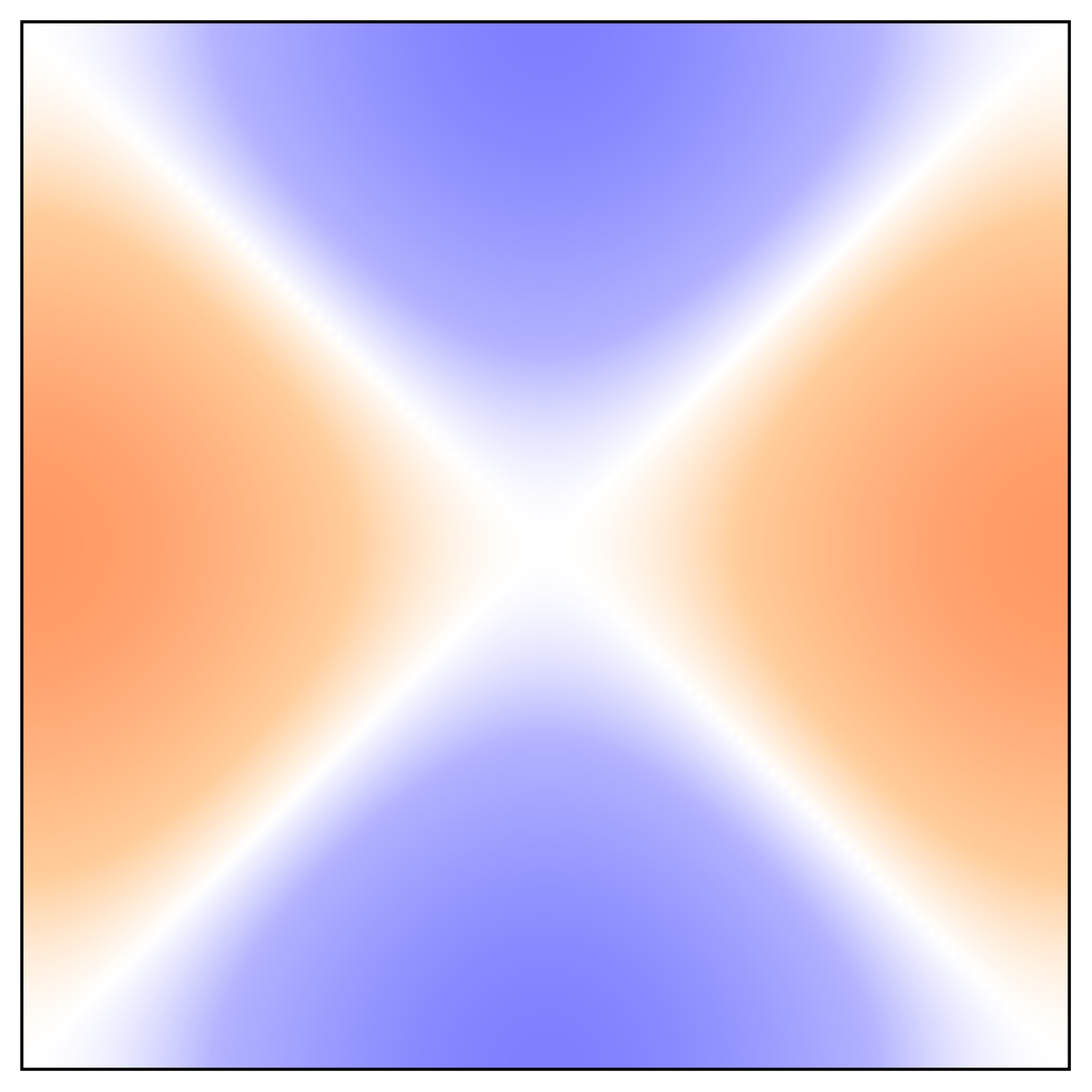}} &   \raisebox{-0.49\height}{\includegraphics[height=40pt]{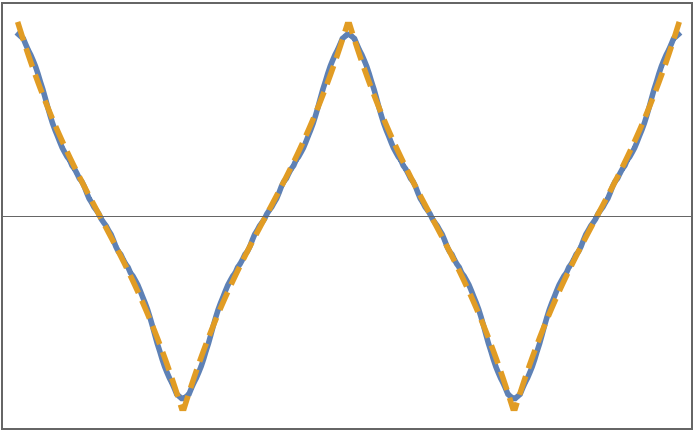}}    &   \raisebox{-0.49\height}{\includegraphics[height=40pt]{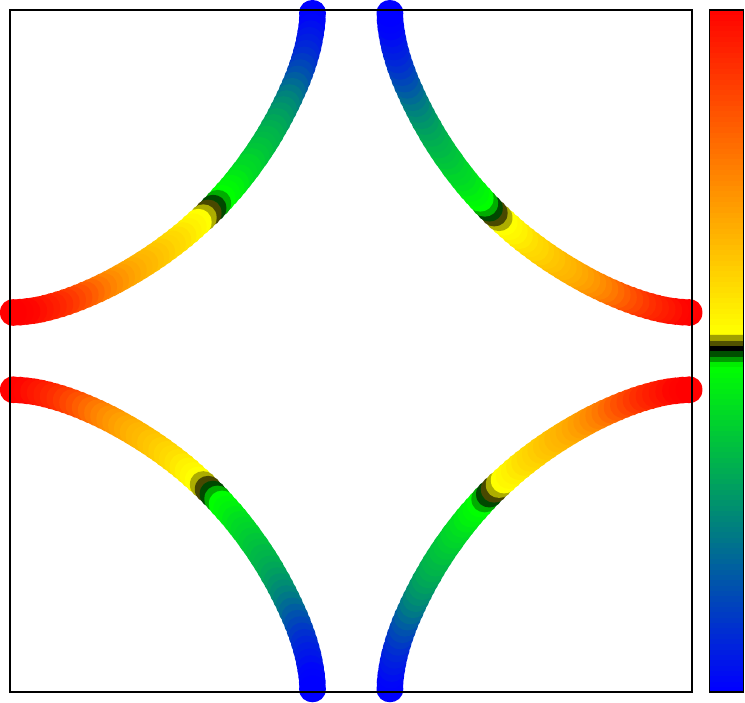}} \\[4pt]
 \hline\\[-12pt]
 B$_{2}$    &   {\small $xy$}   &   \raisebox{-0.49\height}{\includegraphics[height=40pt]{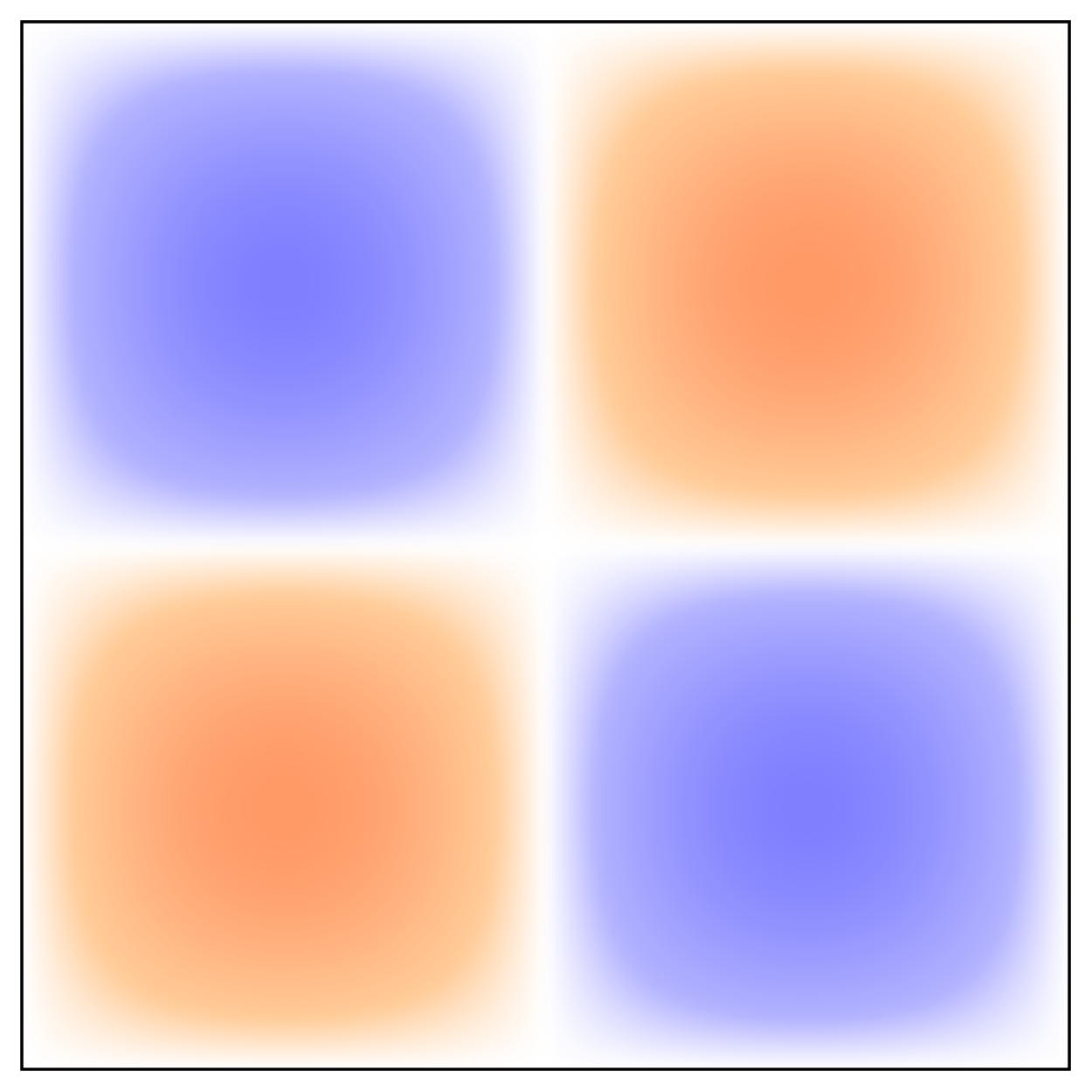}} &   \raisebox{-0.49\height}{\includegraphics[height=40pt]{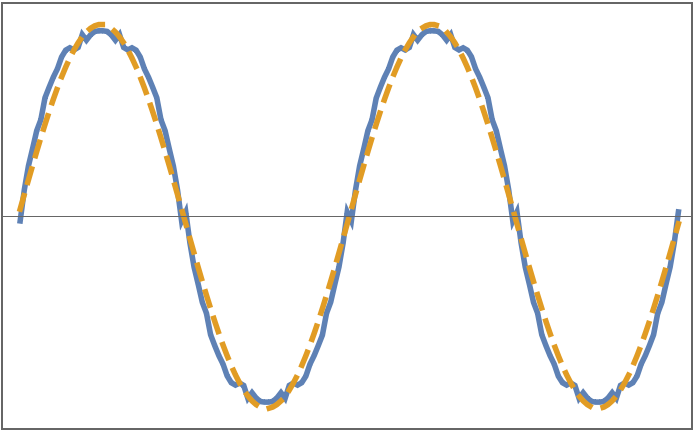}}    &   \raisebox{-0.49\height}{\includegraphics[height=40pt]{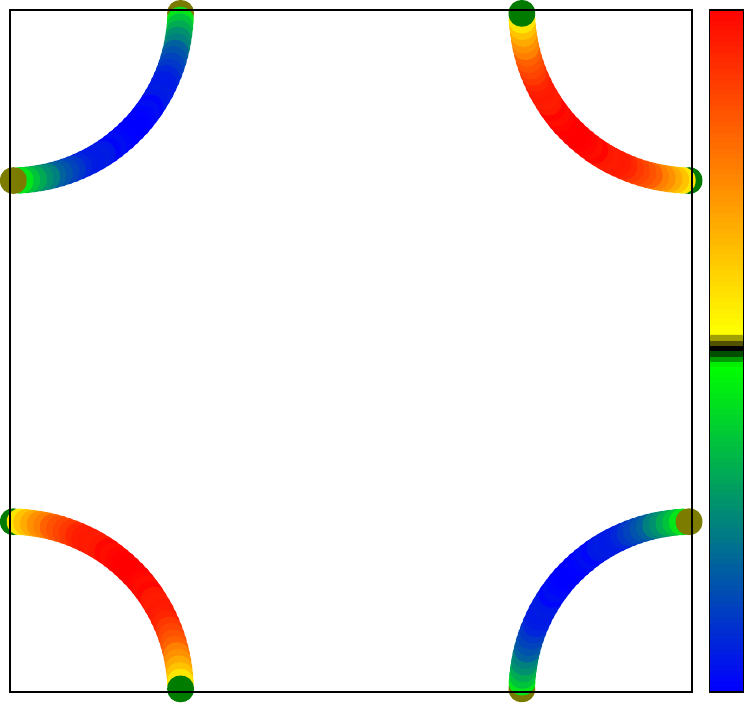}} \\[4pt]
 \hline\\[-12pt]
 \multirow{6}{*}{\vspace{-0.3cm}E}
 & & \multirow{3}{*}{\raisebox{-0.49\height}{\includegraphics[height=40pt,angle=180,origin=c]{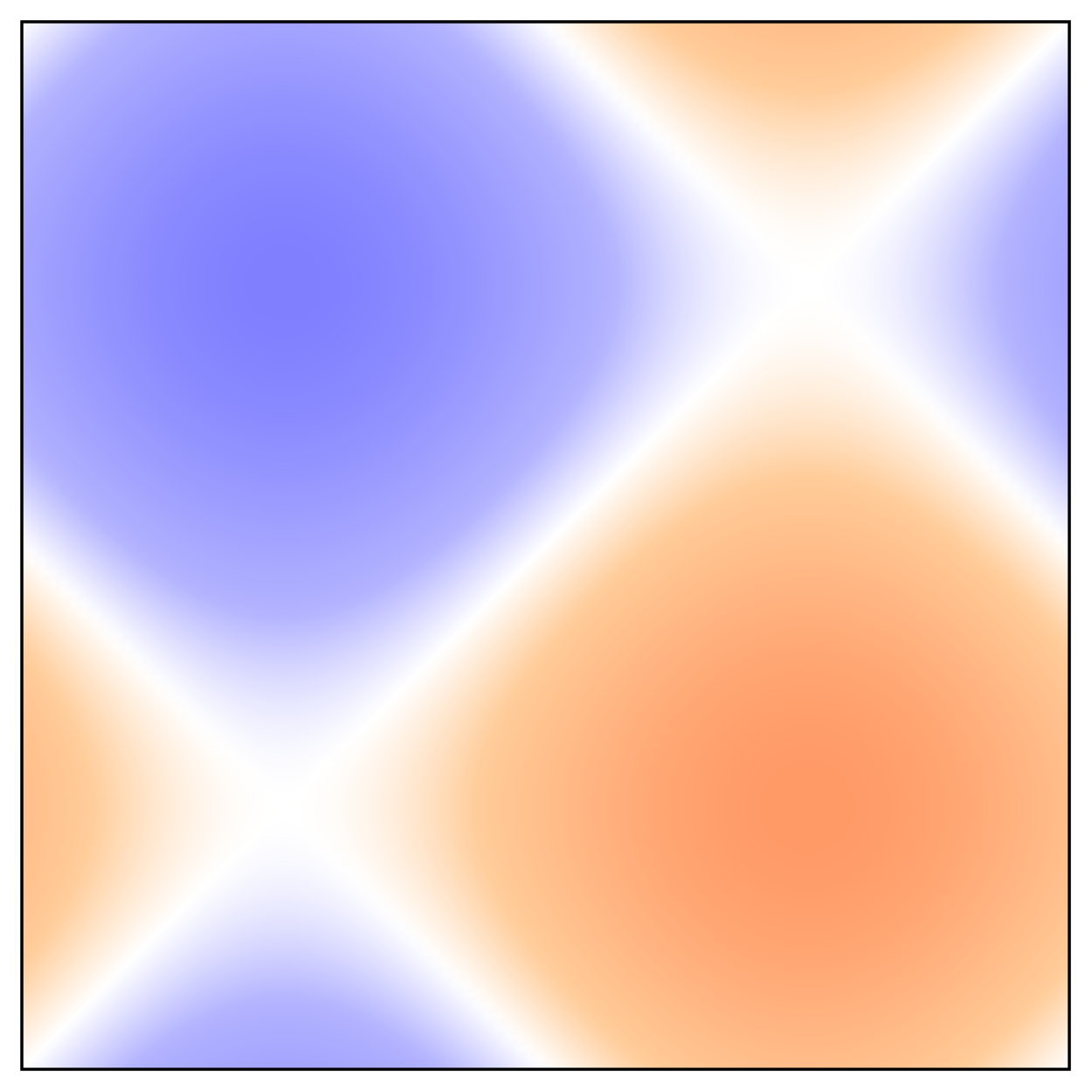}}} &   \multirow{3}{*}{\raisebox{-0.49\height}{\includegraphics[height=40pt]{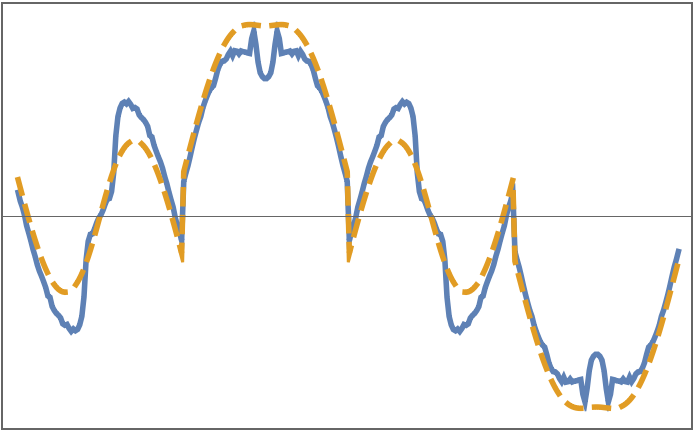}}}    &   \multirow{3}{*}{\raisebox{-0.49\height}{\includegraphics[height=40pt]{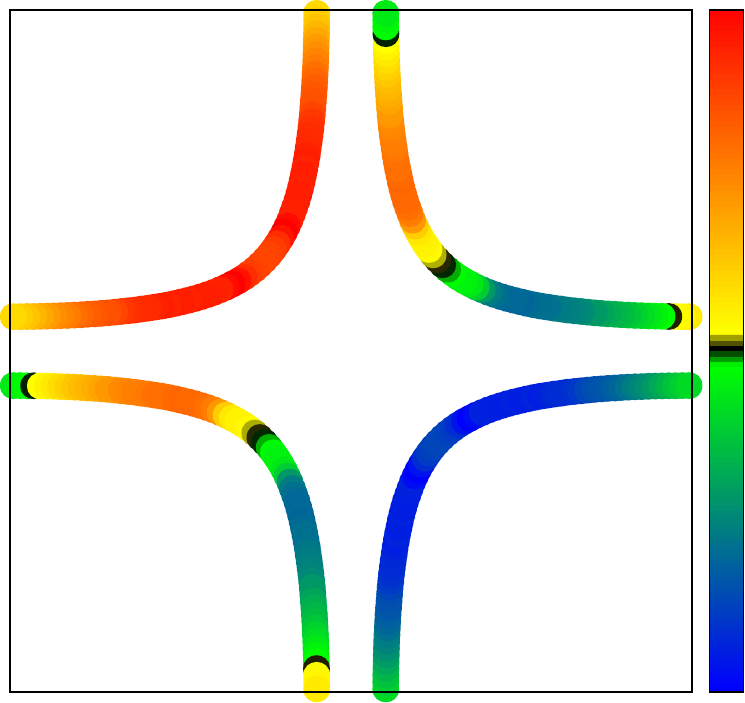}}} \\
 & {\small $x$} \\
 & & \\[4pt]
 & & \multirow{3}{*}{\raisebox{-0.49\height}{\includegraphics[height=40pt,angle=180,origin=c]{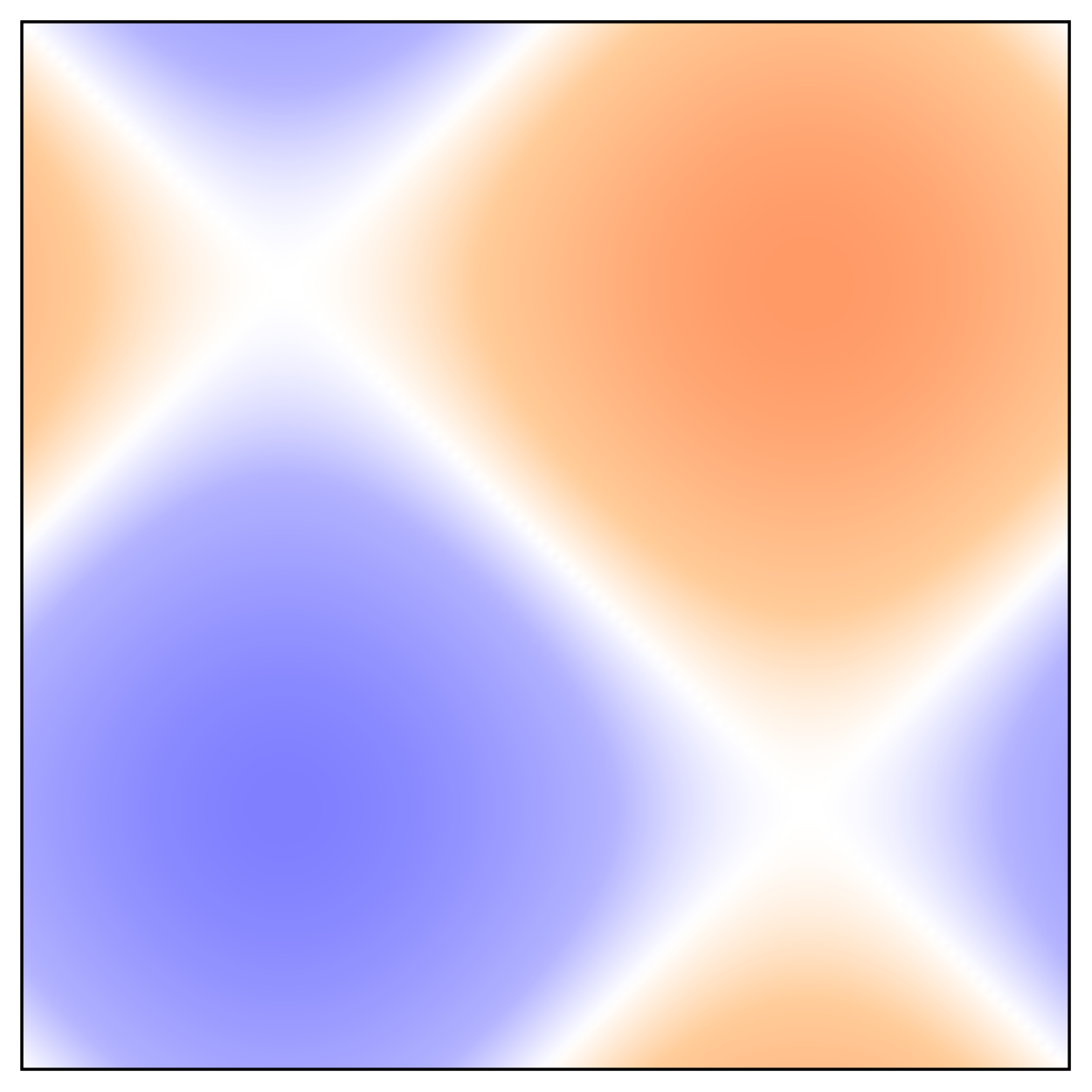}}} &   \multirow{3}{*}{\raisebox{-0.49\height}{\includegraphics[height=40pt]{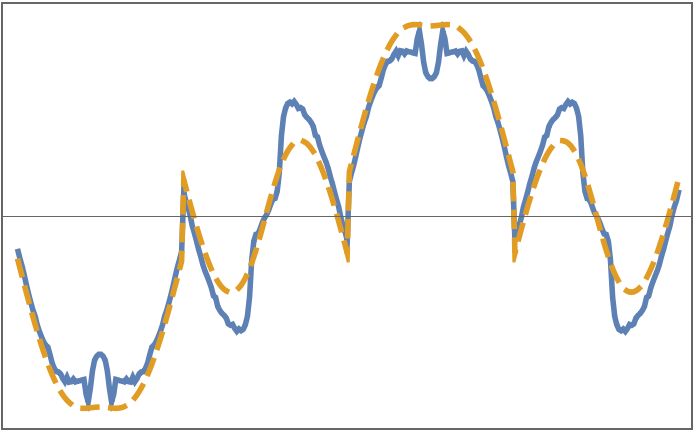}}}    &   \multirow{3}{*}{\raisebox{-0.49\height}{\includegraphics[height=40pt]{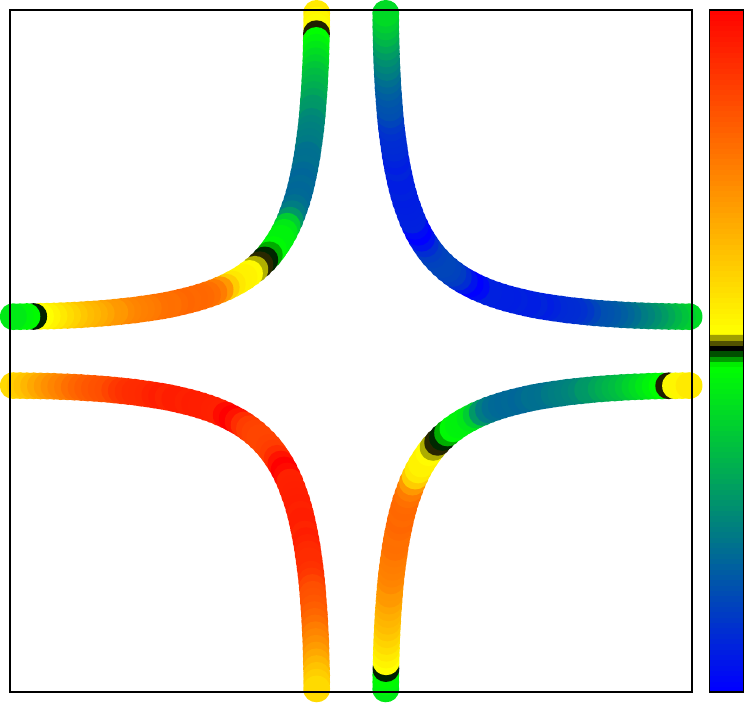}}} \\
 & {\small $y$} & \\
 & & \\[4pt]
 \hline
\end{tabular}
\caption{All irreps of the group $D_{4}$ with their corresponding basis functions. Columns from left to right: label of the irrep, basis function, contour plot of the basis function, plot of the basis function (orange, dashed) and example form factors (blue, solid) along the Fermi surface, and plot of the same form factor on the Fermi surface. Note that the eigenvectors of $g$ may include linear combinations with higher harmonics. $x$ refers to $\sin(x)$ and $x^{2}$ to $\cos(x)$. The basis vectors that span the 2D space of $E$, \ie $x$ and $y$, are rotated by $\pi/4$, such that the nodes and maxima of the form factors are arranged on the Fermi surface to maximize the condensation energy.}
\label{fig:waves_square}
\end{figure}

The square lattice is the canonical choice to test most theories and methods due to its simplicity. In particular, the copper-oxide high-temperature superconductors can be modelled as layered square lattice compounds. The  nearest-neighbor tight-binding model features a parabolic bandstructure, but deforms quickly under increasing influence of second-neighbor hoppings. We consider only one orbital per site, thus there is only one spin-degenerate band and all orbital factors appearing in vertex functions, including Eq.\,\eqref{eq:Gamma-for_U1}, are $M=1$. The bandstructure $E(\vec{k})$ is given by Eq.\,(\ref{eq:general_disp_rel}) and
\begin{align}
 &\varepsilon_{1}(\vec{k})=2\left[\cos(k_{x})+\cos(k_{y})\right], \\
 &\varepsilon_{2}(\vec{k})=4\left[\cos(k_{x})\cos(k_{y})\right]
\end{align}
with $n_r=2$, omitting the band index.
Some examples of Fermi surfaces, \ie constant energy cuts, are shown in the right column of Fig.\,\ref{fig:waves_square}. The point group of the square lattice is $D_{4}$, which contains the irreps $A_{1}$, $A_{2}$, $B_{1}$ and $B_{2}$ in the singlet and the two-dimensional irrep $E$ in the triplet channel. The corresponding basis functions are listed in Fig.\,\ref{fig:waves_square} along with their lowest lattice harmonics. In general, irreps can be realized through several harmonics, \eg  $B_1$ corresponds to a superconducting $d$-wave symmetry but also to $i$-wave (see discussion below). The transformation behavior of $d$- and $i$-wave under symmetry operations is the same, only the number of nodes is different.
\begin{figure}[t!]
\centering
\includegraphics[width=0.99\columnwidth]{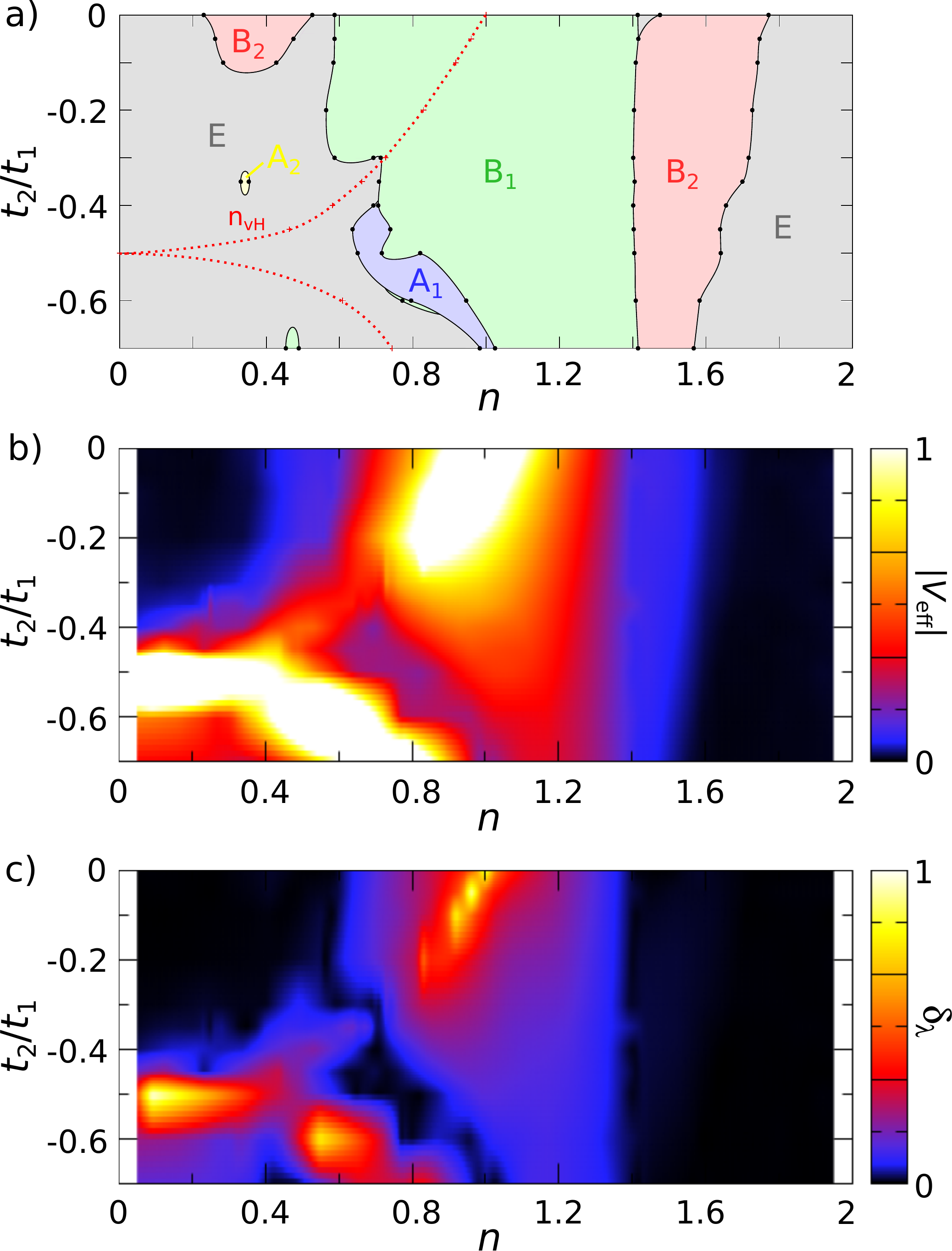}
\caption{a) Phase diagram $t_{2}/t_{1}$ vs.~band filling $n$ of the square lattice. The van Hove filling $n_{\text{vH}}$ is shown as a red dotted line. Note that there is also perfect nesting for $t_{2}/t_{1}=0$ at half filling. Due to the inherent particle-hole symmetry, it is sufficient to consider only negative values for $t_{2}/t_{1}$. b) Effective interaction, $V_{\text{eff}}$, as a function of the band filling $n$ and the second neighbor hopping $t_{2}/t_{1}$. 
c) Difference $\delta_\lambda$ of the two lowest eigenvalues corresonding to the superconducting ground state and the solution which is closest to it. Large differences indicate stable phases; the smaller the difference, the more fragile the superconducting groundstate.
}
\label{fig:square_phased_t2-n}
\end{figure}

The phase diagram for $\alpha=0$ and $-0.7\leq t_{2}/t_{1}\leq0$ as a function of $t_2/t_1$ vs.\ $n$ is shown in Fig.\,\ref{fig:square_phased_t2-n}\,a). It is dominated by $B_1$ in an extended region around half filling $n\approx 1$, which corresponds to $d_{x^2-y^2}$-wave and higher harmonics. The simplest one-band Hubbard model widely accepted to describe the cuprates for $t_2/t_1=-0.3$\,\cite{husslein_quantum_1996} also lies within this phase in agreement with the experimental groundstate of most cuprate materials. We also find a large phase with $B_2$ symmetry corresponding to $d_{xy}$-symmetry for $1.4 \leq n \leq 1.7$. Most interestingly, the remainder of the phase diagram is essentially covered by a superconducting groundstate with $E$ symmetry; the whole region is expected to realize a chiral topological superconductor with $p_x \pm i p_y$-symmetry or higher harmonics. There is a small pocket with $B_2$ symmetry for low fillings at $t_2 \ll 1$ as well as one with $A_1$ symmetry, \ie extended $s$-wave, for larger $t_2$ around half-filling. We also find at $t_2/t_1=-0.35$ and $n=0.35$ a tiny pocket with $g$-wave symmetry (irrep $A_2$); due to the smallness of the phase, we do not expect it to be stable for moderate interactions. We show the filling at which the van Hove singularity appears as a red dotted line; it is given by
\begin{equation}
 n_{\text{vH}}=\int_{\xi^{-}}^{\mu_{\text{vH}}}\rho(E)\text{d}E,\hspace{2mm}\mu_{\text{vH}}=\begin{cases}
           4t_{2}, &\text{if}\,|t_{2}|<0.5,\\
           t_{1}^{2}/t_{2}, &\text{if}\,|t_{2}|\geq0.5.
          \end{cases}
\end{equation}
Note that for $|t_{2}|\geq0.5$, also (imperfect) nesting occurs at the van Hove fillings, \ie Fermi surfaces which, consists of points corresponding to other points on the Fermi surface under translation by a reciprocal {\it nesting} vector $\vec Q_{n,1}=(0,2\arccos[-t_{1}/2t_{2}])$ and $\vec Q_{n,2}=(2\arccos[-t_{1}/2t_{2}],0)$\,\cite{romer_pairing_2015}. The nesting affects $75\,\%$ of the Fermi surface, whereas perfect nesting occurs only in the limit $t_{2}/t_{1}\rightarrow\infty$.

As discussed before, within the WCRG approach we always find superconducting solutions. Thus it is important to know which of these are particularly stable and occur at sufficiently high temperature to be of experimental interest. The effective coupling strength $V_{\rm eff}$ can be directly related to the critical temperature, $T_c \sim \exp{[-1/(\rho V_{\rm eff})]}$. In Fig.\,\ref{fig:square_phased_t2-n}\,b) we show $V_{\rm eff}$ for the same parameters as the phase diagram above. Of interest are the regions which correspond to bright colors. In particular, the white, yellow, and orange spots indicate high critical temperatures. As expected, they are concentrated at van Hove fillings. Note that also the lighter blue regions are interesting and experimentally relevant. Amongst the ground states with promisingly large $V_{\rm eff}$ are those with irreps $B_1$ and $E$, the latter realizing topological superconductivity.

The WCRG approach allows to test the stability of a solution with respect to variation of bandstructure parameters, band filling or variation of interactions. Quite generally, we observe that phases are particularly stable when the lowest eigenvalue $\lambda_{\text{min}}$ of the normalized two-particle vertex $g$ is separated by the second-lowest eigenvalue by a large gap $\delta_\lambda$. In other words, if $\delta_\lambda$ is large, there is no way to destabilize the superconducting ground state. If $\delta_\lambda$ is, however, very small or approaching zero, phase transitions can happen even for the smallest perturbations. In Fig.\,\ref{fig:square_phased_t2-n}(c), we plot $\delta_\lambda$. We observe that large $\delta_\lambda$ regions mostly coincide with large $V_{\rm eff}$ regions in Fig.\,\ref{fig:square_phased_t2-n}(b).

\begin{figure}[t!]
\centering
 \includegraphics[width=0.99\columnwidth]{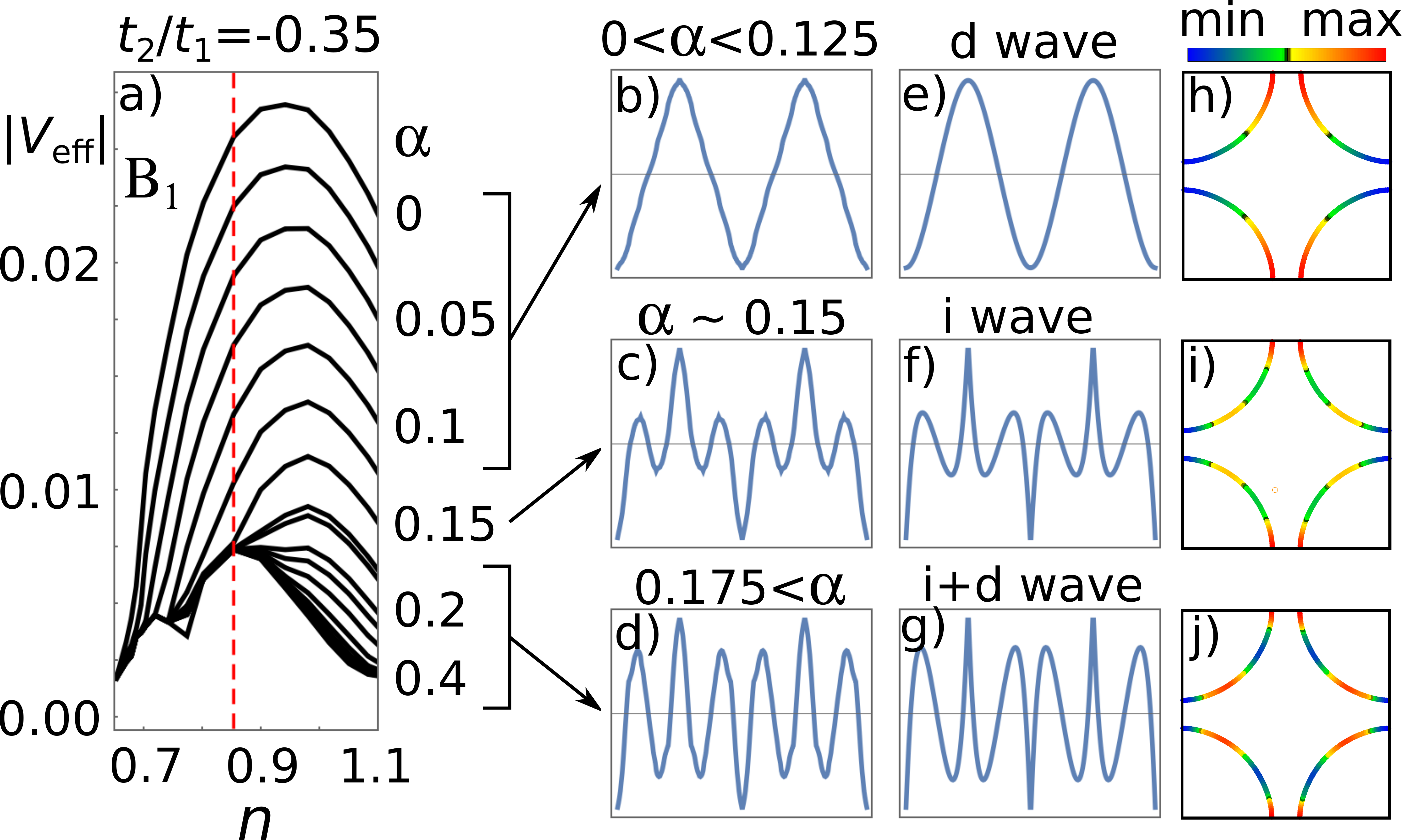}
\caption{(a) Effective interaction, $V_{\text{eff}}$, of the most negative eigenvalue of the irrep $B_{1}$ as a function of the filling $n$ for different nearest-neighbor interaction strengths $\alpha$. The filling $n\approx0.853$, at which the other figures are shown, is indicated as a red, dashed line. (b)-(d) Plot of the form factor corresponding to the eigenvalues plotted in (a). (e)-(g) Form factor of the pure $d$- and $i$-wave for (e) and (f), respectively, and of the linear combination of the two in (g). (h)-(j) Plot of the form factors (e)-(g) now shown on the Fermi surface.}
\label{fig:square_B1_wave_change}
\end{figure}

The second part of this section about the square lattice is devoted to the effect of nearest-neighbor Coulomb repulsions, Eq.\,\eqref{eq:NN-rep}, to the phase diagram, Fig.\,\ref{fig:square_phased_t2-n}(a). The form factor in Eq.\,\eqref{eq:Gamma-for_U1} can be written as
\begin{equation}
  \varepsilon_1(\vec{k}_{2}-\vec{k}_{1})=\sum_{i}\psi_{i}(\vec{k}_{1})\psi_{i}(\vec{k}_{2})\ ,
\end{equation}
where the sum runs over all nearest-neighbor lattice harmonics, \ie $i\in\{A_{1},A_{2},B_{1},B_{2},E\}$.
On the square lattice, these correspond to the following basis functions:
\begin{align}
 &\psi_{A_{1}}^{D_{4},1}(\vec{k})=\cos(k_{x})+\cos(k_{y}), \nonumber\\
 &\psi_{B_{1}}^{D_{4},1}(\vec{k})=\cos(k_{x})-\cos(k_{y}), \nonumber\\
 &\psi_{E,1}^{D_{4},1}(\vec{k})=\sqrt{2}\sin(k_{x}),\hspace{2mm} \psi_{E,2}^{D_{4},1}(\vec{k})=\sqrt{2}\sin(k_{y}), \nonumber\\
 &\psi_{A_{2}}^{D_{4},1}(\vec{k})=\psi_{B_{2}}^{D_{4},1}(\vec{k})=0,
\end{align}
where $\psi_{I}^{G,n}(\vec{k})$ denotes the $n$-th lattice harmonic of irrep $I$ in group $G$.
\begin{figure}[t!]
\centering
 \includegraphics[width=0.99\columnwidth]{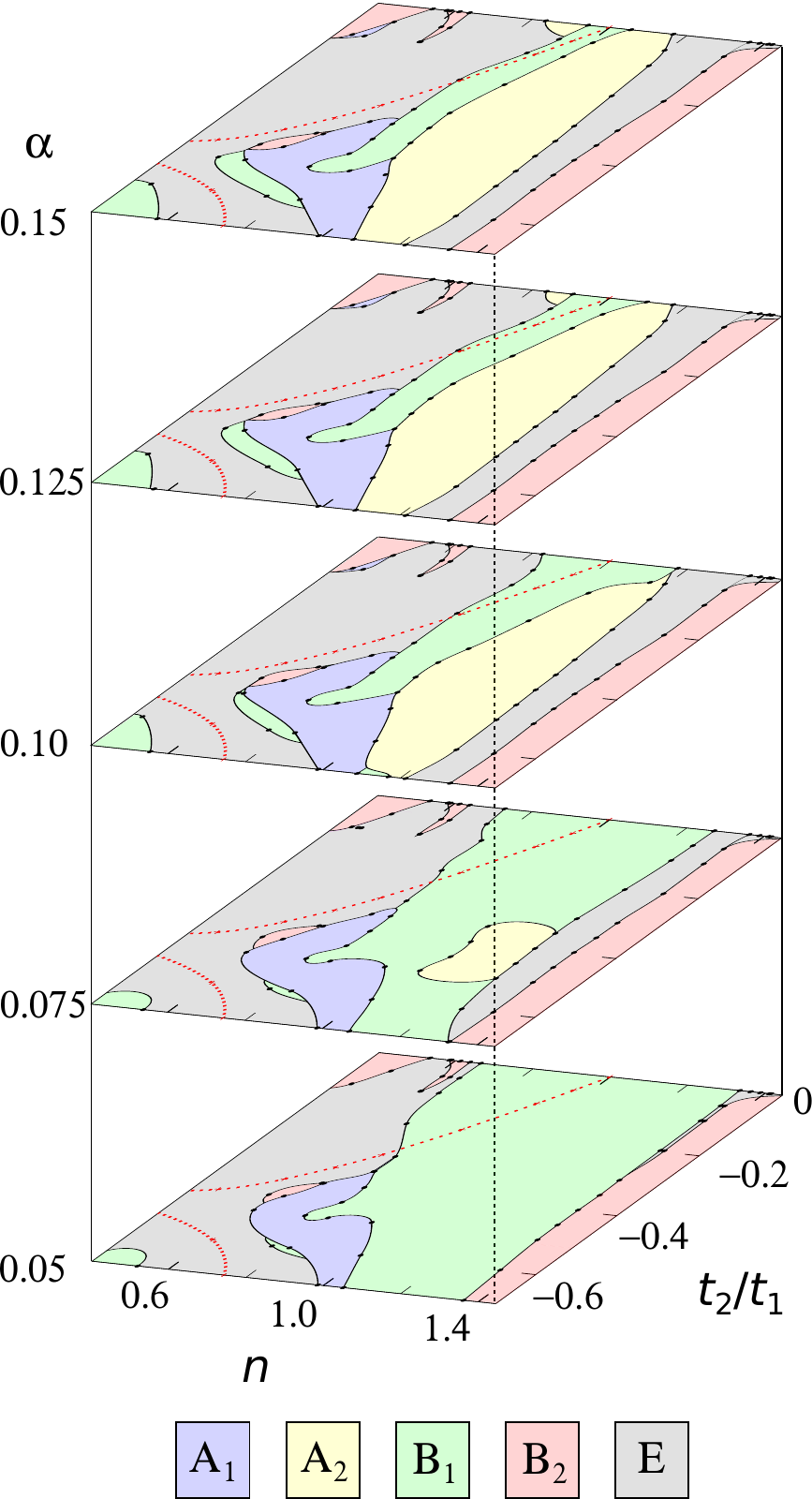}
\caption{Phase diagram showing the irrep of the leading superconducting instability in the square lattice for $0.05\leq\alpha\leq0.15$, $0.4\leq n\leq1.45$ and $-0.7\leq t_{2}/t_{1}\leq0$.}
\label{fig:square_phased_3D}
\end{figure}
Thus, the nearest-neighbor interaction only affects the irreps with nonzero nearest-neighbor lattice harmonics, since the set of all basis functions spans an orthogonal space. It is important to stress that not all superconducting ground states that transform under a given representation are affected in the same way. To illustrate this further, we show in Fig.\,\ref{fig:square_B1_wave_change} an example where the leading instability is affected by $U_{1}$ (parameters used are $t_{2}/t_{1}=-0.35$ and $n\approx0.853$). Here, for onsite interactions $U_{0}$ (\ie $\alpha=0$), the form factor has $d_{x^{2}-y^{2}}$-wave symmetry ($B_{1}$). Adding nearest-neighbor interaction $U_{1}$ (\ie $\alpha >0$), the effective interaction of this form factor becomes smaller the larger $\alpha$ gets. The leading instability can change its symmetry, when the eigenvalue corresponding to the $d_{x^{2}-y^{2}}$-wave becomes larger than the next lowest eigenvalue of $g$. In the case shown here, the form factor of the second eigenvalue is also of the $B_{1}$ irrep, but it forms a linear combination of the $d$- and $i$-waves. Further increasing $\alpha$ will have no effect on the leading instability anymore. In other words, within the region with $B_1$ symmetry, regions with pure $d$-wave symmetry are suppressed by finite $\alpha$ while regions with $i$-wave symmetry are not.

We have mapped out the full three-dimensional $t_2/t_1$ versus $n$ versus $\alpha$ phase diagram. As discussed in detail before, only regions with sufficiently large $V_{\rm eff}$ are of interest. For the sake of clarity, we limited the phase diagram to a relevant range which is $0.5\leq n\leq1.45$ and $0.05\leq\alpha\leq0.15$, see Fig.\,\ref{fig:square_phased_3D}. The main changes due to finite $\alpha$ is the appearance of a pocket with $g$-wave symmetry (irrep $A_2$) around $n\approx 1.2$ which quickly develops into a large phase slightly above half filling. The chiral superconducting phase (irrep $E$) at lower filling essentially persists; in addition, the narrow stripe of $E$ symmetry emerges between the two $d$-wave phases. There are also several small pockets or stripes with $B_1$ and $B_2$ symmetry. We have not shown plots for $V_{\rm eff}$ and $\delta_\lambda$ for the only reason that these plots are very similar to the ones shown for $\alpha=0$.

We will conclude this section with a discussion of the results and comparison to the literature. In Ref.\,\onlinecite{raghu_superconductivity_2010} (see Fig.\,2), the coupling strengths $V_{\rm eff}$ for irreps $B_1$, $B_2$, and $E$ are shown for $\alpha=0$. In Ref.\,\onlinecite{raghu_effects_2012} (see Fig.\,2), the $\alpha$--$n$ phase diagram for $t_2=0$ is shown. In contrast to our results, in these works the $E$ phase is absent, mostly $d_{xy}$-wave is found instead (irrep $B_2$). We further note that a large range of our $t_2/t_1$--$n$ phase diagram Fig.\,\ref{fig:square_phased_t2-n} was discussed within a $T$ matrix approach in Ref.\,\onlinecite{hlubina_phase_1999} (see Fig.\,6); remarkably, it coincides with our findings (and disagrees with Ref.\,\cite{raghu_superconductivity_2010,raghu_effects_2012} regarding the phase with $E$ symmetry, as emphasized in Ref.\,\onlinecite{raghu_effects_2012}). We believe that the sensitivity of the integration grid might be a candidate to explain this discrepancy; we discuss this in detail at the end of the paper in the Section about ``Numerical development and performance analysis'' and explicitly show that there is a phase transition from $B_2$ to $E$ depending on the resolution of the integration grid.

Due to our large parameter range, we could identify superconducting solutions for all irreps belonging to the $D_4$ symmetry group. In particular, we find a large region with $E$ symmetry corresponding to chiral topological superconductivity. In the light of these findings, it becomes apparent that the emergence of $d$- or extended $s$-wave symmetry as present in cuprates or pncitides, respectively, is not a generic feature of the square lattice. Instead, it rather seems to be a consequence of the particular parameters and Fermi surfaces realized in these materials.

%
%
\subsection{Triangular lattice}

\begin{figure}[b!]
\centering
\begin{tabular}{c|c|ccc}
\hline\\[-12pt]
 A$_{1}$    &   {\small $x^{2}+y^{2}$}   &   \raisebox{-0.49\height}{\includegraphics[height=40pt]{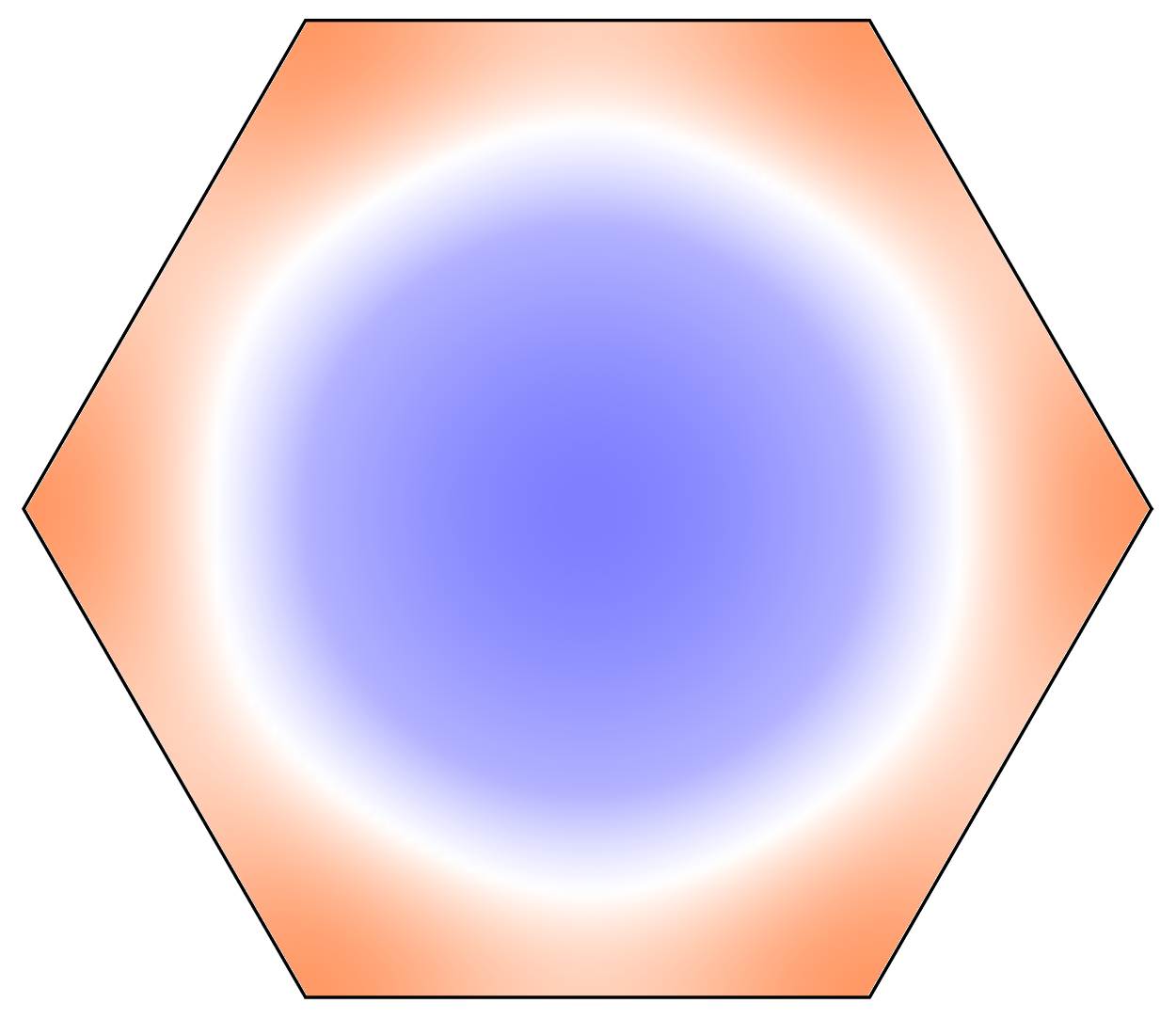}} &   \raisebox{-0.49\height}{\includegraphics[height=40pt]{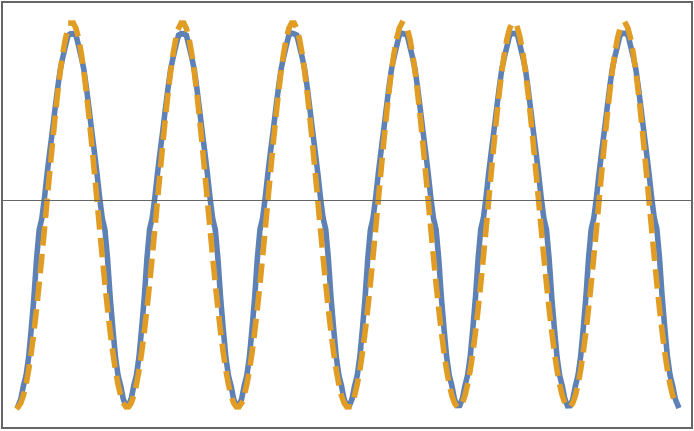}}    &   \raisebox{-0.49\height}{\includegraphics[height=40pt]{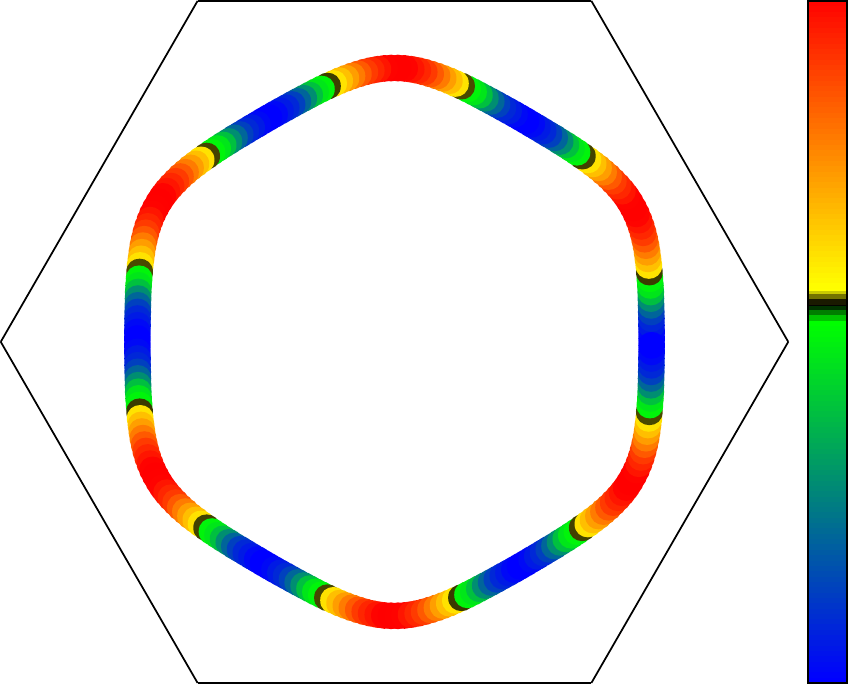}} \\[4pt]
 \hline\\[-12pt]
 A$_{2}$    &   {\small B$_{1}\cdot$B$_{2}$}   &   \raisebox{-0.49\height}{\includegraphics[height=40pt]{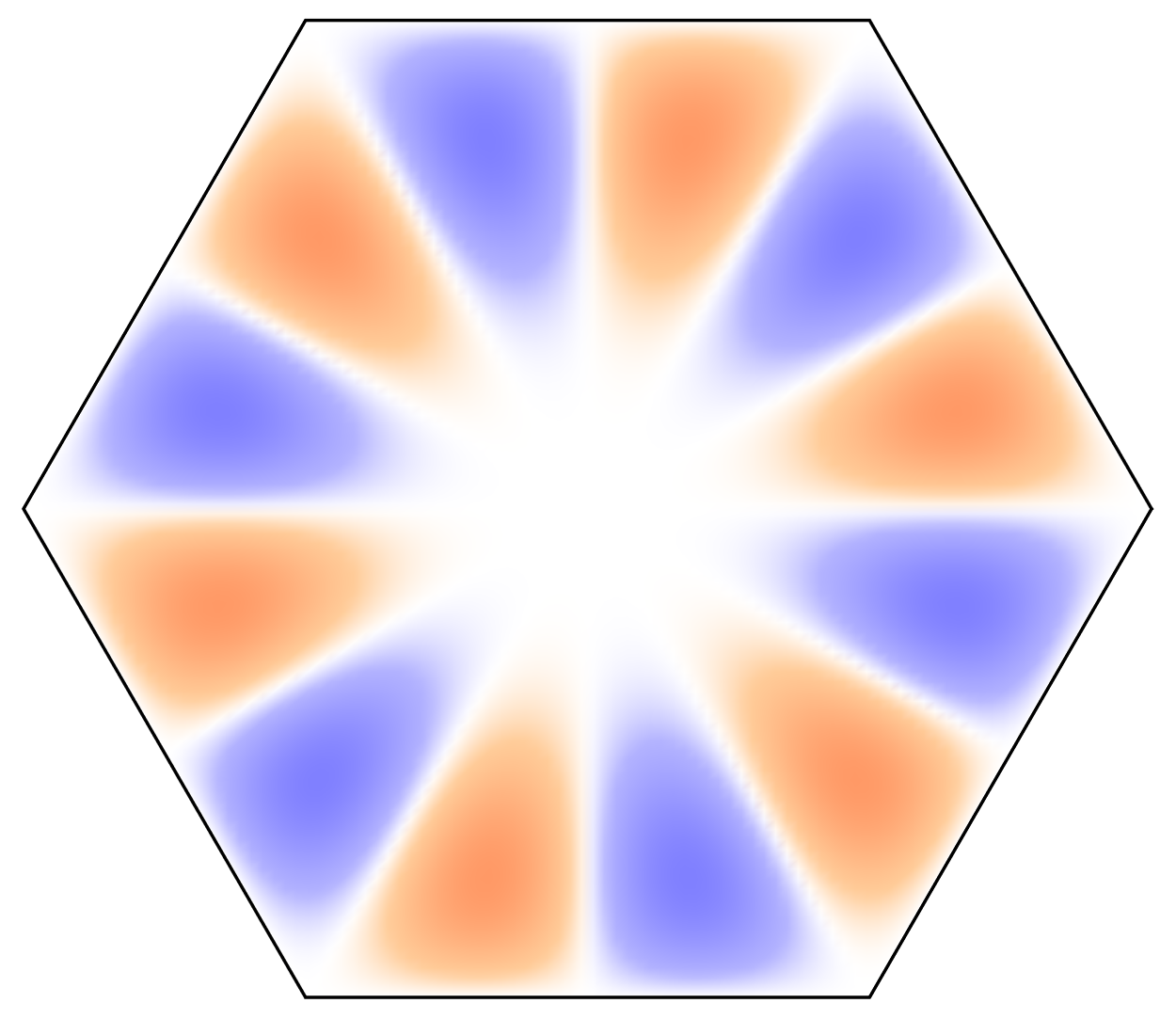}} &   \raisebox{-0.49\height}{\includegraphics[height=40pt]{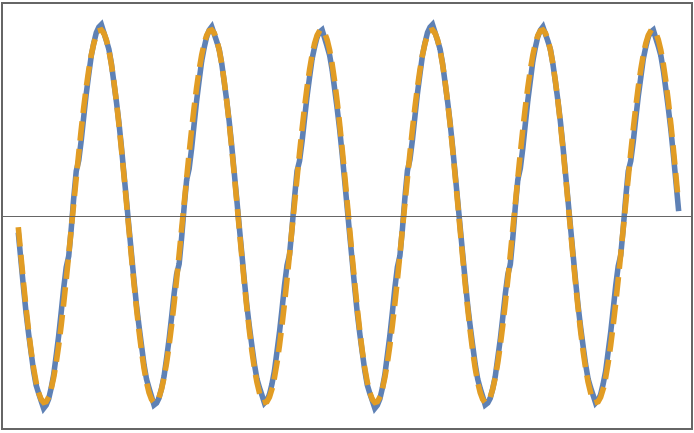}}    &   \raisebox{-0.49\height}{\includegraphics[height=40pt]{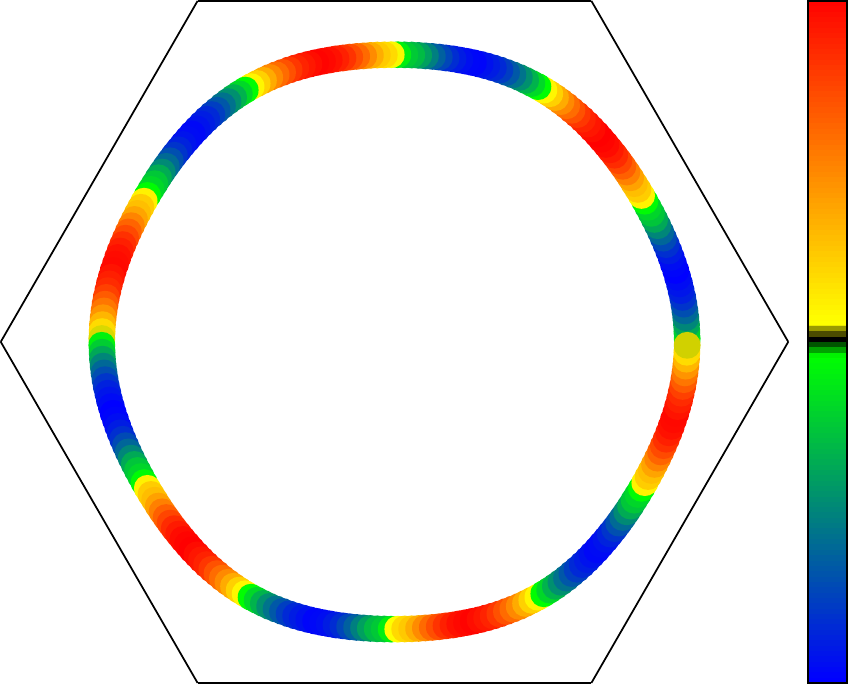}} \\[4pt]
 \hline\\[-12pt]
 B$_{1}$    &   {\small $x(x^{2}-3y^{2})$}   &   \raisebox{-0.49\height}{\includegraphics[height=40pt]{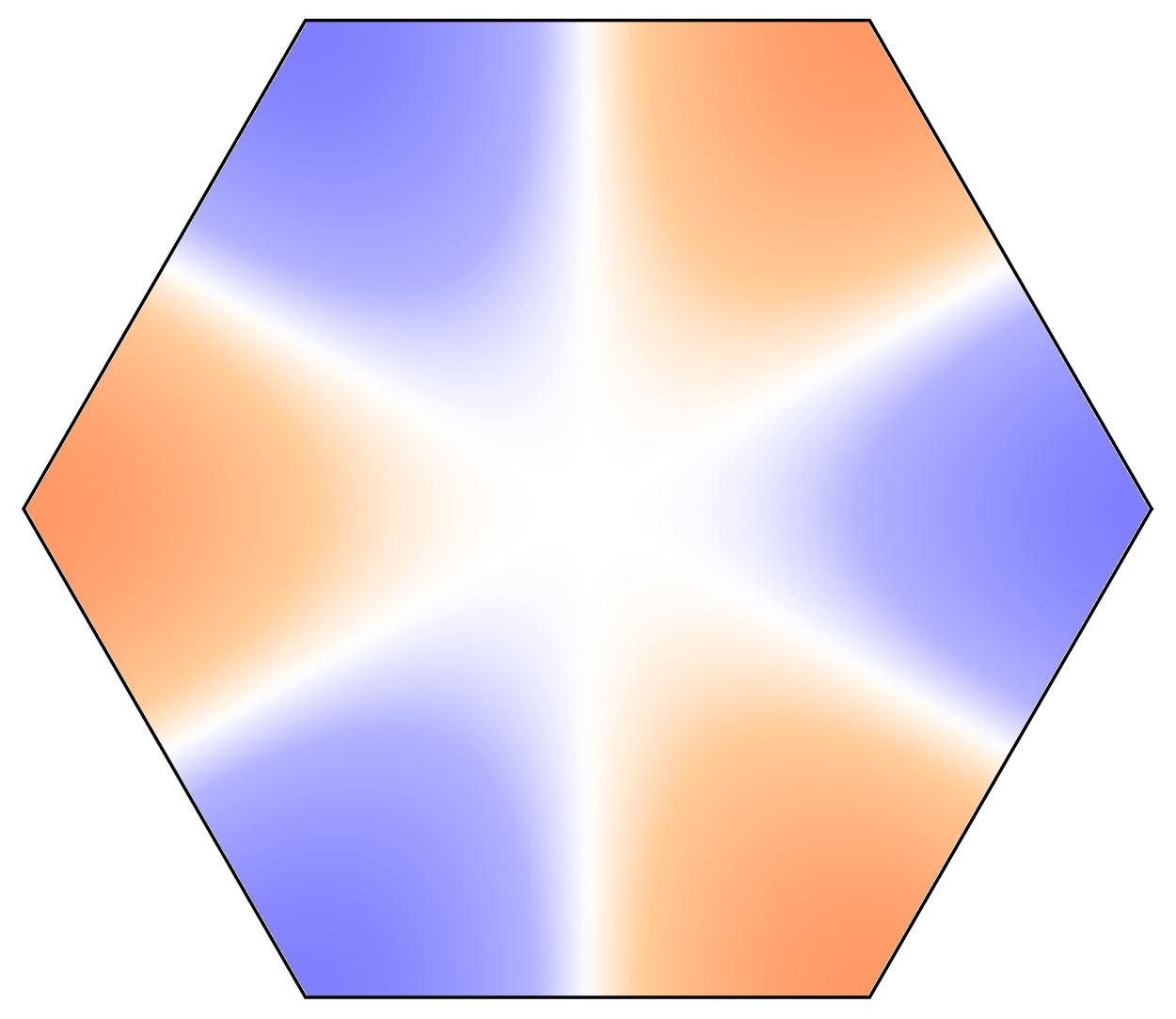}} &   \raisebox{-0.49\height}{\includegraphics[height=40pt]{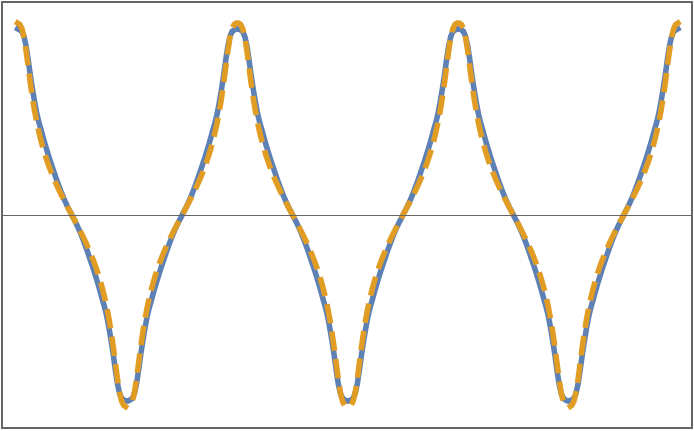}}    &   \raisebox{-0.49\height}{\includegraphics[height=40pt]{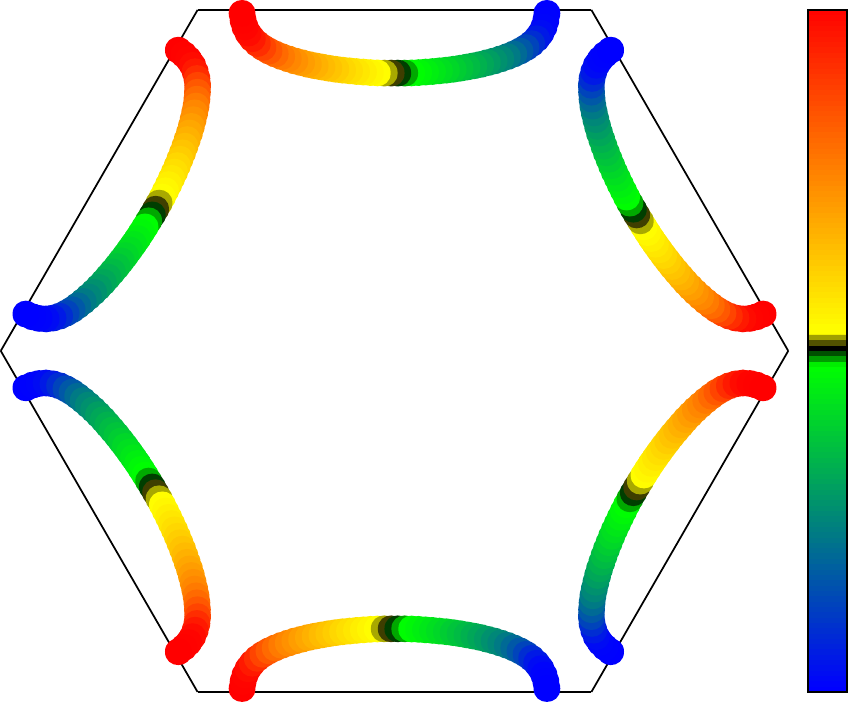}} \\[4pt]
 \hline\\[-12pt]
 B$_{2}$    &   {\small $y(3x^{2}-y^{2})$}   &   \raisebox{-0.49\height}{\includegraphics[height=40pt]{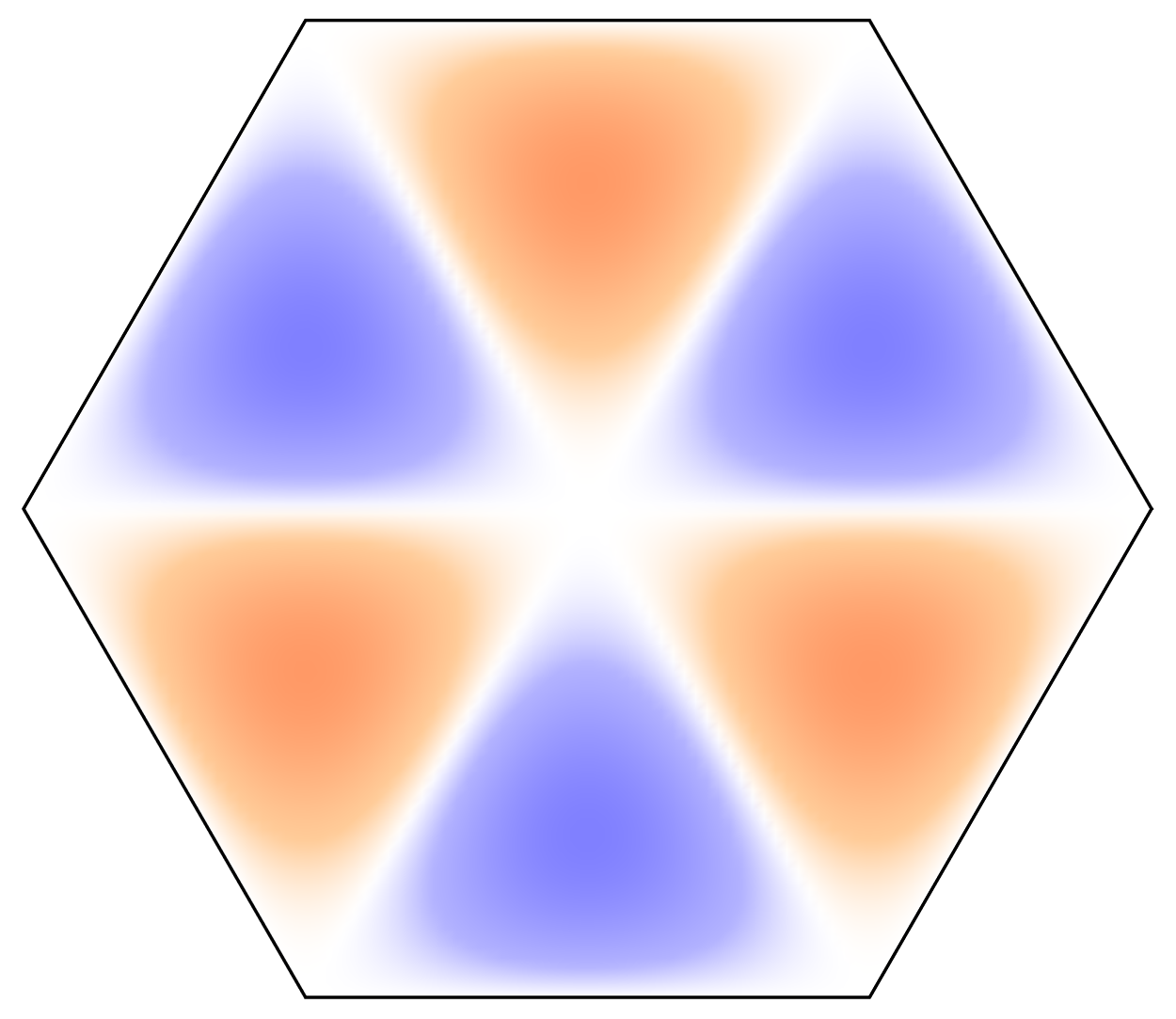}} &   \raisebox{-0.49\height}{\includegraphics[height=40pt]{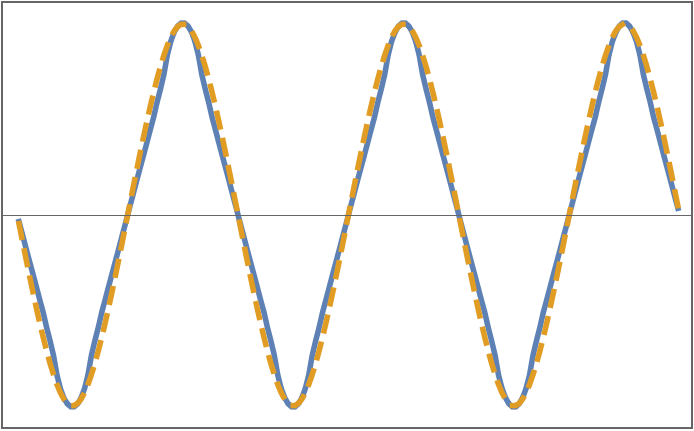}}    &   \raisebox{-0.49\height}{\includegraphics[height=40pt]{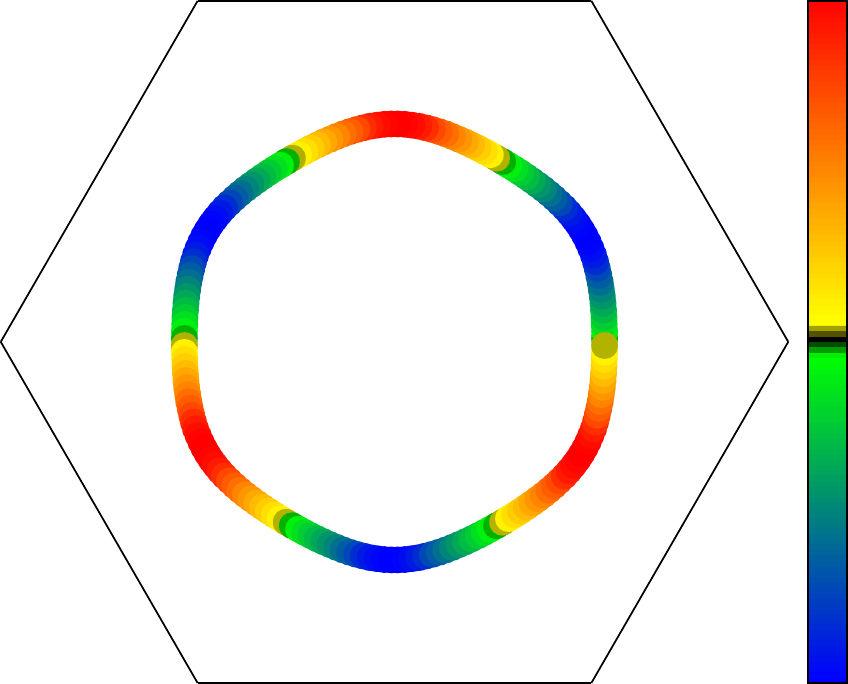}} \\[4pt]
 \hline\\[-12pt]
 \multirow{2}{*}{\vspace{-1cm}E$_{1}$}    &  {\small $x$}   &   \raisebox{-0.49\height}{\includegraphics[height=40pt]{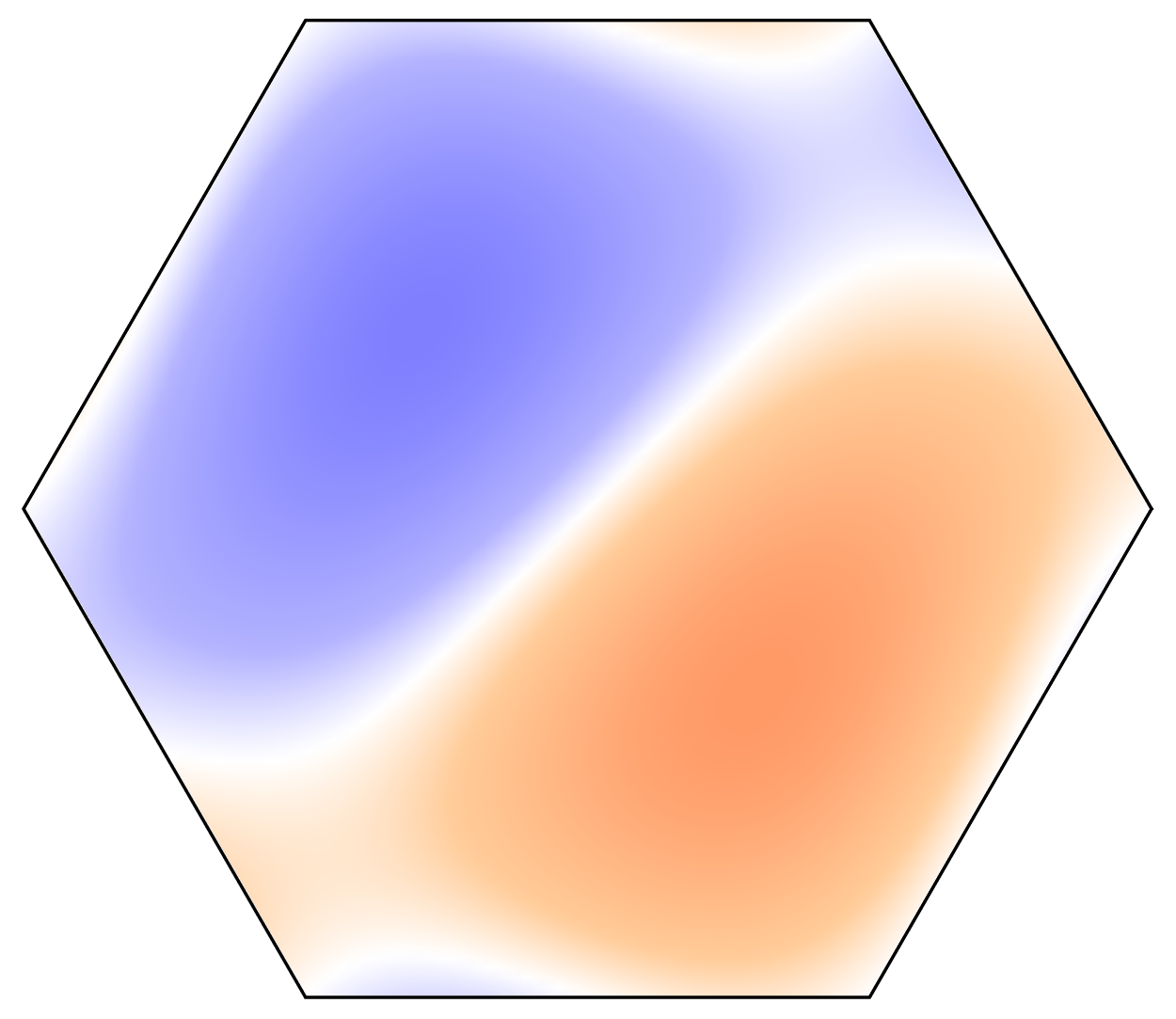}} &   \raisebox{-0.49\height}{\includegraphics[height=40pt]{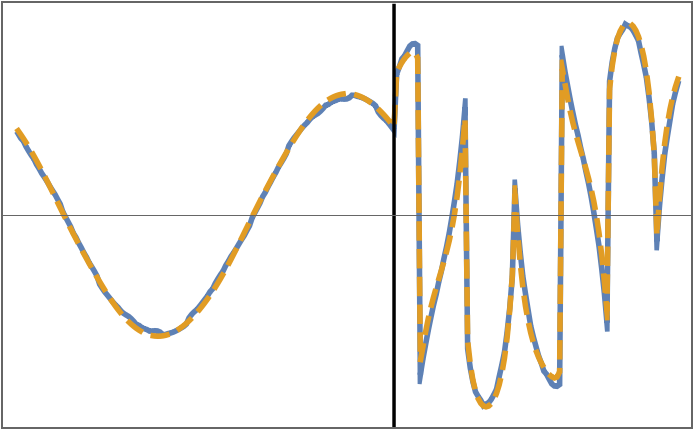}}    &   \raisebox{-0.49\height}{\includegraphics[height=40pt]{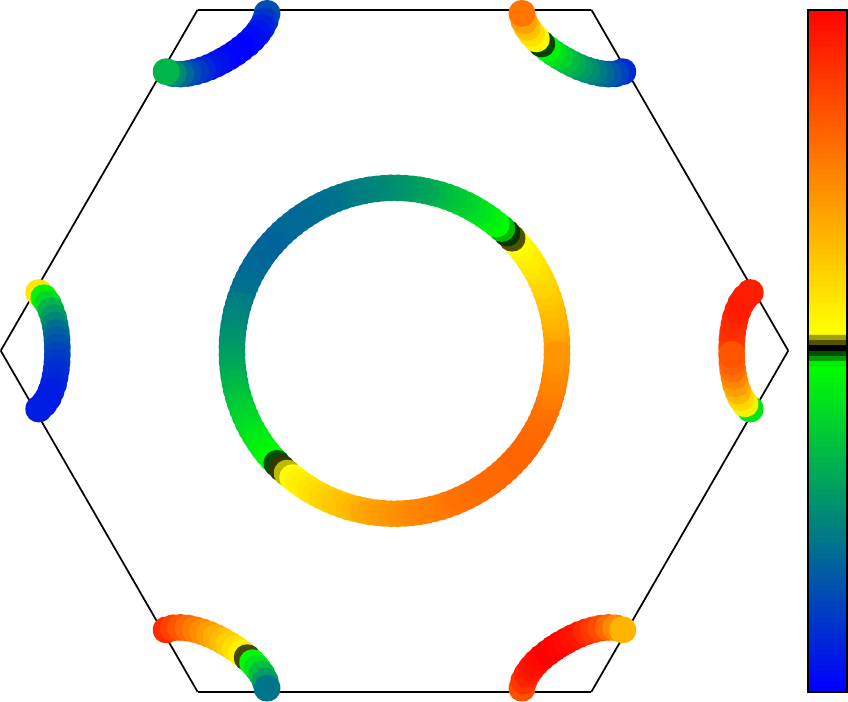}} \\[5pt]
 & {\small $y$}  &   \raisebox{-0.49\height}{\includegraphics[height=40pt]{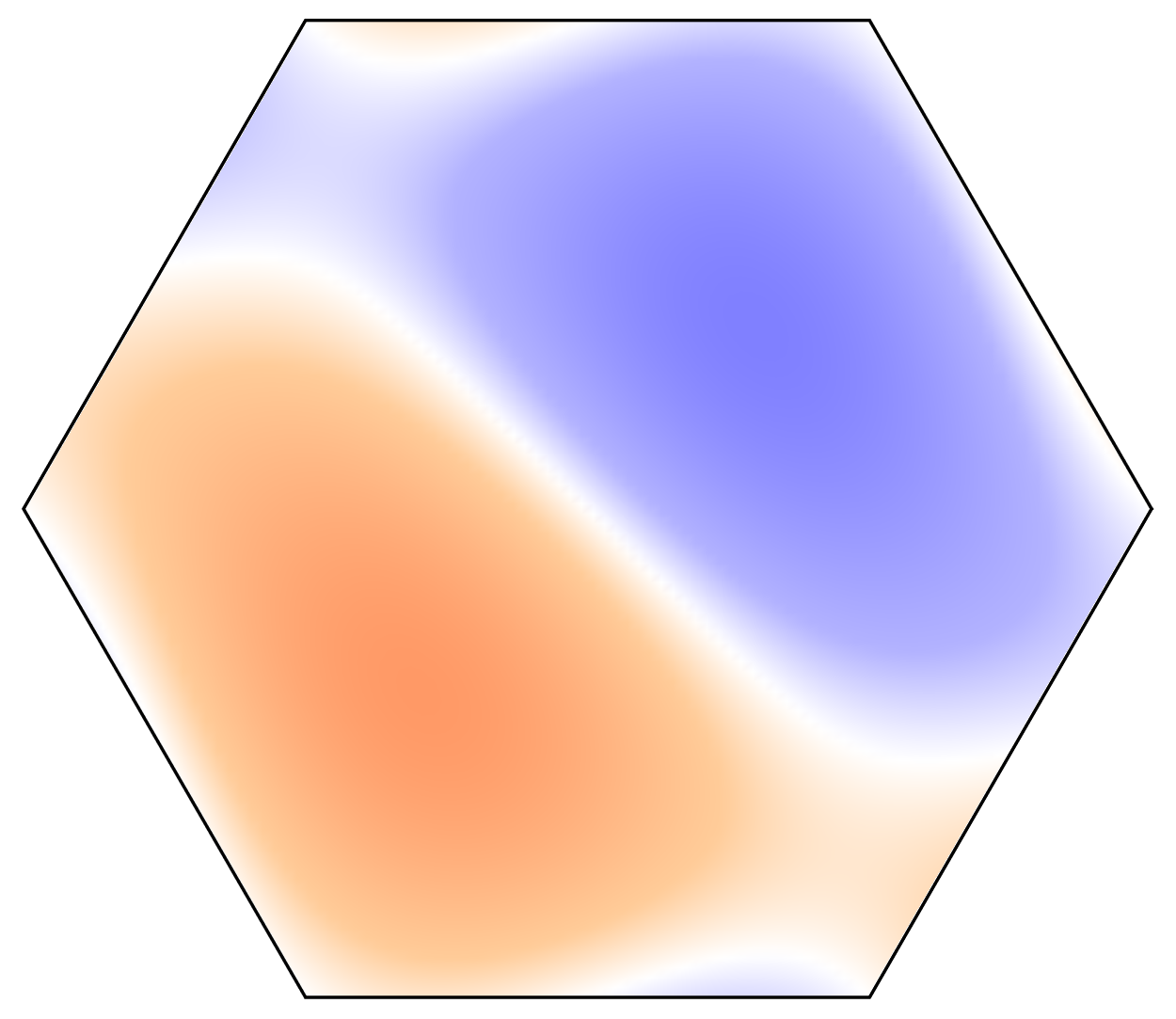}} &   \raisebox{-0.49\height}{\includegraphics[height=40pt]{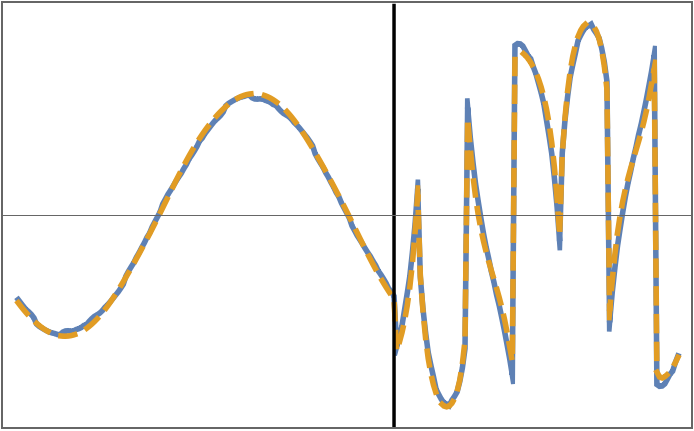}}    &   \raisebox{-0.49\height}{\includegraphics[height=40pt]{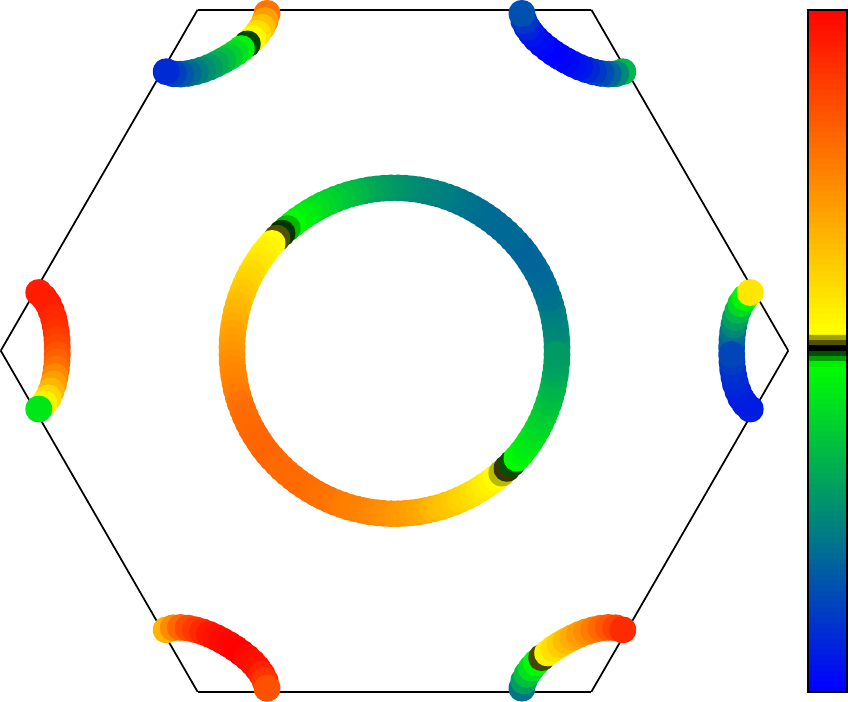}} \\[4pt]
 \hline\\[-12pt]
 \multirow{2}{*}{\vspace{-1cm}E$_{2}$}    &  {\small $x^{2}-y^{2}$}   &   \raisebox{-0.49\height}{\includegraphics[height=40pt]{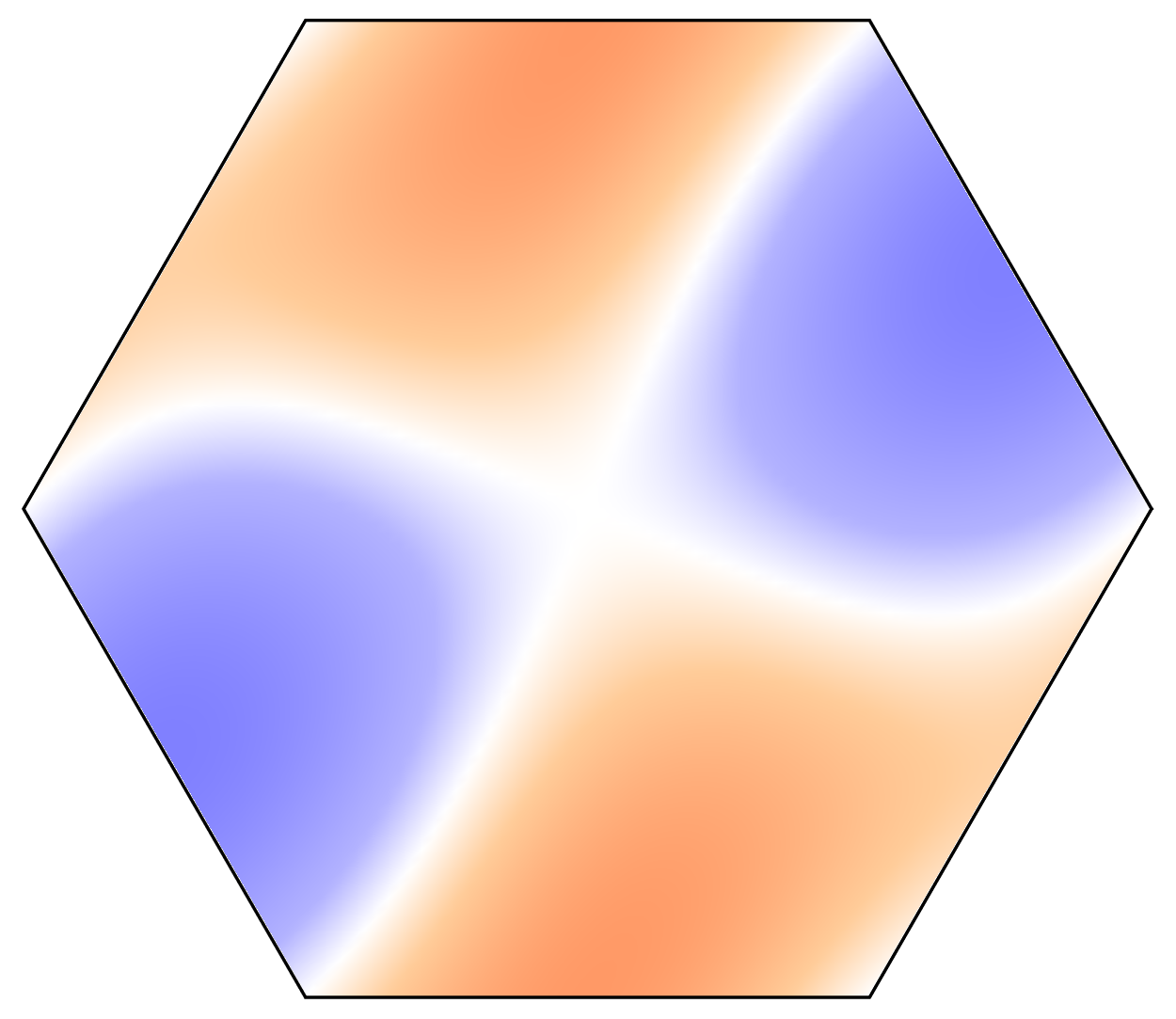}} &   \raisebox{-0.49\height}{\includegraphics[height=40pt]{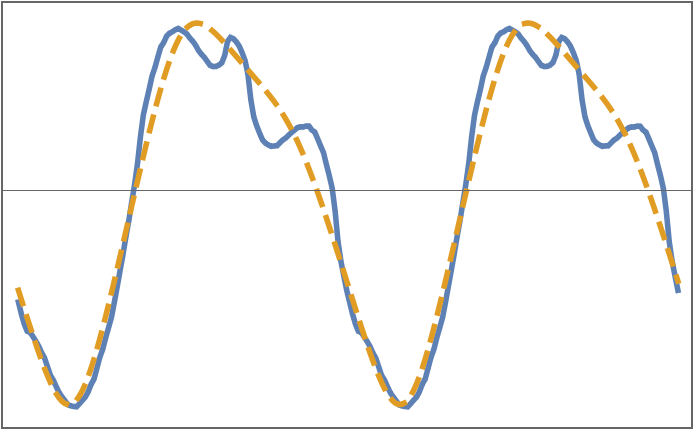}}    &   \raisebox{-0.49\height}{\includegraphics[height=40pt]{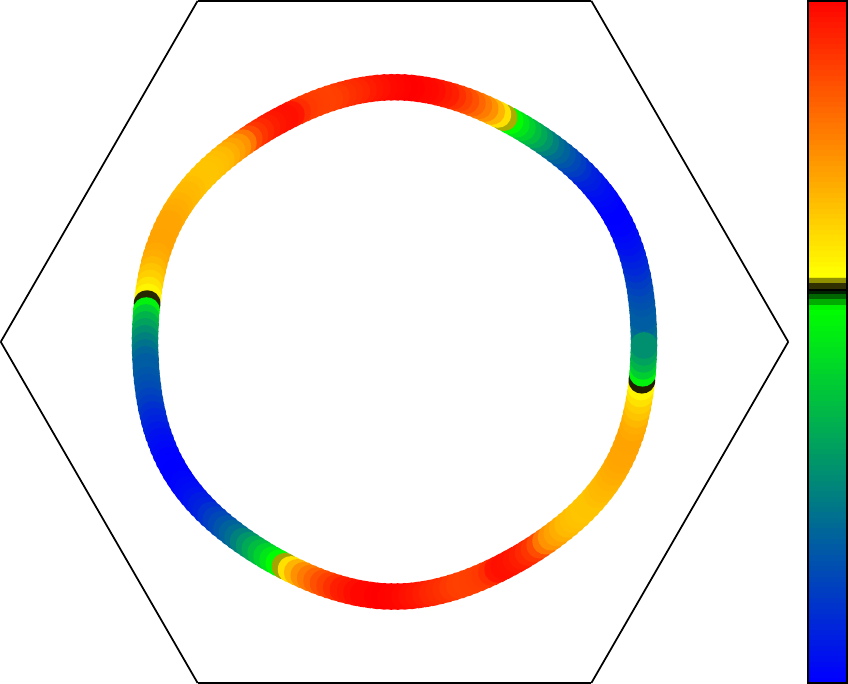}} \\[5pt]
 & {\small $xy$}  &   \raisebox{-0.49\height}{\includegraphics[height=40pt]{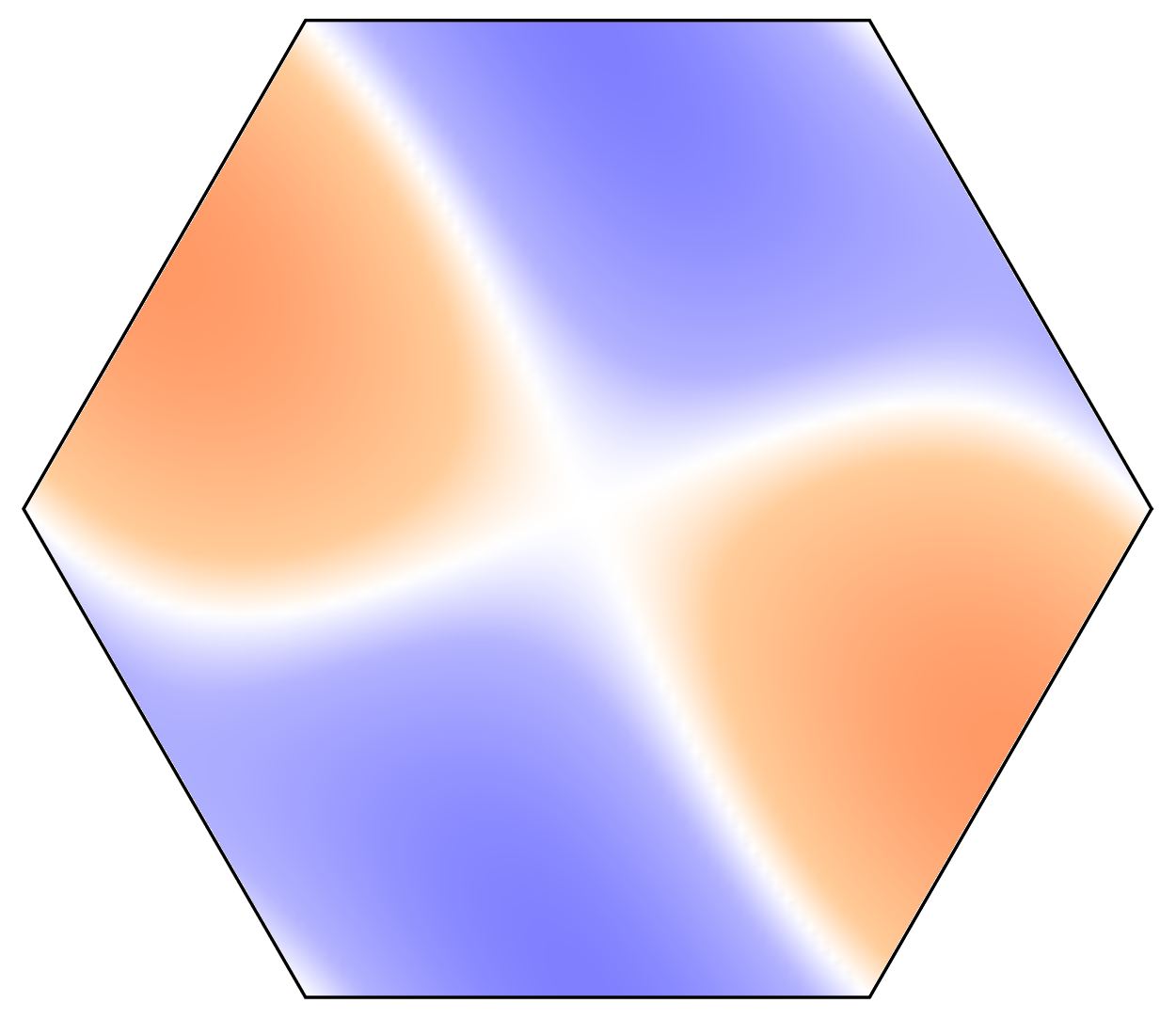}} &   \raisebox{-0.49\height}{\includegraphics[height=40pt]{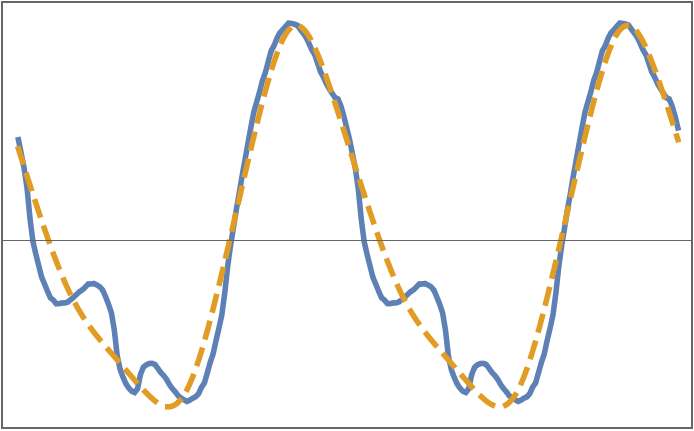}}    &   \raisebox{-0.49\height}{\includegraphics[height=40pt]{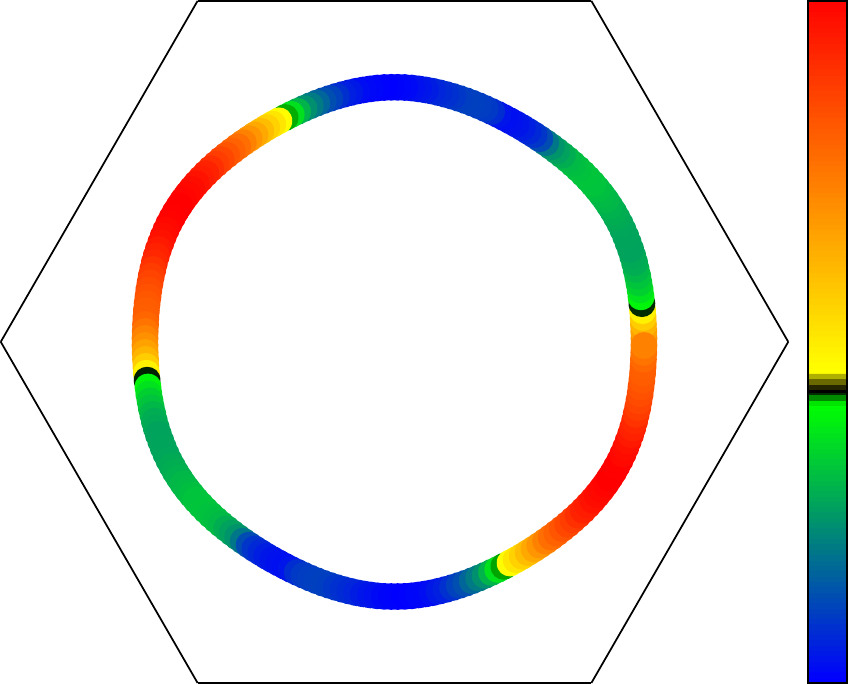}} \\[4pt]
 \hline
\end{tabular}
\caption{All irreps of the group $D_{6}$ with their corresponding basis functions with the lowest appearing angular momentum. Columns from left to right: label of the irrep, basis function, contour plot of the basis function, plot of the basis function (orange, dashed) and example form factors (blue, solid) along the Fermi surface, and plot of the form factor on the Fermi surface. The representative examples shown are taken from the triangular and honeycomb lattices. Here, $x$ and $x^{2}$ are again symbolic short notations for the respective $\sin$ and $\cos$ functions that appear in the perdiodic lattice system. The basis vectors that span the 2D spaces of $E_{1}$ and $E_{2}$ are rotated by $\pi/4$, such that the nodes and maxima of the form factors are arranged on the Fermi surface to maximize the condensation energy.}
\label{fig:waves_tri}
\end{figure}

\begin{figure}[b]
\centering
 \includegraphics[width=0.99\columnwidth]{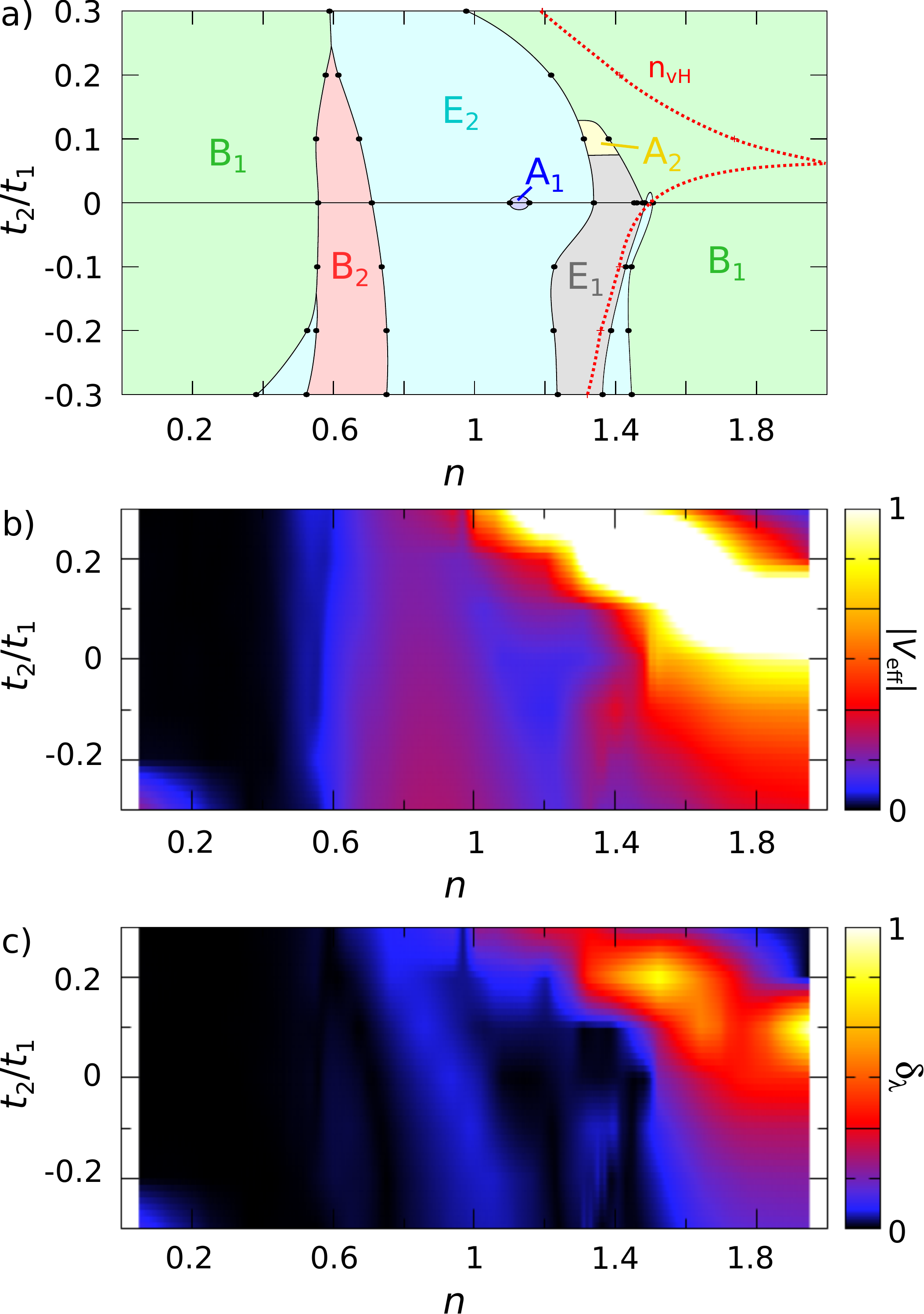}
\caption{(a) Phase diagram $t_{2}/t_{1}$ vs.~band filling $n$ for the triangular lattice. The van Hove filling $n_{vH}$ is drawn as a red dotted line. For fillings $n<0.5$, the effective interaction, $V_{\text{eff}}$, becomes very small, such that the difference between different symmetries reaches the level of numerical noise. (b) Effective interaction, $V_{\text{eff}}$, as a function of the band filling $n$ and the second neighbor hopping $t_{2}/t_{1}$. Regions between calculated points have been interpolated. (c) Difference $\delta_\lambda$ of the two lowest eigenvalues corresonding to the superconducting ground state and the solution that is closest to it.}
\label{fig:triangular_phased_t2-n}
\end{figure}

The triangular lattice is the most fundamental example for a lattice that is not bipartite. The tight-binding bandstructure $E(\vec{k})$ involving nearest and second-nearest-neighbor hoppings ($n_r=2$) is given by Eq.\,(\ref{eq:general_disp_rel}) with
\begin{align}
 &\varepsilon_{1}(\vec{k})=2\cos(k_{x})+4\cos\left(\frac{1}{2}k_{x}\right)\cos\bigg(\frac{\sqrt{3}}{2}k_{y}\bigg), \\
 &\varepsilon_{2}(\vec{k})=2\cos(\sqrt{3}k_{y})+4\cos\left(\frac{3}{2}k_{x}\right)\cos\bigg(\frac{\sqrt{3}}{2}k_{y}\bigg).
\end{align}
Examples of Fermi surfaces are shown in the right column of Fig.\,\ref{fig:triangular_phased_t2-n}.
As for the square lattice, we only consider one orbital per site and have $M=1$ in all vertex functions.
The symmetry group of the triangular lattice is $D_{6}$, which contains the one-dimensional irreps $A_{1}$, $A_{2}$, $B_{1}$, $B_{2}$, and the two-dimensional irreps $E_{1}$ and $E_{2}$. $A_{1}$, $A_{2}$, and $E_{2}$ are the spin singlet representations while $B_{1}$, $B_{2}$, and $E_{1}$ are the triplets. The corresponding basis functions are shown in Fig.~\ref{fig:waves_tri} along with their lowest lattice harmonics. 

The $t_2/t_1$--$n$ phase diagram for $\alpha=0$ and $-0.3\leq t_{2}/t_{1}\leq 0.3$ and the effective interaction of the leading instability are shown in Fig.\,\ref{fig:triangular_phased_t2-n}. For $t_{2}/t_{1}=0$ our results are in good agreement with Ref.\,\onlinecite{raghu_superconductivity_2010}. The largest phase centered around half-filling is the chiral, topological phase with $E_2$ symmetry corresponding to $d+id$ superconductivity. Due to the $D_6$ symmetry, the two $d$-wave states are always degenerate and, following the previously formulated energetic argument, favor to form the chiral state. The $E_2$ phase is surrounded by narrow phases with $E_1$ symmetry for 
larger dopings and with $B_2$ phase for lower dopings. The regions with very high or very low doping, respectively, is occupied by the phase with $B_1$ symmetry. Both the $B_1$ and $B_2$ irreps are odd-parity superconducting states with different $f$-wave symmetries. The two-dimensional $E_1$ irrep is the $p+ip$ state previously discussed for the square lattice.

\begin{figure}[t!]
\centering
 \includegraphics[width=0.60\columnwidth]{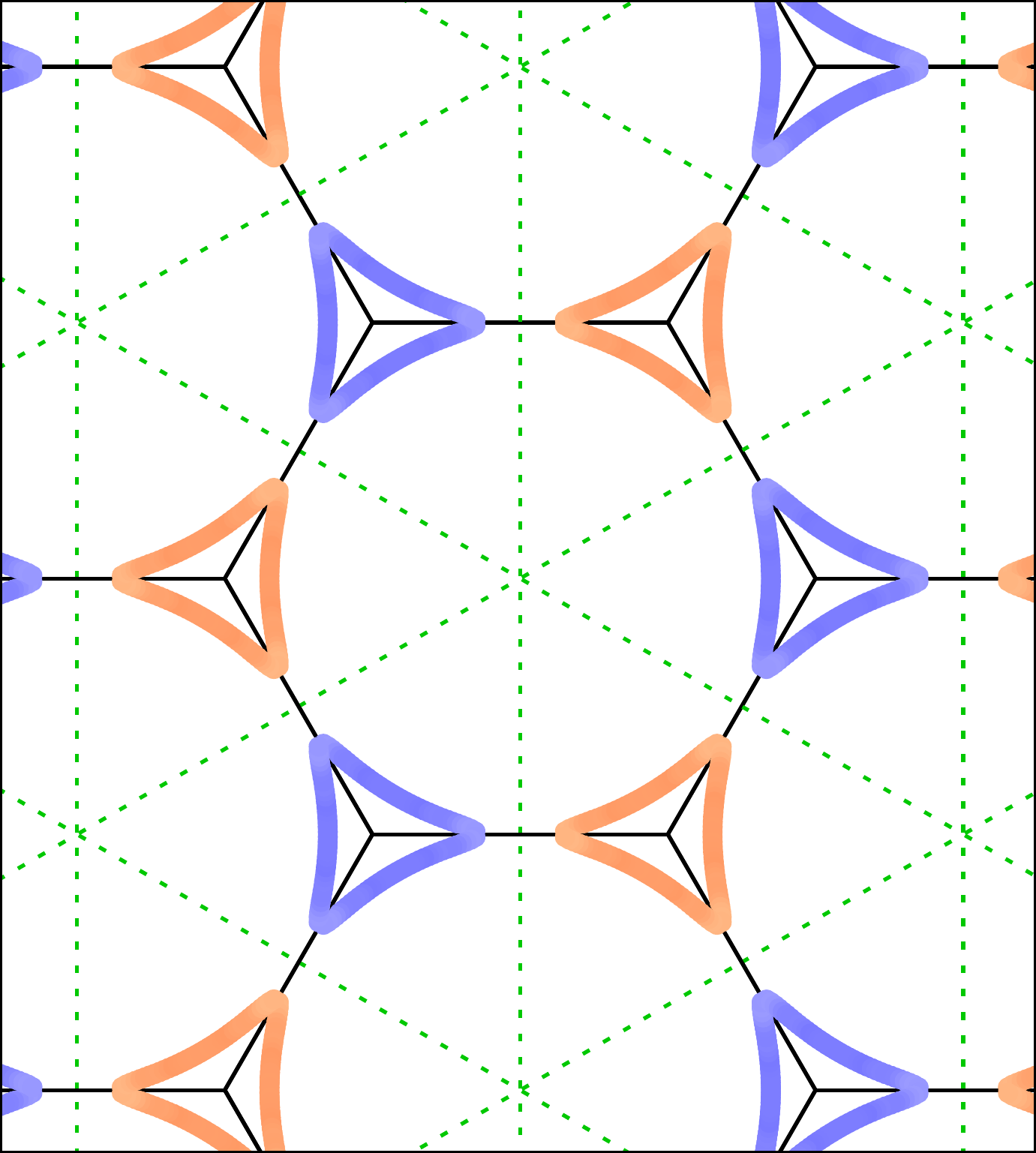}
\caption{Form factor of the superconducting instability on the Fermi surface in the extended Brillouin zone of the triangular lattice. The lattice parameters are $t_{2}/t_{1}=0.1$, $n=1.83$, and $\alpha=0$. The Brillouin zone is drawn in black and the green dashed lines are the nodal line of the $f$-wave form factor of irrep $B_{1}$. Blue and orange colors denote different signs of the form factor.}
\label{fig:triangular_B1_nodeless}
\end{figure}

In Fig.\,\ref{fig:triangular_phased_t2-n} we show the effective coupling strengths of the leading instability, \ie the superconducting groundstate. We observe that for $n<0.5$ superconductivity is generally suppressed. Particularly high critical temperatures can be expected for fillings $n$ above the van Hove singularity, in particular when second-neighbor hoppings are finite and positive. In the regions of largest $V_{\rm eff}$, spin triplet $f$-wave superconductivity (irrep $B_{1}$) is dominating. In Fig.\,\ref{fig:triangular_phased_t2-n}\,c)  the gap $\delta_\lambda$ between lowest and second-lowest eigenvalue of the two-particle vertex function $g$ is shown. As for the square lattice, it resembles the behavior of $V_{\rm eff}$. The reason why the state of irrep $B_{1}$ is strongly dominating for $n>n_{\text{vH}}$ and positive $t_{2}/t_{1}$ is easily understood by considering the geometry of the Fermi surface, shown in Fig.\,\ref{fig:triangular_B1_nodeless}. Here, we see that the nodal lines of the $f$-wave of irrep $B_{1}$ do not touch the Fermi surface. Consequently, the same energetic statement as for the stability of the chiral topological states in the two dimensional irreps holds: the superconducting state is particularly stable when nodes can be avoided. Note, however, that for purely repulsive electron-electron interactions this case is only possible when there are at least two Fermi surface pockets.

For larger values of $t_{2}/t_{1}$, the Fermi surface will move into the nodal lines of the $f$-wave form factor of $B_{1}$ again. The geometry of the Fermi surface in this region is captured in the last column for $B_{1}$ in Fig.\,\ref{fig:waves_tri}. However, this $f$-wave state still dominates, since it is the state that maximizes the superconducting order parameter close to the van Hove points.

\begin{figure}[t!]
\centering
 \includegraphics[width=0.99\columnwidth]{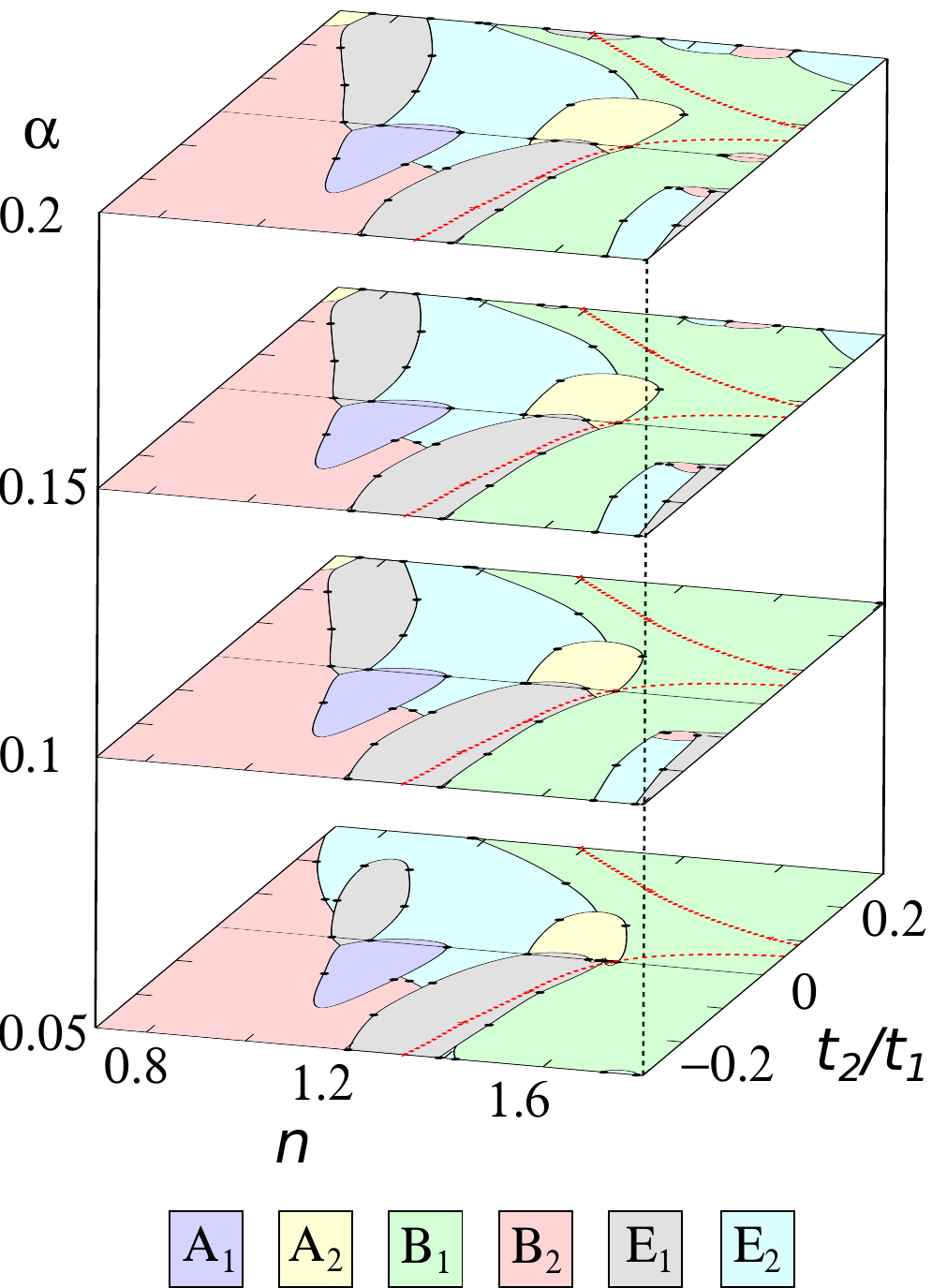}
\caption{Phase diagram showing the irrep of the leading superconducting instability in the triangular lattice for $0.05\leq\alpha\leq0.2$, $0.7\leq n\leq1.8$ and $-0.3\leq t_{2}/t_{1}\leq0.3$.}
\label{fig:triangular_phased_3D}
\end{figure}

In the following, we will discuss the effect of longer-ranged interactions. The form factor of \eqref{eq:NN-rep} can be written as
\begin{equation}
  \varepsilon_1(\vec{k}_{2}-\vec{k}_{1})=\sum_{i}\psi_{i}^{D_{6},1}(\vec{k}_{1})\psi_{i}^{D_{6},1}(\vec{k}_{2}),
\end{equation}
where the sum runs over all nearest neighbor lattice harmonics, \ie $i\in\{A_{1},A_{2},B_{1},B_{2},E_{1},E_{2}\}$.
On the triangular lattice, these correspond to the following basis functions:
\begin{align}
 &\psi_{A_{1}}^{D_{6},1}(\vec{k})=\sqrt{\frac{2}{3}}\left[\cos(k_{x})+2\cos\left(\frac{1}{2}k_{x}\right)\cos\left(\frac{\sqrt{3}}{2}k_{y}\right)\right], \nonumber\\
 &\psi_{B_{1}}^{D_{6},1}(\vec{k})=\sqrt{\frac{2}{3}}\left[\sin\left(k_{x}\right)-2\sin\left( \frac{1}{2}k_{x} \right)\cos\left( \frac{\sqrt{3}}{2}k_{y} \right)\right], \nonumber\\
 &\psi_{E_{1},1}^{D_{6},1}(\vec{k})=\frac{2}{\sqrt{3}}\left[\sin(k_{x})+\sin\left(\frac{1}{2}k_{x}\right)\cos\left(\frac{\sqrt{3}}{2}k_{y}\right)\right], \nonumber\\
 &\psi_{E_{1},2}^{D_{6},1}(\vec{k})=2\cos\left(\frac{1}{2}k_{x}\right)\sin\left(\frac{\sqrt{3}}{2}k_{y}\right), \nonumber\\
 &\psi_{E_{2},1}^{D_{6},1}(\vec{k})=\frac{2}{\sqrt{3}}\left[\cos(k_{x})-\cos\left(\frac{1}{2}k_{x}\right)\cos\left(\frac{\sqrt{3}}{2}k_{y}\right)\right], \nonumber\\
 &\psi_{E_{2},2}^{D_{6},1}(\vec{k})=2\sin\left(\frac{1}{2}k_{x}\right)\sin\left(\frac{\sqrt{3}}{2}k_{y}\right), \nonumber\\
 &\psi_{A_{2}}^{D_{6},1}(\vec{k})=\psi_{B_{2}}^{D_{6},1}(\vec{k})=0. 
\end{align}

 The phase diagram for $0.05\leq\alpha\leq0.2$, $-0.3\leq t_{2}/t_{1}\leq0.3$, and $0.7\leq n\leq1.8$ is shown in Fig.\,\ref{fig:triangular_phased_3D}. Without nearest-neighbor interactions, the leading superconducting instabilities are mostly of triplet type ($B_{1}$) or chiral topological states ($E_{1}$ and $E_{2}$). The biggest change happens for small $\alpha$: especially the chiral singlet states (irrep $E_2$) for $n<1$ are suppressed, whereas new domains of chiral triplet states appear (irrep $E_1$) and the pocket with extended $s$-wave ($A_1$) becomes large.
We note, however, that there are no dramatic changes in the range $0.1<\alpha<0.2$ in the phase diagram. In the regime of moderate negative $t_2$ changes due to $\alpha$ are almost absent; in contrast, for positive moderate $t_2$ several small pockets with $f$-wave, $i$-wave or $E_{1/2}$ symmetry are induced. In addition, we observe that the $i$-wave phase for positive $t_2$ around $n\approx 1.4$ becomes stabilized and increased towards smaller $t_2$. At $\alpha=0.2$, the $i$-wave phase is even present at $t_2=0$.

As for the square lattice, we could identify superconducting solutions for all irreps belonging to the $D_6$ symmetry group. It is important to stress that essentially all unconventional superconducting phases on the triangular lattice are either of chiral topological type or spin-triplet states, which are also topologically nontrivial as discussed in the introduction. While one could have guessed this already from the list of irreps, we find that the chance to obtain a topologically non-trivial state on the triangular lattice is extremely high.

%
%
\subsection{Honeycomb lattice}

The honeycomb lattice is the third paradigmatic lattice we are studying. Like the square lattice it is bipartite. Like the triangular lattice it is hexagonal. Unlike both square and triangular lattices, it features a two-atomic unit cell yielding two bands. Both bands touch each other at special points, the so-called Dirac points, located at the corners of the Brillouin zone. In the vicinity of these points, the dispersion relation corresponds to the spectrum of a Dirac Hamiltonian. The tight-binding band structure $E(\vec{k})$ involving nearest and second-nearest-neighbor hoppings ($n_r=2$) is given by Eq.\,(\ref{eq:general_disp_rel}) with
\begin{align}\label{hc-nn-band}
 &\varepsilon_{1}(\vec{k})=\pm\sqrt{3+2\cos(\sqrt{3}k_{x})+4\cos\bigg(\frac{\sqrt{3}}{2}k_{x}\bigg)\cos\left(\frac{3}{2}k_{y}\right)}, \\
 &\varepsilon_{2}(\vec{k})=2\cos(\sqrt{3}k_{x})+4\cos\bigg(\frac{\sqrt{3}}{2}k_{x}\bigg)\cos\bigg(\frac{3}{2}k_{y}\bigg)\ .\label{hc-nnn-band}
\end{align}
The upper (lower) band corresponds to the plus (minus) sign in Eq.\,\eqref{hc-nn-band}.
Examples of Fermi surfaces are similar to those shown in the right column of Fig.\,\ref{fig:triangular_phased_t2-n}.
In contrast to the other lattices, now we deal with two orbitals per unit cell and $M(1,2,3,4)$ appearing in the vertex functions.
The symmetry group of the honeycomb lattice still is $D_{6}$ and the discussion of the triangular lattice applies. Thus we can look up the corresponding basis functions in Fig.\,\ref{fig:waves_tri}.

\begin{figure}[t]
\centering
 \includegraphics[width=0.99\columnwidth]{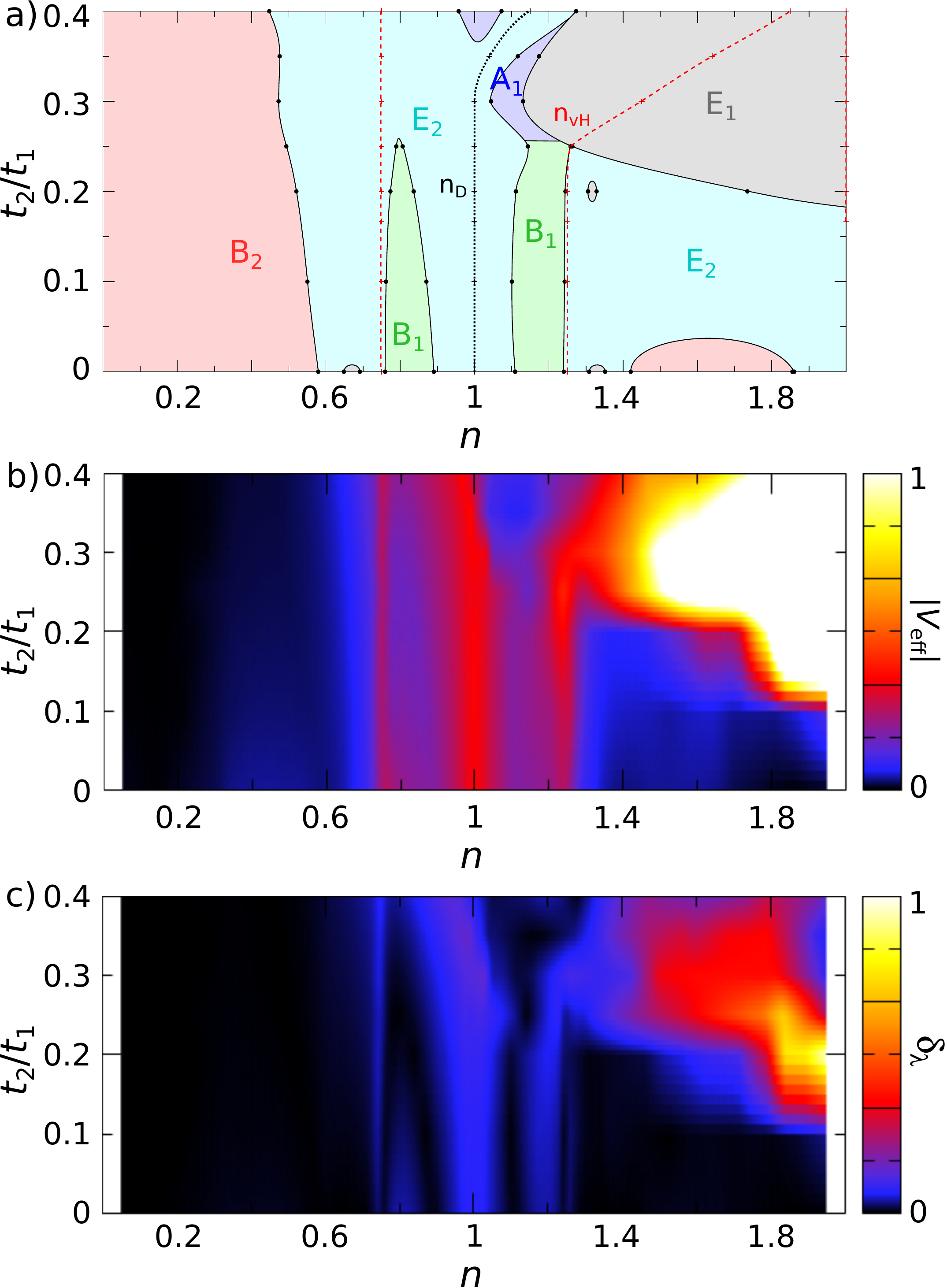}
\caption{(a) Phase diagram $t_{2}/t_{1}$ vs.~band filling $n$ for the honeycomb lattice. The van Hove fillings $n_{\text{vH}}$ are drawn as a red dotted line, the filling $n_{D}$ at which the Dirac points are located is drawn as a black dotted line. (b) Effective interaction, $V_{\text{eff}}$, as a function of the band filling $n$ and the second neighbor hopping $t_{2}/t_{1}$. Regions between calculated points have been interpolated. (c) Difference $\delta_\lambda$ of the two lowest eigenvalues corresonding to the superconducting ground state and the solution which is closest to it.}
\label{fig:honeycomb_phased_t2-n}
\end{figure}

The superconducting $t_2/t_1$--$n$ phase diagram, Fig.\,\ref{fig:honeycomb_phased_t2-n}, is somewhat different from the previous cases: now we have two bands and for $|t_{2}/t_{1}|<1/3$ the ``half-filled'' case, $n=1$, corresponds to a completely filled lower band. For $t_{2}=0$, the honeycomb lattice for $0<n<1$ is somewhat similar to the triangular lattice for $0<n<2$. In the honeycomb lattice, we find that due to the nonconstant orbital factors, $M(1,2,3,4)$, superconductivity for $n\lesssim 0.6$ is suppressed (\ie $V_{\rm eff}$ is very small), and chiral topological singlet states (irrep $E_{2}$) become the leading instability close to the Dirac filling, $n=1$.
As for the triangular lattice, we find that mostly spin-triplet phases ($f$-wave with irreps $B_1$ or $B_2$) or chiral topological phases ($p+ip$ or $d+id$ with irreps $E_1$ or $E_2$, respectively) are present. In particular, the region around van Hove filling is dominated by chiral $d$- and $p$-wave superconductivity. For large $t_2/t_1$, also a pocket with extended $s$-wave symmetry (irrep $A_1$) shows up. In agreement with van Hove fillings, the effective coupling strength is much larger for fillings $n>1$ and finite $t_2/t_{1}$. Especially the chiral $p$-wave state for large $n$ and $t_{2}$ shows an extraordinarily large effective coupling strength.

Eventually we consider the effect of longer-ranged interactions, $\alpha>0$. The form factor of Eq.\,\eqref{eq:NN-rep} can be written as
\begin{align}
  &h_1(\vec{k}_{1}-\vec{k}_{2})=\begin{pmatrix}
                                 0 & T(\vec{k}_{1}-\vec{k}_{2}) \\
                                 T^{*}(\vec{k}_{1}-\vec{k}_{2}) & 0
                                \end{pmatrix},\\
  &T(\vec{k})=e^{-ik_{y}}+e^{i\frac{1}{2}k_{y}}\left(e^{i\frac{\sqrt{3}}{2}k_{x}}+e^{-i\frac{\sqrt{3}}{2}k_{x}}\right),
\end{align}
with
\begin{align}
 T(\vec{k}_{1}-\vec{k}_{2})=\frac{1}{\sqrt{3}}\sum_{i}\psi_{i}^{D_{3},1}(\vec{k}_{1})\left(\psi_{i}^{D_{3},1}(\vec{k}_{2})\right)^{*},
\end{align}
where the sum runs over the nearest neighbor lattice harmonics, \ie $i\in\{A_{1},A_{2},E\}$.
On the honeycomb lattice, these correspond to the following basis functions:
\begin{align}
 &\psi_{A_{1}}^{D_{3},1}(\vec{k})=\frac{1}{\sqrt{3}}\left[e^{-ik_{y}}+e^{i\frac{1}{2}k_{y}}\left(e^{i\frac{\sqrt{3}}{2}k_{x}}+e^{-i\frac{\sqrt{3}}{2}k_{x}}\right)\right], \nonumber\\
 &\psi_{E,1}^{D_{3},1}(\vec{k})=\frac{1}{\sqrt{6}}\left[2e^{-ik_{y}}-e^{i\frac{1}{2}k_{y}}\left(e^{i\frac{\sqrt{3}}{2}k_{x}}+e^{-i\frac{\sqrt{3}}{2}k_{x}}\right)\right], \nonumber\\
 &\psi_{E,2}^{D_{3},1}(\vec{k})=\frac{1}{\sqrt{2}}\,e^{i\frac{1}{2}k_{y}}\left(e^{i\frac{\sqrt{3}}{2}k_{x}}-e^{-i\frac{\sqrt{3}}{2}k_{x}}\right), \nonumber\\
 &\psi_{A_{2}}^{D_{3},1}(\vec{k})=0
\end{align}
Note that the set of nearest-neighbor vectors on the honeycomb lattice has lower symmetry ($D_{3}$) than the point group of the lattice itself ($D_{6}$). Also, $D_{3}$ is a subgroup of $D_{6}$, hence, the symmetry of the superconducting order parameters is still fully described by the $D_{6}$ group.

Comparing the possible basis functions of $D_{3}$ and $D_{6}$, we can observe the correspondence
\begin{align}
 &A_{1}^{D_{3}}\rightarrow A_{1}^{D_{6}},B_{1}^{D_{6}},\nonumber\\
 &A_{2}^{D_{3}}\rightarrow A_{2}^{D_{6}},B_{2}^{D_{6}},\nonumber\\
 &E^{D_{3}}\rightarrow E_{1}^{D_{6}},E_{2}^{D_{6}},\nonumber
\end{align}
\ie we can expect the nearest-neighbor interaction on the honeycomb lattice to affect only the lowest order basis functions of the irreps $A_{1}$, $B_{1}$, $E_{1}$, and $E_{2}$, similar to the triangular lattice. However, we do not observe changes for $B_{1}$.

\begin{figure}[t!]
\centering
 \includegraphics[width=0.99\columnwidth]{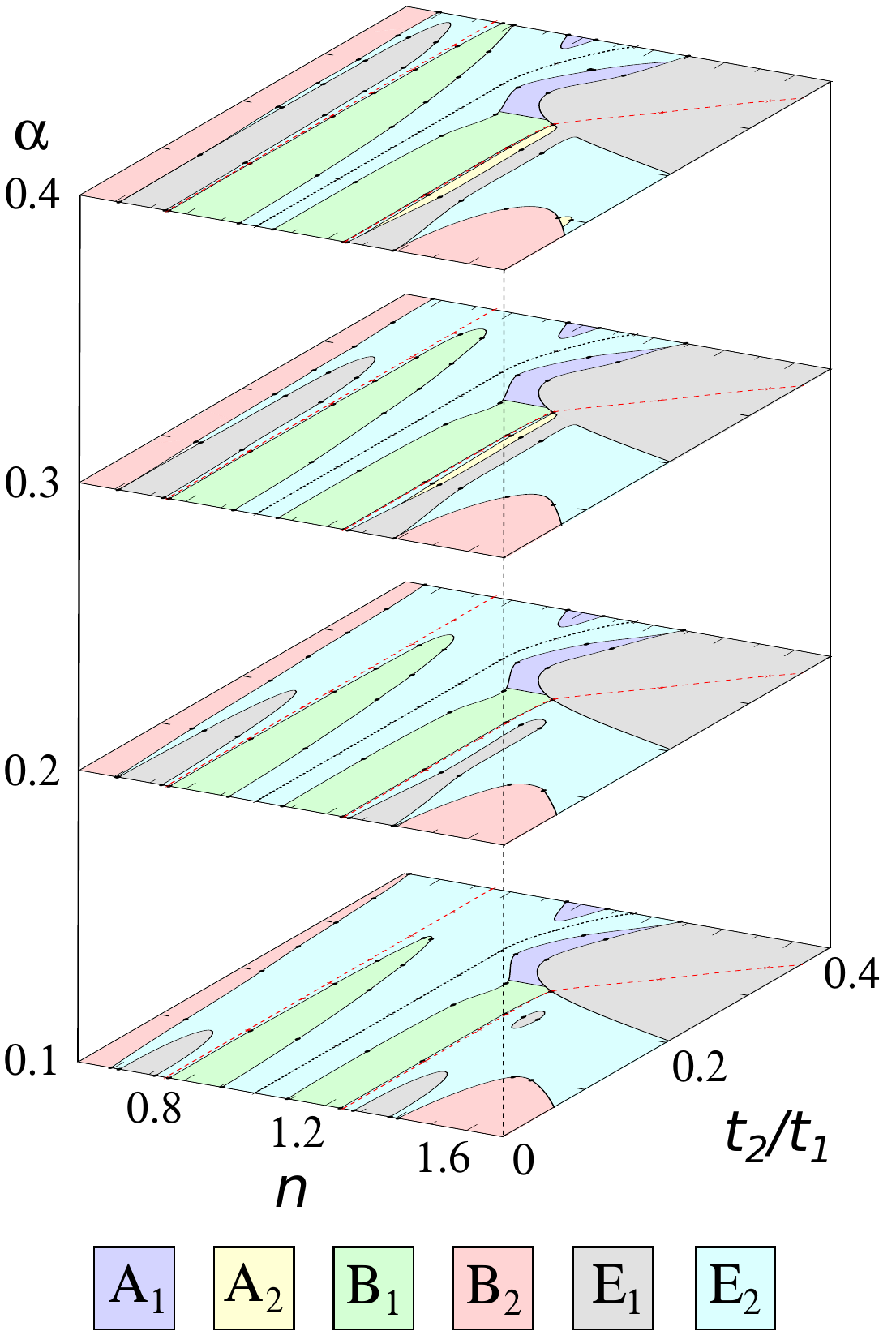} 
\caption{Phase diagram showing the irrep of the leading superconducting instability in the honeycomb lattice for $0.1\leq\alpha\leq0.4$, $0.5\leq n\leq1.7$ and $0\leq t_{2}/t_{1}\leq0.4$.}
\label{fig:honeycomb_phased_3D}
\end{figure}

The main effect of including the nearest-neighbor interactions is the suppression of the chiral singlet states ($E_2$) in favor of the chiral triplet states ($E_{1}$), the $f$-wave states ($B_{1}$ and $B_{2}$), and a small pocket of $i$-wave ($A_{2}$) close to the van Hove singularity at $n\approx1.25$, as shown in the phase diagram for $0.1\leq\alpha\leq0.4$ in Fig.\,\ref{fig:honeycomb_phased_3D}. However, the $i$-wave state is almost degenerate to the chiral singlet state and thus susceptible to small perturbations. As for the case $\alpha=0$, the effective interaction $V_{\rm eff}$ is almost negligible for $n<1$ but particularly strong for large $n$ and finite $t_2/t_1$.

Again we could identify superconducting solutions for all irreps belonging to the $D_6$ symmetry group.
As for the triangular lattice case, almost all superconducting groundstates are topologically nontrivial -- either they constitute spin-triplet superconductivity or they realize a chiral topological superconductor (or both in case of $E_1$ irrep). The honeycomb lattice is thus as good as the triangular lattice to search for topological superconductivity.

\subsection{Summary of results and discussion}
For all paradigmatic 2D lattices studied in this section, we find that the effective interaction, $V_{\text{eff}}$, is especially high close to fillings with a van Hove singularity in the density of states, as expected. Also, we find all pairing symmetries that are possible within the symmetry group of the respective lattices by varying the nearest-neighbor hopping, $t_{2}$, and nearest-neighbor interaction, $U_{1}$ (quantified through $\alpha$). All the phase diagrams we find have large domains of chiral superconducting states: chiral triplet states such as $p_x \pm i p_y$, but the hexagonal lattices can also host chiral singlet states such as $d_{x^2-y^2} \pm i d_{xy}$.

Concerning the magnitude of $T_{c}$ ($\sim V_{\text{eff}}$), our results suggest that chiral superconducting phases with reasonably high $T_{c}$ may be found in all the studied lattice systems, as shown in Figs.\,\ref{fig:square_phased_t2-n}, \ref{fig:triangular_phased_t2-n}, and \ref{fig:honeycomb_phased_t2-n}. The honeycomb lattice is the most promising candidate to search for high temperature topological superconductivity.

Our results presented above are only exact in the limit $U\rightarrow0$. Note, however, that comparison with other methods (see for instance Refs.\,\onlinecite{metzner_functional_2012,platt_functional_2013,vucicevic_trilex_2017,cao_chiral_2018}) that yield promising results even for stronger interactions shows that the superconducting instabilities found within the WCRG approach often coincides with those found for stronger interactions. 

Quite generally, there is no reason to assume that a weak-coupling instability coincides with a strong-coupling instability. In principle, there might be one or even several phase transitions under increasing coupling strength. When we have a closer look, however, at some of the most important material classes, namely cuprates, pnictides, ruthenates but also the honeycomb-lattice Hubbard model systems, we find a remarkable correspondence between weak-coupling and strong-coupling instabilities. Let us first consider the cuprates: in the weak-coupling regime, the dominating instability is located at momentum $(\pi,\pi)$ with superconducting $d_{x^2-y^2}$ symmetry; in the strong-coupling regime, \ie for the corresponding spin model, again the dominating instability is located at $(\pi,\pi)$ leading to antiferromagnetic Neel order (for an extensive review see Ref.\,\onlinecite{lee_doping_2006}). The second prominent example are the pnictides, where the dominating superconducting fluctuations are at $(\pi,0)$ and $(0,\pi)$\,\cite{wang_functional_2009,wang_electron-pairing_2011}; the strong-coupling analysis leads to a spin Hamiltonian featuring columnar antiferromagnetic order with magnetic Bragg peaks located at $(\pi,0)$ and $(0,\pi)$\,\cite{seo_pairing_2008}. A similar reasoning is valid for the ruthenates and a class of honeycomb-lattice Hubbard models, showing that this line of argument is not exclusive for the square lattice. While there is no exact method at intermediate coupling strength, the above-mentioned examples are so extensively studied that it is widely accepted that there are no (superconducting) intermediate phases and phase transitions from the superconducting weak-coupling phase into another superconducting intermediate-coupling phase do not occur. Thus we can conclude that, at least empirically, the weak-coupling RG method finds superconducting solutions which are not only exact in the limit $U\to 0$ but which also provide often a guiding principle for stronger-correlated regimes or materials.

%
%
\section{Numerical development and performance analysis}

At some regions in the $t_{2}/t_1$--$n$ phase space, the second lowest eigenvalue of the scattering amplitude matrix $g$ is very close to the lowest one, see panels c) in Figs.\,\ref{fig:square_phased_t2-n}, \ref{fig:triangular_phased_t2-n}, and \ref{fig:honeycomb_phased_t2-n}. These regions are particularly prone to the smallest variations in the parameters or perturbations, including variations in the numerical resolution.
In other words, it is more challenging to obtain the true groundstate. 

\begin{figure}[t]
\centering
 \includegraphics[width=0.99\columnwidth]{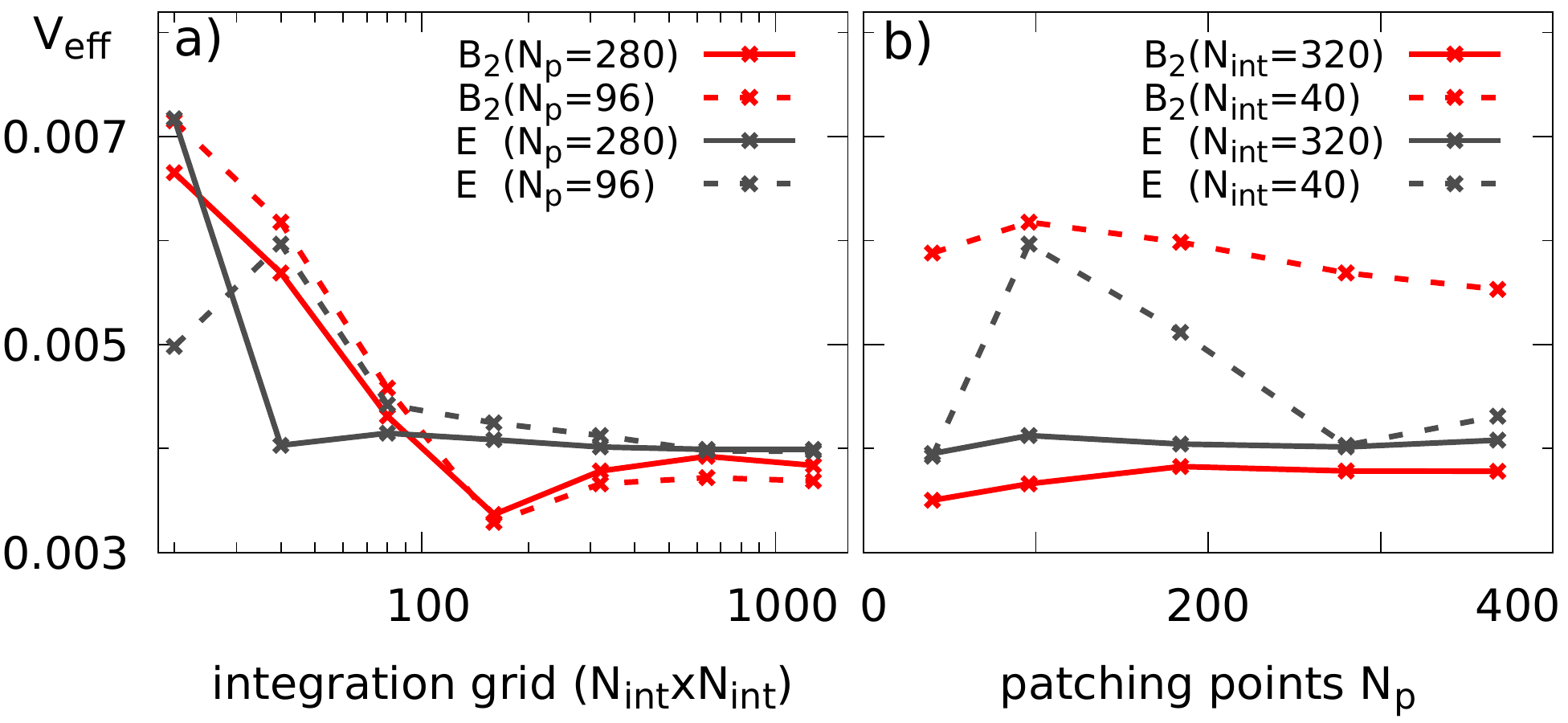}
\caption{Effects of low accuracy of the integration grid, $N_{\text{int}}$, and of the number of patching points, $N_{p}$, on the effective interaction $V_{\text{eff}}$. The irrep with the largest $V_{\text{eff}}$ is the leading superconducting instability. The figures shown are for the square lattice with $t_{2}=\alpha=0$ and $n\approx0.54$. (a) Effective interaction, $V_{\text{eff}}$, of the $B_{2}$ and $E$ irreps as a function of $N_{\text{int}}$. (b) $V_{\text{eff}}$ as a function of $N_{p}$.}
\label{fig:accuracy_sql}
\end{figure}

In this section, we demonstrate the importance of the accuracy in terms of the number of patching points, $N_{p}$, used to discretize the eigenvalue equation \eqref{eq:gmat}, and the number of points for the integration grid, $N_{\text{int}}$, which refers to the numerical evaluation of Eq.\,\eqref{eq:Gamma_2b}, \ie the number of discretization points in each momentum dimension of the integral in Eq.\,\eqref{eq:short_int_1}. As an example, we take the regime $0.5\leq n \leq 0.6$ and $t_{2}=\alpha=0$ on the square lattice, where the $d_{xy}$ ($B_{2}$) and $p_{x}+ip_{y}$ ($E$) symmetries are almost degenerate. Figure~\ref{fig:accuracy_sql} shows the impact of the accuracy of the integration grid and of the number of patching points on the effective interaction $V_{\text{eff}}$ of $B_{2}$ and $E$. For an integration grid with $40\times40$ grid points, the leading superconducting instability is the $B_{2}$ irrep, independent of $N_{p}$. This is true for up to $N_{\text{int}}\approx80$. However, for more dense integration grids, the $E$ representation becomes the leading instability. Thus, too low accuracy in the integration may clearly lead to wrong results for the symmetry of the superconducting state.

Thus we come back to the earlier discussion about the correct groundstate for the square lattice phase diagram. The discrepancy between our result on the one hand and the findings of Ref.\,\cite{raghu_superconductivity_2010} on the other hand  (see section ``Results'') might be understood as an issue of numerical resolution as demonstrated in Fig.\,\ref{fig:accuracy_sql}.
Reference\,\cite{raghu_superconductivity_2010} finds the leading instability with $B_2$ symmetry, while we find it with $E$ symmetry. As explicitly shown in Fig.\,\ref{fig:accuracy_sql}, increase of $N_p$ does not resolve this issue. Only a sufficiently large $N_{\rm int}$ renders the state with $E$ symmetry to be the groundstate.

\begin{figure}[t!]
\centering
 \includegraphics[width=0.3\columnwidth]{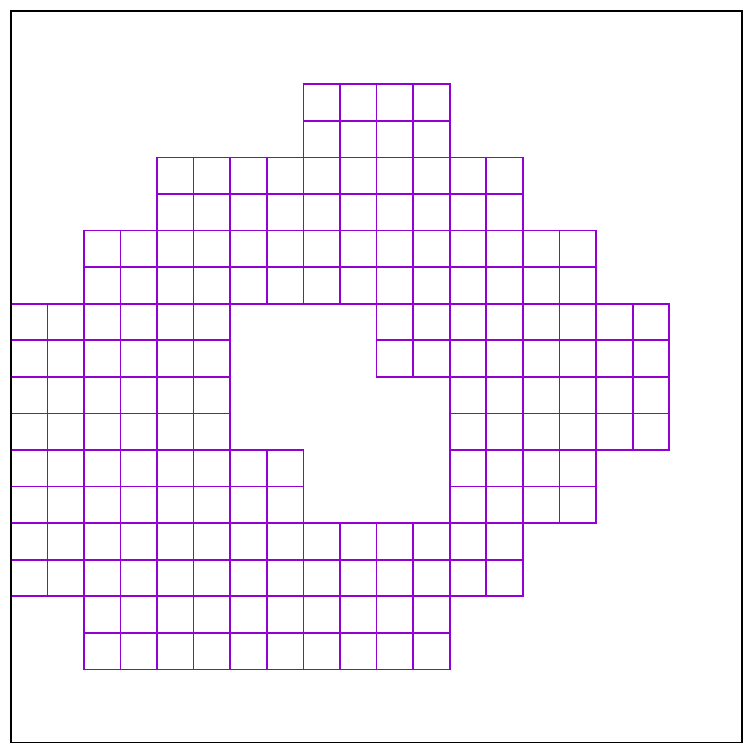}
 \includegraphics[width=0.3\columnwidth]{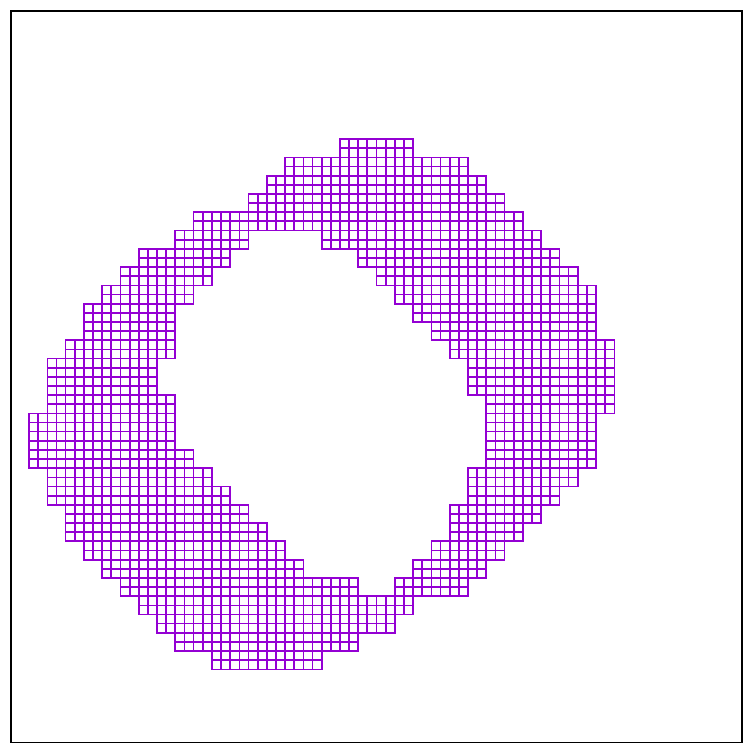}
 \includegraphics[width=0.3\columnwidth]{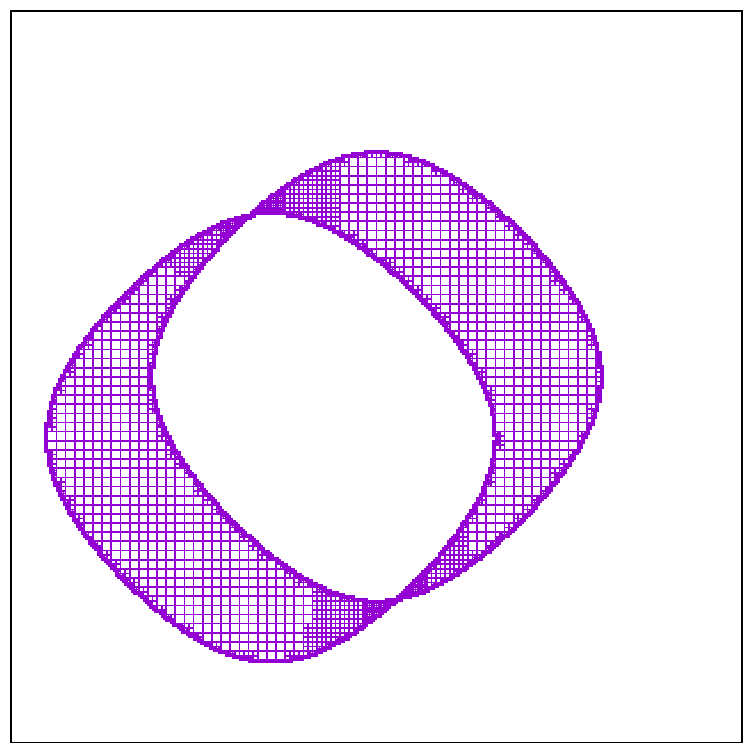}\\
 \includegraphics[width=0.3\columnwidth]{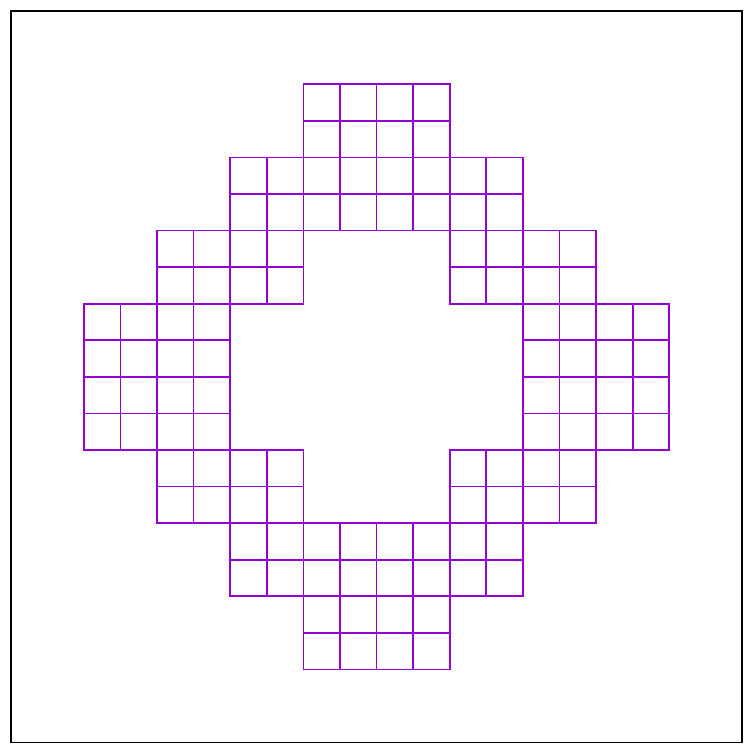}
 \includegraphics[width=0.3\columnwidth]{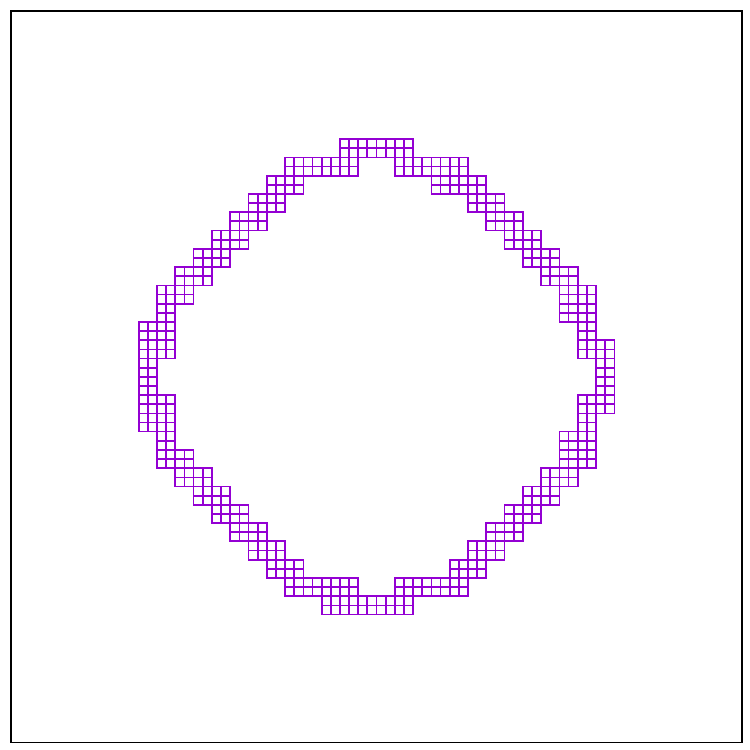}
 \includegraphics[width=0.3\columnwidth]{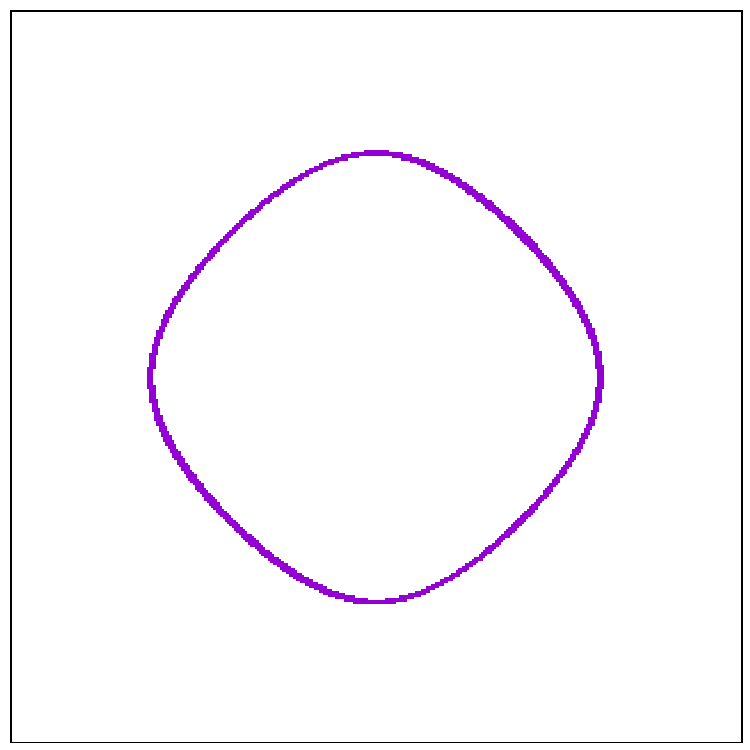}\\
 \includegraphics[width=0.3\columnwidth]{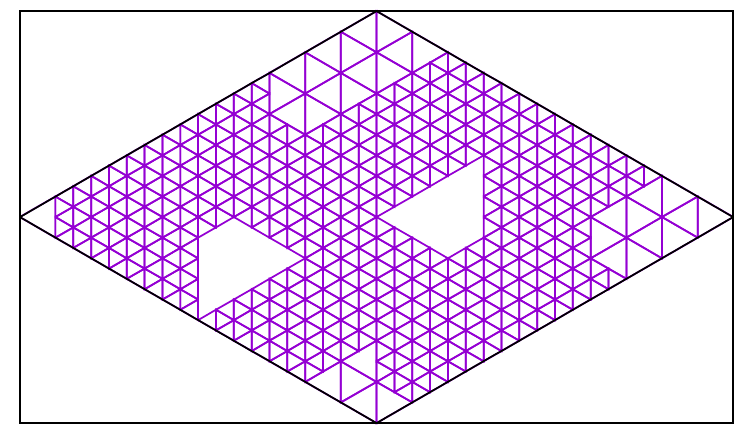}
 \includegraphics[width=0.3\columnwidth]{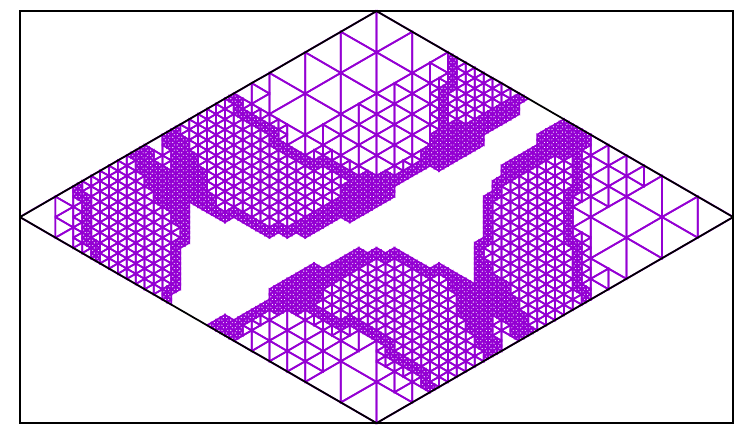}
 \includegraphics[width=0.3\columnwidth]{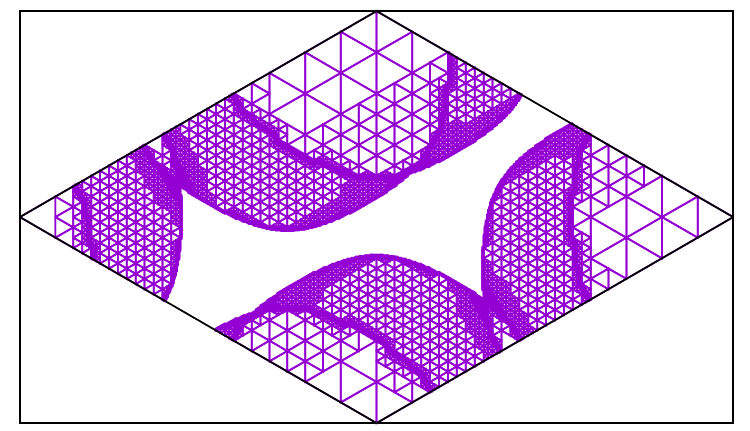}\\
 \includegraphics[width=0.3\columnwidth]{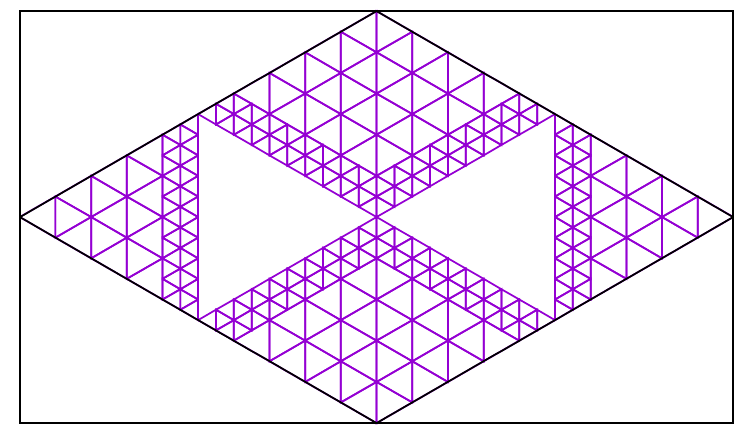}
 \includegraphics[width=0.3\columnwidth]{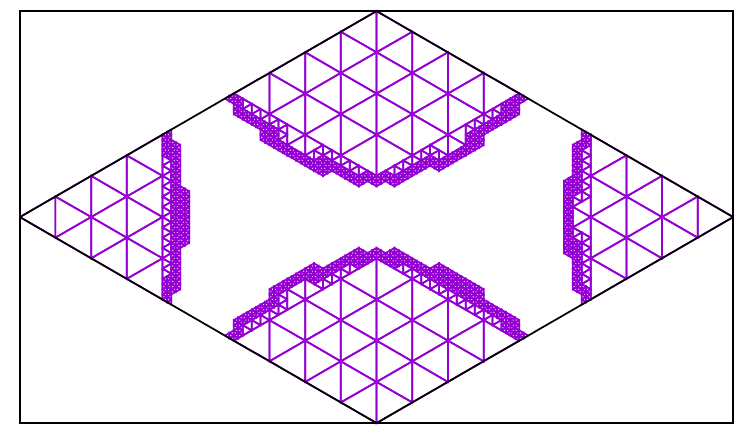}
 \includegraphics[width=0.3\columnwidth]{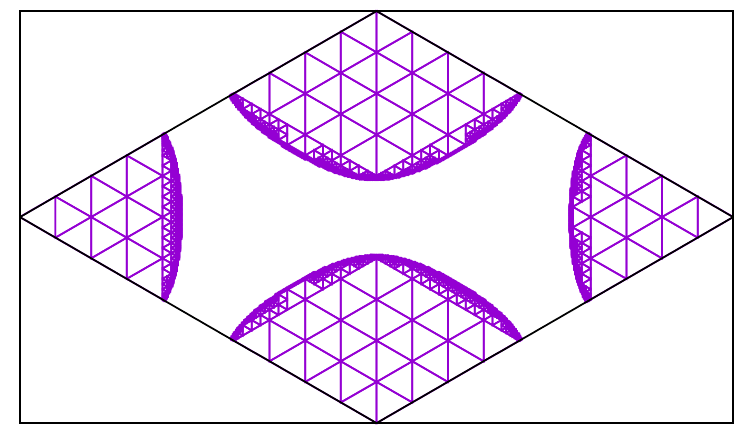}
\caption{Progression of the dynamic grid after 1, 3 and 6 iterations for a starting grid size $N_{0}=10$. First row: square lattice with large $|\vec{k}|$, $t_{2}=0$ and $n\approx0.54$. Second row: square lattice with the same parameters and $|\vec{k}|$ close to zero. Third row: honeycomb lattice with large $|\vec{k}|$, $t_{2}=0$ and $n\approx1.4$. Fourth row: honeycomb lattice with the same system parameters and $|\vec{k}|$ close to zero.}
\label{fig:dynamic_grid}
\end{figure}
%
However, keeping $N_{\text{int}}$ and $N_{p}$ moderately large, also considerably increases the runtime of the routine, which grows $\propto (N_{\text{int}}N_{p})^{2}$. Studying the integrand
\begin{equation}
 \tilde{X}_{ph}(\vec{k},\vec{p})=\sum_{n_{1},n_{2}}X_{\rm ph}(n_{1},\vec{k}+\vec{p};n_{2},\vec{p}),
\end{equation}
where $X_{\rm ph}$ is given by Eq.\,(\ref{eq:chi_integrand}) and $\vec{p}$ is the integration variable, one finds that it consists mainly of areas with little change, narrow regions in which divergencies can be found and large areas where it is zero, all depending on the momentum $\vec{k}$. The integration is done in the $p_{x}$-$p_{y}$ plane. Using a standard quadrature integration routine leaves us with the following dilemma: if we cover the divergencies and narrow areas with fast changing integrand with appropriate accuracy, we also add up unecessarily many zeros and integrate the slow changing regions of the integrand with much higher accuracy than needed. If we take too low accuracy, we just cut out the regions around the divergencies, where the integrand is very large.

\begin{figure}[t]
\centering
 \includegraphics[width=0.99\columnwidth]{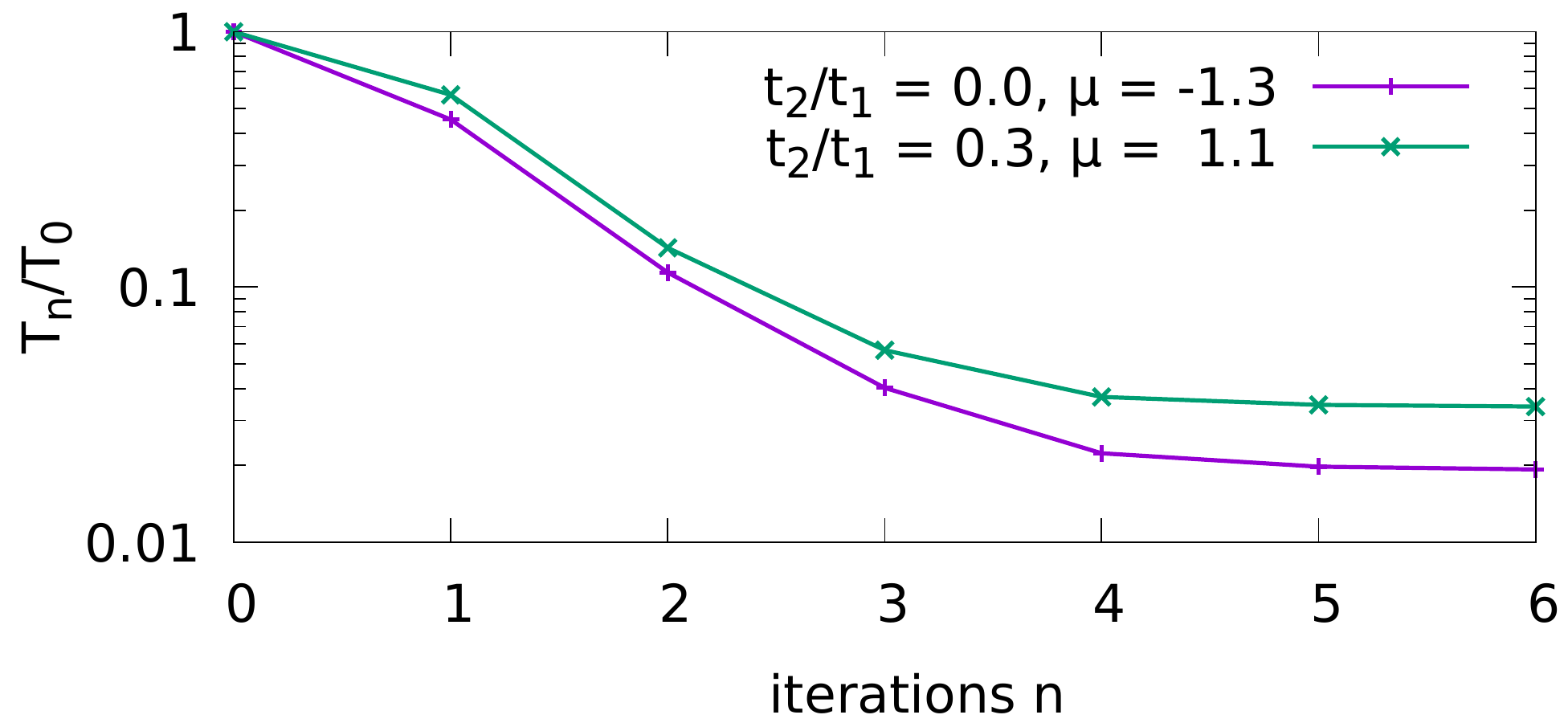}
\caption{Relative runtime, $T_{n}/T_{0}$, as a function of the number of iterations, $n$, while leaving the effective gridsize constant at $N_{\text{eff}}=1280$. $T_{n}$ denotes the runtime when using $n$ iterations to calculate $g$ from Eq.\,(\ref{eq:gmat}) for a fixed set of parameters in each curve.}
\label{fig:time_iterations}
\end{figure}

Thus, we have developed a dynamical grid method, a 2D analog the tetrahedron method\,\cite{jepson_electronic_1971,lehmann_numerical_1972,blochl_improved_1994}, which puts a more dense integration grid around the divergencies (more generally, around regions where the gradient of the integrand is large), and ignores areas where the integrand is zero anyway. In this routine, we start with a low-density grid of polygons appropriate for the Brillouin zone (e.g., squares for a rectangular BZ and triangles for a hexagonal one). Then, we gradually split each polygon of neighboring grid points into smaller polygons where the integrand changes fast and drop all points where it is zero. For instance, starting with a $20\times20$ grid implies that the first iteration effectively computes  a $40\times40$ grid, where all grid points with zero integrand are omitted. This yields the advantage that in regions with zero integrand, we only calculate the integrand at very few points. Figure\,\ref{fig:dynamic_grid} shows the progression of the dynamic grid for the square and honeycomb lattices. After the $n$-th iteration, we achieve, hence, an effective grid size of $N_{\text{eff}}=N_{0}\times2^{n}$, where $N_{0}$ is the starting gridsize and $n$ is the number of iterations, \ie the highest density of grid points is the same as for a $N_{\text{eff}}\times N_{\text{eff}}$ grid. As Fig.\,\ref{fig:time_iterations} shows, with the dynamic grid method we can achieve the same accuracy as the standard quadrature method with an increase in speed by a factor of up to 50.

%
%
\section{Conclusion} 

We have studied the extended Hubbard model on different two-dimensional lattices within the weak-coupling renormalization group approach and investigated the unconventional superconducting ground states. 
We find a variety of higher-angular momentum superconducting phases as expected from repulsive interactions. 
By tuning not only longer-ranged hoppings, but also nonlocal electron-electron interactions, we are able to identify superconducting solutions for all irreducible representations on the square and on the hexagonal lattices.
For the square, triangular, and honeycomb lattices, we provide detailed superconducting phase diagrams as well as coupling strengths which quantify the corresponding critical temperatures 
depending on the band-structure parameters, band filling, and interaction parameters. We have also computed the gap size between the two strongest instabilities, which can be seen as a criteria for the robustness of the superconducting ground state. For large parameter spaces, we find either spin-triplet superconductivity or chiral topological superconducting phases. 
We have discussed the sensitivity of the method with respect to the numerical resolution of the integration grid and the patching scheme. Eventually, we have demonstrated how to efficiently reach a high numerical accuracy.

\begin{acknowledgments}
We acknowledge instructive discussions with R.\ Thomale, M.\ Fink, and M.\ Klett. 
SR acknowledges an ARC Future Fellowship (FT180100211).
TLS acknowledges support by the National Research Fund Luxembourg under grants ATTRACT 7556175 and INTER 11223315.
This research was undertaken using the HPC facility Spartan hosted at the University of Melbourne.
\end{acknowledgments}

%
%
\section{Appendix: Construction of lattice harmonics}

Here, we discuss the aforementioned construction of $n$-th neighbor lattice harmonics\,\cite{platt_functional_2013}. We show only the calculation for the nearest-neighbor lattice harmonics on the square lattice as an instructive example, since the calculations for higher $n$ and different lattices are analogous.

First, we write down the relative positions of the first-neighbor atoms in the lattice, \ie
\begin{equation}
 \delta\vec{r}=\{\hat{x},\hat{y},-\hat{x},-\hat{y}\},
\end{equation}
where $\hat{x}$ is the unit vector in $x$-direction. In the next step, we pick one of the vectors $\delta\vec{r}$, \eg $\hat{x}$, and let each element of the symmetry group act on it separately, where we write down the Fourier components $F(R\cdot\delta\vec{r})$ of the results, \ie
\begin{align}
 F(R_{e}\cdot\delta\vec{r})=&\hspace{2mm}e^{ik_{x}},\nonumber\\
 F(R_{c_{2}}\cdot\delta\vec{r})=&\hspace{2mm}e^{-ik_{x}},\nonumber\\
 F(R_{c_{4}}\cdot\delta\vec{r})=&\hspace{2mm}\{e^{ik_{y}},\hspace{1mm}e^{-ik_{y}}\},\nonumber\\
 F(R_{s}\cdot\delta\vec{r})=&\hspace{2mm}\{e^{ik_{x}},\hspace{1mm}e^{-ik_{x}}\},\nonumber\\
 F(R_{s'}\cdot\delta\vec{r})=&\hspace{2mm}\{e^{ik_{y}},\hspace{1mm}e^{-ik_{y}}\}.\nonumber
\end{align}
Then, the lattice harmonic of an irrep $I$ is given by the sum
\begin{equation}
 \phi_{I}^{D_{4},1}(\vec{k})=\sum_{i}\chi_{I}(R_{i})F(R_{i}\cdot\delta\vec{r}),
\end{equation}
where $\chi_{I}(R_{i})$ denotes the character of irrep $I$ and conjugacy class $R_{i}$. For $n$-dimensional irreps, we obtain $n$ linearly independent lattice harmonic by starting the calculation with $n$ linearly independent lattice vectors $\delta\vec{r}$. An orthonormal basis of lattice harmonics can then be constructed from the found $\phi$'s.

The orthonormality of the lattice harmonics can be checked using the scalar product defined on the Brillouin zone (BZ) by:
\begin{equation}
 \left<\phi_{1}(\vec{k}),\phi_{2}(\vec{k})\right>=\frac{1}{V_{\text{BZ}}}\int_{\text{BZ}}\phi_{1}^{\ps}(\vec{k})\phi_{2}^{*}(\vec{k})\text{d}V,
\end{equation}
where $V_{\text{BZ}}$ is the volume of the BZ.

\bibliography{wcrg1}

\end{document}